\documentclass[longauth]{aa}

\usepackage{graphicx}
\usepackage{natbib}
\usepackage{scalerel}
\usepackage[table]{xcolor}
\usepackage{wrapfig}

\usepackage[table]{xcolor}

\usepackage{txfonts}
\usepackage{hyperref}
\hypersetup{
    colorlinks=true,
    linkcolor=blue,
    filecolor=magenta,      
    urlcolor=blue,
    citecolor=blue
}
\makeatletter
\renewcommand*\aa@pageof{, page \thepage{} of \pageref*{LastPage}}
\makeatother

\usepackage[utf8]{inputenc}
\usepackage[modulo, switch]{lineno}

\usepackage{orcidlink}
\usepackage{euclid}
\newcommand{\lensmc}{\textsc{LensMC}\xspace}

\newcommand{\orcid}[1]{\orcidlink{#1}}

\begin{document}

%
%
\title{\Euclid preparation}
\subtitle{LXVII. Deep learning true galaxy morphologies for weak lensing shear bias calibration}

\author{Euclid Collaboration: B.~Csizi\orcid{0000-0003-3227-6581}\thanks{\email{benjamin.csizi@uibk.ac.at}}\inst{\ref{aff1}}
\and T.~Schrabback\orcid{0000-0002-6987-7834}\inst{\ref{aff1}}
\and S.~Grandis\orcid{0000-0002-4577-8217}\inst{\ref{aff1}}
\and H.~Hoekstra\orcid{0000-0002-0641-3231}\inst{\ref{aff2}}
\and H.~Jansen\orcid{0009-0002-1332-7742}\inst{\ref{aff1}}
\and L.~Linke\orcid{0000-0002-2622-8113}\inst{\ref{aff1}}
\and G.~Congedo\orcid{0000-0003-2508-0046}\inst{\ref{aff3}}
\and A.~N.~Taylor\inst{\ref{aff3}}
\and A.~Amara\inst{\ref{aff4}}
\and S.~Andreon\orcid{0000-0002-2041-8784}\inst{\ref{aff5}}
\and C.~Baccigalupi\orcid{0000-0002-8211-1630}\inst{\ref{aff6},\ref{aff7},\ref{aff8},\ref{aff9}}
\and M.~Baldi\orcid{0000-0003-4145-1943}\inst{\ref{aff10},\ref{aff11},\ref{aff12}}
\and S.~Bardelli\orcid{0000-0002-8900-0298}\inst{\ref{aff11}}
\and P.~Battaglia\orcid{0000-0002-7337-5909}\inst{\ref{aff11}}
\and R.~Bender\orcid{0000-0001-7179-0626}\inst{\ref{aff13},\ref{aff14}}
\and C.~Bodendorf\inst{\ref{aff13}}
\and D.~Bonino\orcid{0000-0002-3336-9977}\inst{\ref{aff15}}
\and E.~Branchini\orcid{0000-0002-0808-6908}\inst{\ref{aff16},\ref{aff17},\ref{aff5}}
\and M.~Brescia\orcid{0000-0001-9506-5680}\inst{\ref{aff18},\ref{aff19},\ref{aff20}}
\and J.~Brinchmann\orcid{0000-0003-4359-8797}\inst{\ref{aff21},\ref{aff22}}
\and S.~Camera\orcid{0000-0003-3399-3574}\inst{\ref{aff23},\ref{aff24},\ref{aff15}}
\and V.~Capobianco\orcid{0000-0002-3309-7692}\inst{\ref{aff15}}
\and C.~Carbone\orcid{0000-0003-0125-3563}\inst{\ref{aff25}}
\and J.~Carretero\orcid{0000-0002-3130-0204}\inst{\ref{aff26},\ref{aff27}}
\and S.~Casas\orcid{0000-0002-4751-5138}\inst{\ref{aff28}}
\and F.~J.~Castander\orcid{0000-0001-7316-4573}\inst{\ref{aff29},\ref{aff30}}
\and M.~Castellano\orcid{0000-0001-9875-8263}\inst{\ref{aff31}}
\and G.~Castignani\orcid{0000-0001-6831-0687}\inst{\ref{aff11}}
\and S.~Cavuoti\orcid{0000-0002-3787-4196}\inst{\ref{aff19},\ref{aff20}}
\and A.~Cimatti\inst{\ref{aff32}}
\and C.~Colodro-Conde\inst{\ref{aff33}}
\and C.~J.~Conselice\orcid{0000-0003-1949-7638}\inst{\ref{aff34}}
\and L.~Conversi\orcid{0000-0002-6710-8476}\inst{\ref{aff35},\ref{aff36}}
\and Y.~Copin\orcid{0000-0002-5317-7518}\inst{\ref{aff37}}
\and F.~Courbin\orcid{0000-0003-0758-6510}\inst{\ref{aff38},\ref{aff39},\ref{aff40}}
\and H.~M.~Courtois\orcid{0000-0003-0509-1776}\inst{\ref{aff41}}
\and M.~Cropper\orcid{0000-0003-4571-9468}\inst{\ref{aff42}}
\and A.~Da~Silva\orcid{0000-0002-6385-1609}\inst{\ref{aff43},\ref{aff44}}
\and H.~Degaudenzi\orcid{0000-0002-5887-6799}\inst{\ref{aff45}}
\and G.~De~Lucia\orcid{0000-0002-6220-9104}\inst{\ref{aff7}}
\and J.~Dinis\orcid{0000-0001-5075-1601}\inst{\ref{aff43},\ref{aff44}}
\and M.~Douspis\orcid{0000-0003-4203-3954}\inst{\ref{aff46}}
\and F.~Dubath\orcid{0000-0002-6533-2810}\inst{\ref{aff45}}
\and X.~Dupac\inst{\ref{aff36}}
\and S.~Dusini\orcid{0000-0002-1128-0664}\inst{\ref{aff47}}
\and M.~Farina\orcid{0000-0002-3089-7846}\inst{\ref{aff48}}
\and S.~Farrens\orcid{0000-0002-9594-9387}\inst{\ref{aff49}}
\and F.~Faustini\orcid{0000-0001-6274-5145}\inst{\ref{aff50},\ref{aff31}}
\and S.~Ferriol\inst{\ref{aff37}}
\and S.~Fotopoulou\orcid{0000-0002-9686-254X}\inst{\ref{aff51}}
\and M.~Frailis\orcid{0000-0002-7400-2135}\inst{\ref{aff7}}
\and E.~Franceschi\orcid{0000-0002-0585-6591}\inst{\ref{aff11}}
\and S.~Galeotta\orcid{0000-0002-3748-5115}\inst{\ref{aff7}}
\and B.~Gillis\orcid{0000-0002-4478-1270}\inst{\ref{aff3}}
\and C.~Giocoli\orcid{0000-0002-9590-7961}\inst{\ref{aff11},\ref{aff52}}
\and A.~Grazian\orcid{0000-0002-5688-0663}\inst{\ref{aff53}}
\and F.~Grupp\inst{\ref{aff13},\ref{aff14}}
\and L.~Guzzo\orcid{0000-0001-8264-5192}\inst{\ref{aff54},\ref{aff5}}
\and S.~V.~H.~Haugan\orcid{0000-0001-9648-7260}\inst{\ref{aff55}}
\and W.~Holmes\inst{\ref{aff56}}
\and I.~Hook\orcid{0000-0002-2960-978X}\inst{\ref{aff57}}
\and F.~Hormuth\inst{\ref{aff58}}
\and A.~Hornstrup\orcid{0000-0002-3363-0936}\inst{\ref{aff59},\ref{aff60}}
\and P.~Hudelot\inst{\ref{aff61}}
\and S.~Ili\'c\orcid{0000-0003-4285-9086}\inst{\ref{aff62},\ref{aff63}}
\and K.~Jahnke\orcid{0000-0003-3804-2137}\inst{\ref{aff64}}
\and M.~Jhabvala\inst{\ref{aff65}}
\and B.~Joachimi\orcid{0000-0001-7494-1303}\inst{\ref{aff66}}
\and E.~Keih\"anen\orcid{0000-0003-1804-7715}\inst{\ref{aff67}}
\and S.~Kermiche\orcid{0000-0002-0302-5735}\inst{\ref{aff68}}
\and A.~Kiessling\orcid{0000-0002-2590-1273}\inst{\ref{aff56}}
\and M.~Kilbinger\orcid{0000-0001-9513-7138}\inst{\ref{aff49}}
\and B.~Kubik\orcid{0009-0006-5823-4880}\inst{\ref{aff37}}
\and K.~Kuijken\orcid{0000-0002-3827-0175}\inst{\ref{aff2}}
\and M.~K\"ummel\orcid{0000-0003-2791-2117}\inst{\ref{aff14}}
\and M.~Kunz\orcid{0000-0002-3052-7394}\inst{\ref{aff69}}
\and H.~Kurki-Suonio\orcid{0000-0002-4618-3063}\inst{\ref{aff70},\ref{aff71}}
\and S.~Ligori\orcid{0000-0003-4172-4606}\inst{\ref{aff15}}
\and P.~B.~Lilje\orcid{0000-0003-4324-7794}\inst{\ref{aff55}}
\and V.~Lindholm\orcid{0000-0003-2317-5471}\inst{\ref{aff70},\ref{aff71}}
\and I.~Lloro\inst{\ref{aff72}}
\and D.~Maino\inst{\ref{aff54},\ref{aff25},\ref{aff73}}
\and E.~Maiorano\orcid{0000-0003-2593-4355}\inst{\ref{aff11}}
\and O.~Mansutti\orcid{0000-0001-5758-4658}\inst{\ref{aff7}}
\and S.~Marcin\inst{\ref{aff74}}
\and O.~Marggraf\orcid{0000-0001-7242-3852}\inst{\ref{aff75}}
\and K.~Markovic\orcid{0000-0001-6764-073X}\inst{\ref{aff56}}
\and M.~Martinelli\orcid{0000-0002-6943-7732}\inst{\ref{aff31},\ref{aff76}}
\and N.~Martinet\orcid{0000-0003-2786-7790}\inst{\ref{aff77}}
\and F.~Marulli\orcid{0000-0002-8850-0303}\inst{\ref{aff78},\ref{aff11},\ref{aff12}}
\and R.~Massey\orcid{0000-0002-6085-3780}\inst{\ref{aff79}}
\and E.~Medinaceli\orcid{0000-0002-4040-7783}\inst{\ref{aff11}}
\and S.~Mei\orcid{0000-0002-2849-559X}\inst{\ref{aff80}}
\and M.~Melchior\inst{\ref{aff74}}
\and Y.~Mellier\inst{\ref{aff81},\ref{aff61}}
\and M.~Meneghetti\orcid{0000-0003-1225-7084}\inst{\ref{aff11},\ref{aff12}}
\and G.~Meylan\inst{\ref{aff38}}
\and M.~Moresco\orcid{0000-0002-7616-7136}\inst{\ref{aff78},\ref{aff11}}
\and L.~Moscardini\orcid{0000-0002-3473-6716}\inst{\ref{aff78},\ref{aff11},\ref{aff12}}
\and S.-M.~Niemi\inst{\ref{aff82}}
\and C.~Padilla\orcid{0000-0001-7951-0166}\inst{\ref{aff83}}
\and S.~Paltani\orcid{0000-0002-8108-9179}\inst{\ref{aff45}}
\and F.~Pasian\orcid{0000-0002-4869-3227}\inst{\ref{aff7}}
\and K.~Pedersen\inst{\ref{aff84}}
\and V.~Pettorino\inst{\ref{aff82}}
\and S.~Pires\orcid{0000-0002-0249-2104}\inst{\ref{aff49}}
\and G.~Polenta\orcid{0000-0003-4067-9196}\inst{\ref{aff50}}
\and M.~Poncet\inst{\ref{aff85}}
\and L.~A.~Popa\inst{\ref{aff86}}
\and F.~Raison\orcid{0000-0002-7819-6918}\inst{\ref{aff13}}
\and A.~Renzi\orcid{0000-0001-9856-1970}\inst{\ref{aff87},\ref{aff47}}
\and J.~Rhodes\orcid{0000-0002-4485-8549}\inst{\ref{aff56}}
\and G.~Riccio\inst{\ref{aff19}}
\and E.~Romelli\orcid{0000-0003-3069-9222}\inst{\ref{aff7}}
\and M.~Roncarelli\orcid{0000-0001-9587-7822}\inst{\ref{aff11}}
\and E.~Rossetti\orcid{0000-0003-0238-4047}\inst{\ref{aff10}}
\and R.~Saglia\orcid{0000-0003-0378-7032}\inst{\ref{aff14},\ref{aff13}}
\and Z.~Sakr\orcid{0000-0002-4823-3757}\inst{\ref{aff88},\ref{aff63},\ref{aff89}}
\and A.~G.~S\'anchez\orcid{0000-0003-1198-831X}\inst{\ref{aff13}}
\and B.~Sartoris\orcid{0000-0003-1337-5269}\inst{\ref{aff14},\ref{aff7}}
\and P.~Schneider\orcid{0000-0001-8561-2679}\inst{\ref{aff75}}
\and A.~Secroun\orcid{0000-0003-0505-3710}\inst{\ref{aff68}}
\and G.~Seidel\orcid{0000-0003-2907-353X}\inst{\ref{aff64}}
\and S.~Serrano\orcid{0000-0002-0211-2861}\inst{\ref{aff30},\ref{aff90},\ref{aff29}}
\and C.~Sirignano\orcid{0000-0002-0995-7146}\inst{\ref{aff87},\ref{aff47}}
\and G.~Sirri\orcid{0000-0003-2626-2853}\inst{\ref{aff12}}
\and L.~Stanco\orcid{0000-0002-9706-5104}\inst{\ref{aff47}}
\and J.~Steinwagner\orcid{0000-0001-7443-1047}\inst{\ref{aff13}}
\and P.~Tallada-Cresp\'{i}\orcid{0000-0002-1336-8328}\inst{\ref{aff26},\ref{aff27}}
\and D.~Tavagnacco\orcid{0000-0001-7475-9894}\inst{\ref{aff7}}
\and H.~I.~Teplitz\orcid{0000-0002-7064-5424}\inst{\ref{aff91}}
\and I.~Tereno\inst{\ref{aff43},\ref{aff92}}
\and R.~Toledo-Moreo\orcid{0000-0002-2997-4859}\inst{\ref{aff93}}
\and F.~Torradeflot\orcid{0000-0003-1160-1517}\inst{\ref{aff27},\ref{aff26}}
\and I.~Tutusaus\orcid{0000-0002-3199-0399}\inst{\ref{aff63}}
\and E.~A.~Valentijn\inst{\ref{aff94}}
\and L.~Valenziano\orcid{0000-0002-1170-0104}\inst{\ref{aff11},\ref{aff95}}
\and T.~Vassallo\orcid{0000-0001-6512-6358}\inst{\ref{aff14},\ref{aff7}}
\and G.~Verdoes~Kleijn\orcid{0000-0001-5803-2580}\inst{\ref{aff94}}
\and A.~Veropalumbo\orcid{0000-0003-2387-1194}\inst{\ref{aff5},\ref{aff17},\ref{aff96}}
\and Y.~Wang\orcid{0000-0002-4749-2984}\inst{\ref{aff91}}
\and J.~Weller\orcid{0000-0002-8282-2010}\inst{\ref{aff14},\ref{aff13}}
\and G.~Zamorani\orcid{0000-0002-2318-301X}\inst{\ref{aff11}}
\and E.~Zucca\orcid{0000-0002-5845-8132}\inst{\ref{aff11}}
\and A.~Biviano\orcid{0000-0002-0857-0732}\inst{\ref{aff7},\ref{aff6}}
\and M.~Bolzonella\orcid{0000-0003-3278-4607}\inst{\ref{aff11}}
\and E.~Bozzo\orcid{0000-0002-8201-1525}\inst{\ref{aff45}}
\and C.~Burigana\orcid{0000-0002-3005-5796}\inst{\ref{aff97},\ref{aff95}}
\and M.~Calabrese\orcid{0000-0002-2637-2422}\inst{\ref{aff98},\ref{aff25}}
\and D.~Di~Ferdinando\inst{\ref{aff12}}
\and J.~A.~Escartin~Vigo\inst{\ref{aff13}}
\and R.~Farinelli\inst{\ref{aff11}}
\and J.~Gracia-Carpio\inst{\ref{aff13}}
\and S.~Matthew\orcid{0000-0001-8448-1697}\inst{\ref{aff3}}
\and N.~Mauri\orcid{0000-0001-8196-1548}\inst{\ref{aff32},\ref{aff12}}
\and A.~Pezzotta\orcid{0000-0003-0726-2268}\inst{\ref{aff13}}
\and M.~P\"ontinen\orcid{0000-0001-5442-2530}\inst{\ref{aff70}}
\and V.~Scottez\inst{\ref{aff81},\ref{aff99}}
\and M.~Tenti\orcid{0000-0002-4254-5901}\inst{\ref{aff12}}
\and M.~Viel\orcid{0000-0002-2642-5707}\inst{\ref{aff6},\ref{aff7},\ref{aff9},\ref{aff8},\ref{aff100}}
\and M.~Wiesmann\orcid{0009-0000-8199-5860}\inst{\ref{aff55}}
\and Y.~Akrami\orcid{0000-0002-2407-7956}\inst{\ref{aff101},\ref{aff102}}
\and V.~Allevato\orcid{0000-0001-7232-5152}\inst{\ref{aff19}}
\and S.~Anselmi\orcid{0000-0002-3579-9583}\inst{\ref{aff47},\ref{aff87},\ref{aff103}}
\and M.~Archidiacono\orcid{0000-0003-4952-9012}\inst{\ref{aff54},\ref{aff73}}
\and F.~Atrio-Barandela\orcid{0000-0002-2130-2513}\inst{\ref{aff104}}
\and M.~Ballardini\orcid{0000-0003-4481-3559}\inst{\ref{aff105},\ref{aff11},\ref{aff106}}
\and A.~Blanchard\orcid{0000-0001-8555-9003}\inst{\ref{aff63}}
\and L.~Blot\orcid{0000-0002-9622-7167}\inst{\ref{aff107},\ref{aff103}}
\and S.~Borgani\orcid{0000-0001-6151-6439}\inst{\ref{aff108},\ref{aff6},\ref{aff7},\ref{aff8}}
\and S.~Bruton\orcid{0000-0002-6503-5218}\inst{\ref{aff109}}
\and R.~Cabanac\orcid{0000-0001-6679-2600}\inst{\ref{aff63}}
\and A.~Calabro\orcid{0000-0003-2536-1614}\inst{\ref{aff31}}
\and G.~Ca\~nas-Herrera\orcid{0000-0003-2796-2149}\inst{\ref{aff82},\ref{aff110}}
\and A.~Cappi\inst{\ref{aff11},\ref{aff111}}
\and F.~Caro\inst{\ref{aff31}}
\and C.~S.~Carvalho\inst{\ref{aff92}}
\and T.~Castro\orcid{0000-0002-6292-3228}\inst{\ref{aff7},\ref{aff8},\ref{aff6},\ref{aff100}}
\and K.~C.~Chambers\orcid{0000-0001-6965-7789}\inst{\ref{aff112}}
\and S.~Contarini\orcid{0000-0002-9843-723X}\inst{\ref{aff13}}
\and A.~R.~Cooray\orcid{0000-0002-3892-0190}\inst{\ref{aff113}}
\and G.~Desprez\orcid{0000-0001-8325-1742}\inst{\ref{aff114}}
\and A.~D\'iaz-S\'anchez\orcid{0000-0003-0748-4768}\inst{\ref{aff115}}
\and J.~J.~Diaz\inst{\ref{aff116}}
\and S.~Di~Domizio\orcid{0000-0003-2863-5895}\inst{\ref{aff16},\ref{aff17}}
\and H.~Dole\orcid{0000-0002-9767-3839}\inst{\ref{aff46}}
\and S.~Escoffier\orcid{0000-0002-2847-7498}\inst{\ref{aff68}}
\and A.~G.~Ferrari\orcid{0009-0005-5266-4110}\inst{\ref{aff32},\ref{aff12}}
\and P.~G.~Ferreira\orcid{0000-0002-3021-2851}\inst{\ref{aff117}}
\and I.~Ferrero\orcid{0000-0002-1295-1132}\inst{\ref{aff55}}
\and A.~Finoguenov\orcid{0000-0002-4606-5403}\inst{\ref{aff70}}
\and A.~Fontana\orcid{0000-0003-3820-2823}\inst{\ref{aff31}}
\and F.~Fornari\orcid{0000-0003-2979-6738}\inst{\ref{aff95}}
\and L.~Gabarra\orcid{0000-0002-8486-8856}\inst{\ref{aff117}}
\and K.~Ganga\orcid{0000-0001-8159-8208}\inst{\ref{aff80}}
\and J.~Garc\'ia-Bellido\orcid{0000-0002-9370-8360}\inst{\ref{aff101}}
\and T.~Gasparetto\orcid{0000-0002-7913-4866}\inst{\ref{aff7}}
\and E.~Gaztanaga\orcid{0000-0001-9632-0815}\inst{\ref{aff29},\ref{aff30},\ref{aff118}}
\and F.~Giacomini\orcid{0000-0002-3129-2814}\inst{\ref{aff12}}
\and F.~Gianotti\orcid{0000-0003-4666-119X}\inst{\ref{aff11}}
\and G.~Gozaliasl\orcid{0000-0002-0236-919X}\inst{\ref{aff119},\ref{aff70}}
\and C.~M.~Gutierrez\orcid{0000-0001-7854-783X}\inst{\ref{aff120}}
\and A.~Hall\orcid{0000-0002-3139-8651}\inst{\ref{aff3}}
\and H.~Hildebrandt\orcid{0000-0002-9814-3338}\inst{\ref{aff121}}
\and J.~Hjorth\orcid{0000-0002-4571-2306}\inst{\ref{aff84}}
\and A.~Jimenez~Mu\~noz\orcid{0009-0004-5252-185X}\inst{\ref{aff122}}
\and S.~Joudaki\orcid{0000-0001-8820-673X}\inst{\ref{aff118}}
\and J.~J.~E.~Kajava\orcid{0000-0002-3010-8333}\inst{\ref{aff123},\ref{aff124}}
\and V.~Kansal\orcid{0000-0002-4008-6078}\inst{\ref{aff125},\ref{aff126}}
\and D.~Karagiannis\orcid{0000-0002-4927-0816}\inst{\ref{aff127},\ref{aff128}}
\and C.~C.~Kirkpatrick\inst{\ref{aff67}}
\and A.~M.~C.~Le~Brun\orcid{0000-0002-0936-4594}\inst{\ref{aff103}}
\and J.~Le~Graet\orcid{0000-0001-6523-7971}\inst{\ref{aff68}}
\and L.~Legrand\orcid{0000-0003-0610-5252}\inst{\ref{aff129}}
\and J.~Lesgourgues\orcid{0000-0001-7627-353X}\inst{\ref{aff28}}
\and T.~I.~Liaudat\orcid{0000-0002-9104-314X}\inst{\ref{aff130}}
\and A.~Loureiro\orcid{0000-0002-4371-0876}\inst{\ref{aff131},\ref{aff132}}
\and J.~Macias-Perez\orcid{0000-0002-5385-2763}\inst{\ref{aff122}}
\and G.~Maggio\orcid{0000-0003-4020-4836}\inst{\ref{aff7}}
\and M.~Magliocchetti\orcid{0000-0001-9158-4838}\inst{\ref{aff48}}
\and C.~Mancini\orcid{0000-0002-4297-0561}\inst{\ref{aff25}}
\and F.~Mannucci\orcid{0000-0002-4803-2381}\inst{\ref{aff133}}
\and R.~Maoli\orcid{0000-0002-6065-3025}\inst{\ref{aff134},\ref{aff31}}
\and J.~Mart\'{i}n-Fleitas\orcid{0000-0002-8594-569X}\inst{\ref{aff135}}
\and C.~J.~A.~P.~Martins\orcid{0000-0002-4886-9261}\inst{\ref{aff136},\ref{aff21}}
\and L.~Maurin\orcid{0000-0002-8406-0857}\inst{\ref{aff46}}
\and R.~B.~Metcalf\orcid{0000-0003-3167-2574}\inst{\ref{aff78},\ref{aff11}}
\and M.~Miluzio\inst{\ref{aff36},\ref{aff137}}
\and P.~Monaco\orcid{0000-0003-2083-7564}\inst{\ref{aff108},\ref{aff7},\ref{aff8},\ref{aff6}}
\and A.~Montoro\orcid{0000-0003-4730-8590}\inst{\ref{aff29},\ref{aff30}}
\and A.~Mora\orcid{0000-0002-1922-8529}\inst{\ref{aff135}}
\and C.~Moretti\orcid{0000-0003-3314-8936}\inst{\ref{aff9},\ref{aff100},\ref{aff7},\ref{aff6},\ref{aff8}}
\and G.~Morgante\inst{\ref{aff11}}
\and Nicholas~A.~Walton\orcid{0000-0003-3983-8778}\inst{\ref{aff138}}
\and L.~Pagano\orcid{0000-0003-1820-5998}\inst{\ref{aff105},\ref{aff106}}
\and L.~Patrizii\inst{\ref{aff12}}
\and V.~Popa\orcid{0000-0002-9118-8330}\inst{\ref{aff86}}
\and D.~Potter\orcid{0000-0002-0757-5195}\inst{\ref{aff139}}
\and I.~Risso\orcid{0000-0003-2525-7761}\inst{\ref{aff140}}
\and P.-F.~Rocci\inst{\ref{aff46}}
\and M.~Sahl\'en\orcid{0000-0003-0973-4804}\inst{\ref{aff141}}
\and E.~Sarpa\orcid{0000-0002-1256-655X}\inst{\ref{aff9},\ref{aff100},\ref{aff8}}
\and A.~Schneider\orcid{0000-0001-7055-8104}\inst{\ref{aff139}}
\and M.~Sereno\orcid{0000-0003-0302-0325}\inst{\ref{aff11},\ref{aff12}}
\and P.~Simon\inst{\ref{aff75}}
\and A.~Spurio~Mancini\orcid{0000-0001-5698-0990}\inst{\ref{aff142},\ref{aff42}}
\and J.~Stadel\orcid{0000-0001-7565-8622}\inst{\ref{aff139}}
\and K.~Tanidis\inst{\ref{aff117}}
\and C.~Tao\orcid{0000-0001-7961-8177}\inst{\ref{aff68}}
\and N.~Tessore\orcid{0000-0002-9696-7931}\inst{\ref{aff66}}
\and G.~Testera\inst{\ref{aff17}}
\and R.~Teyssier\orcid{0000-0001-7689-0933}\inst{\ref{aff143}}
\and S.~Toft\orcid{0000-0003-3631-7176}\inst{\ref{aff60},\ref{aff145}}
\and S.~Tosi\orcid{0000-0002-7275-9193}\inst{\ref{aff16},\ref{aff17}}
\and A.~Troja\orcid{0000-0003-0239-4595}\inst{\ref{aff87},\ref{aff47}}
\and M.~Tucci\inst{\ref{aff45}}
\and C.~Valieri\inst{\ref{aff12}}
\and J.~Valiviita\orcid{0000-0001-6225-3693}\inst{\ref{aff70},\ref{aff71}}
\and D.~Vergani\orcid{0000-0003-0898-2216}\inst{\ref{aff11}}
\and G.~Verza\orcid{0000-0002-1886-8348}\inst{\ref{aff146},\ref{aff147}}
\and P.~Vielzeuf\orcid{0000-0003-2035-9339}\inst{\ref{aff68}}}
										   
\institute{Universit\"at Innsbruck, Institut f\"ur Astro- und Teilchenphysik, Technikerstr. 25/8, 6020 Innsbruck, Austria\label{aff1}
\and
Leiden Observatory, Leiden University, Einsteinweg 55, 2333 CC Leiden, The Netherlands\label{aff2}
\and
Institute for Astronomy, University of Edinburgh, Royal Observatory, Blackford Hill, Edinburgh EH9 3HJ, UK\label{aff3}
\and
School of Mathematics and Physics, University of Surrey, Guildford, Surrey, GU2 7XH, UK\label{aff4}
\and
INAF-Osservatorio Astronomico di Brera, Via Brera 28, 20122 Milano, Italy\label{aff5}
\and
IFPU, Institute for Fundamental Physics of the Universe, via Beirut 2, 34151 Trieste, Italy\label{aff6}
\and
INAF-Osservatorio Astronomico di Trieste, Via G. B. Tiepolo 11, 34143 Trieste, Italy\label{aff7}
\and
INFN, Sezione di Trieste, Via Valerio 2, 34127 Trieste TS, Italy\label{aff8}
\and
SISSA, International School for Advanced Studies, Via Bonomea 265, 34136 Trieste TS, Italy\label{aff9}
\and
Dipartimento di Fisica e Astronomia, Universit\`a di Bologna, Via Gobetti 93/2, 40129 Bologna, Italy\label{aff10}
\and
INAF-Osservatorio di Astrofisica e Scienza dello Spazio di Bologna, Via Piero Gobetti 93/3, 40129 Bologna, Italy\label{aff11}
\and
INFN-Sezione di Bologna, Viale Berti Pichat 6/2, 40127 Bologna, Italy\label{aff12}
\and
Max Planck Institute for Extraterrestrial Physics, Giessenbachstr. 1, 85748 Garching, Germany\label{aff13}
\and
Universit\"ats-Sternwarte M\"unchen, Fakult\"at f\"ur Physik, Ludwig-Maximilians-Universit\"at M\"unchen, Scheinerstrasse 1, 81679 M\"unchen, Germany\label{aff14}
\and
INAF-Osservatorio Astrofisico di Torino, Via Osservatorio 20, 10025 Pino Torinese (TO), Italy\label{aff15}
\and
Dipartimento di Fisica, Universit\`a di Genova, Via Dodecaneso 33, 16146, Genova, Italy\label{aff16}
\and
INFN-Sezione di Genova, Via Dodecaneso 33, 16146, Genova, Italy\label{aff17}
\and
Department of Physics "E. Pancini", University Federico II, Via Cinthia 6, 80126, Napoli, Italy\label{aff18}
\and
INAF-Osservatorio Astronomico di Capodimonte, Via Moiariello 16, 80131 Napoli, Italy\label{aff19}
\and
INFN section of Naples, Via Cinthia 6, 80126, Napoli, Italy\label{aff20}
\and
Instituto de Astrof\'isica e Ci\^encias do Espa\c{c}o, Universidade do Porto, CAUP, Rua das Estrelas, PT4150-762 Porto, Portugal\label{aff21}
\and
Faculdade de Ci\^encias da Universidade do Porto, Rua do Campo de Alegre, 4150-007 Porto, Portugal\label{aff22}
\and
Dipartimento di Fisica, Universit\`a degli Studi di Torino, Via P. Giuria 1, 10125 Torino, Italy\label{aff23}
\and
INFN-Sezione di Torino, Via P. Giuria 1, 10125 Torino, Italy\label{aff24}
\and
INAF-IASF Milano, Via Alfonso Corti 12, 20133 Milano, Italy\label{aff25}
\and
Centro de Investigaciones Energ\'eticas, Medioambientales y Tecnol\'ogicas (CIEMAT), Avenida Complutense 40, 28040 Madrid, Spain\label{aff26}
\and
Port d'Informaci\'{o} Cient\'{i}fica, Campus UAB, C. Albareda s/n, 08193 Bellaterra (Barcelona), Spain\label{aff27}
\and
Institute for Theoretical Particle Physics and Cosmology (TTK), RWTH Aachen University, 52056 Aachen, Germany\label{aff28}
\and
Institute of Space Sciences (ICE, CSIC), Campus UAB, Carrer de Can Magrans, s/n, 08193 Barcelona, Spain\label{aff29}
\and
Institut d'Estudis Espacials de Catalunya (IEEC),  Edifici RDIT, Campus UPC, 08860 Castelldefels, Barcelona, Spain\label{aff30}
\and
INAF-Osservatorio Astronomico di Roma, Via Frascati 33, 00078 Monteporzio Catone, Italy\label{aff31}
\and
Dipartimento di Fisica e Astronomia "Augusto Righi" - Alma Mater Studiorum Universit\`a di Bologna, Viale Berti Pichat 6/2, 40127 Bologna, Italy\label{aff32}
\and
Instituto de Astrof\'{\i}sica de Canarias, V\'{\i}a L\'actea, 38205 La Laguna, Tenerife, Spain\label{aff33}
\and
Jodrell Bank Centre for Astrophysics, Department of Physics and Astronomy, University of Manchester, Oxford Road, Manchester M13 9PL, UK\label{aff34}
\and
European Space Agency/ESRIN, Largo Galileo Galilei 1, 00044 Frascati, Roma, Italy\label{aff35}
\and
ESAC/ESA, Camino Bajo del Castillo, s/n., Urb. Villafranca del Castillo, 28692 Villanueva de la Ca\~nada, Madrid, Spain\label{aff36}
\and
Universit\'e Claude Bernard Lyon 1, CNRS/IN2P3, IP2I Lyon, UMR 5822, Villeurbanne, F-69100, France\label{aff37}
\and
Institute of Physics, Laboratory of Astrophysics, Ecole Polytechnique F\'ed\'erale de Lausanne (EPFL), Observatoire de Sauverny, 1290 Versoix, Switzerland\label{aff38}
\and
Institut de Ci\`{e}ncies del Cosmos (ICCUB), Universitat de Barcelona (IEEC-UB), Mart\'{i} i Franqu\`{e}s 1, 08028 Barcelona, Spain\label{aff39}
\and
Instituci\'o Catalana de Recerca i Estudis Avan\c{c}ats (ICREA), Passeig de Llu\'{\i}s Companys 23, 08010 Barcelona, Spain\label{aff40}
\and
UCB Lyon 1, CNRS/IN2P3, IUF, IP2I Lyon, 4 rue Enrico Fermi, 69622 Villeurbanne, France\label{aff41}
\and
Mullard Space Science Laboratory, University College London, Holmbury St Mary, Dorking, Surrey RH5 6NT, UK\label{aff42}
\and
Departamento de F\'isica, Faculdade de Ci\^encias, Universidade de Lisboa, Edif\'icio C8, Campo Grande, PT1749-016 Lisboa, Portugal\label{aff43}
\and
Instituto de Astrof\'isica e Ci\^encias do Espa\c{c}o, Faculdade de Ci\^encias, Universidade de Lisboa, Campo Grande, 1749-016 Lisboa, Portugal\label{aff44}
\and
Department of Astronomy, University of Geneva, ch. d'Ecogia 16, 1290 Versoix, Switzerland\label{aff45}
\and
Universit\'e Paris-Saclay, CNRS, Institut d'astrophysique spatiale, 91405, Orsay, France\label{aff46}
\and
INFN-Padova, Via Marzolo 8, 35131 Padova, Italy\label{aff47}
\and
INAF-Istituto di Astrofisica e Planetologia Spaziali, via del Fosso del Cavaliere, 100, 00100 Roma, Italy\label{aff48}
\and
Universit\'e Paris-Saclay, Universit\'e Paris Cit\'e, CEA, CNRS, AIM, 91191, Gif-sur-Yvette, France\label{aff49}
\and
Space Science Data Center, Italian Space Agency, via del Politecnico snc, 00133 Roma, Italy\label{aff50}
\and
School of Physics, HH Wills Physics Laboratory, University of Bristol, Tyndall Avenue, Bristol, BS8 1TL, UK\label{aff51}
\and
Istituto Nazionale di Fisica Nucleare, Sezione di Bologna, Via Irnerio 46, 40126 Bologna, Italy\label{aff52}
\and
INAF-Osservatorio Astronomico di Padova, Via dell'Osservatorio 5, 35122 Padova, Italy\label{aff53}
\and
Dipartimento di Fisica "Aldo Pontremoli", Universit\`a degli Studi di Milano, Via Celoria 16, 20133 Milano, Italy\label{aff54}
\and
Institute of Theoretical Astrophysics, University of Oslo, P.O. Box 1029 Blindern, 0315 Oslo, Norway\label{aff55}
\and
Jet Propulsion Laboratory, California Institute of Technology, 4800 Oak Grove Drive, Pasadena, CA, 91109, USA\label{aff56}
\and
Department of Physics, Lancaster University, Lancaster, LA1 4YB, UK\label{aff57}
\and
Felix Hormuth Engineering, Goethestr. 17, 69181 Leimen, Germany\label{aff58}
\and
Technical University of Denmark, Elektrovej 327, 2800 Kgs. Lyngby, Denmark\label{aff59}
\and
Cosmic Dawn Center (DAWN), Copenhagen, Denmark\label{aff60}
\and
Institut d'Astrophysique de Paris, UMR 7095, CNRS, and Sorbonne Universit\'e, 98 bis boulevard Arago, 75014 Paris, France\label{aff61}
\and
Universit\'e Paris-Saclay, CNRS/IN2P3, IJCLab, 91405 Orsay, France\label{aff62}
\and
Institut de Recherche en Astrophysique et Plan\'etologie (IRAP), Universit\'e de Toulouse, CNRS, UPS, CNES, 14 Av. Edouard Belin, 31400 Toulouse, France\label{aff63}
\and
Max-Planck-Institut f\"ur Astronomie, K\"onigstuhl 17, 69117 Heidelberg, Germany\label{aff64}
\and
NASA Goddard Space Flight Center, Greenbelt, MD 20771, USA\label{aff65}
\and
Department of Physics and Astronomy, University College London, Gower Street, London WC1E 6BT, UK\label{aff66}
\and
Department of Physics and Helsinki Institute of Physics, Gustaf H\"allstr\"omin katu 2, 00014 University of Helsinki, Finland\label{aff67}
\and
Aix-Marseille Universit\'e, CNRS/IN2P3, CPPM, Marseille, France\label{aff68}
\and
Universit\'e de Gen\`eve, D\'epartement de Physique Th\'eorique and Centre for Astroparticle Physics, 24 quai Ernest-Ansermet, CH-1211 Gen\`eve 4, Switzerland\label{aff69}
\and
Department of Physics, P.O. Box 64, 00014 University of Helsinki, Finland\label{aff70}
\and
Helsinki Institute of Physics, Gustaf H{\"a}llstr{\"o}min katu 2, University of Helsinki, Helsinki, Finland\label{aff71}
\and
NOVA optical infrared instrumentation group at ASTRON, Oude Hoogeveensedijk 4, 7991PD, Dwingeloo, The Netherlands\label{aff72}
\and
INFN-Sezione di Milano, Via Celoria 16, 20133 Milano, Italy\label{aff73}
\and
University of Applied Sciences and Arts of Northwestern Switzerland, School of Engineering, 5210 Windisch, Switzerland\label{aff74}
\and
Universit\"at Bonn, Argelander-Institut f\"ur Astronomie, Auf dem H\"ugel 71, 53121 Bonn, Germany\label{aff75}
\and
INFN-Sezione di Roma, Piazzale Aldo Moro, 2 - c/o Dipartimento di Fisica, Edificio G. Marconi, 00185 Roma, Italy\label{aff76}
\and
Aix-Marseille Universit\'e, CNRS, CNES, LAM, Marseille, France\label{aff77}
\and
Dipartimento di Fisica e Astronomia "Augusto Righi" - Alma Mater Studiorum Universit\`a di Bologna, via Piero Gobetti 93/2, 40129 Bologna, Italy\label{aff78}
\and
Department of Physics, Institute for Computational Cosmology, Durham University, South Road, Durham, DH1 3LE, UK\label{aff79}
\and
Universit\'e Paris Cit\'e, CNRS, Astroparticule et Cosmologie, 75013 Paris, France\label{aff80}
\and
Institut d'Astrophysique de Paris, 98bis Boulevard Arago, 75014, Paris, France\label{aff81}
\and
European Space Agency/ESTEC, Keplerlaan 1, 2201 AZ Noordwijk, The Netherlands\label{aff82}
\and
Institut de F\'{i}sica d'Altes Energies (IFAE), The Barcelona Institute of Science and Technology, Campus UAB, 08193 Bellaterra (Barcelona), Spain\label{aff83}
\and
DARK, Niels Bohr Institute, University of Copenhagen, Jagtvej 155, 2200 Copenhagen, Denmark\label{aff84}
\and
Centre National d'Etudes Spatiales -- Centre spatial de Toulouse, 18 avenue Edouard Belin, 31401 Toulouse Cedex 9, France\label{aff85}
\and
Institute of Space Science, Str. Atomistilor, nr. 409 M\u{a}gurele, Ilfov, 077125, Romania\label{aff86}
\and
Dipartimento di Fisica e Astronomia "G. Galilei", Universit\`a di Padova, Via Marzolo 8, 35131 Padova, Italy\label{aff87}
\and
Institut f\"ur Theoretische Physik, University of Heidelberg, Philosophenweg 16, 69120 Heidelberg, Germany\label{aff88}
\and
Universit\'e St Joseph; Faculty of Sciences, Beirut, Lebanon\label{aff89}
\and
Satlantis, University Science Park, Sede Bld 48940, Leioa-Bilbao, Spain\label{aff90}
\and
Infrared Processing and Analysis Center, California Institute of Technology, Pasadena, CA 91125, USA\label{aff91}
\and
Instituto de Astrof\'isica e Ci\^encias do Espa\c{c}o, Faculdade de Ci\^encias, Universidade de Lisboa, Tapada da Ajuda, 1349-018 Lisboa, Portugal\label{aff92}
\and
Universidad Polit\'ecnica de Cartagena, Departamento de Electr\'onica y Tecnolog\'ia de Computadoras,  Plaza del Hospital 1, 30202 Cartagena, Spain\label{aff93}
\and
Kapteyn Astronomical Institute, University of Groningen, PO Box 800, 9700 AV Groningen, The Netherlands\label{aff94}
\and
INFN-Bologna, Via Irnerio 46, 40126 Bologna, Italy\label{aff95}
\and
Dipartimento di Fisica, Universit\`a degli studi di Genova, and INFN-Sezione di Genova, via Dodecaneso 33, 16146, Genova, Italy\label{aff96}
\and
INAF, Istituto di Radioastronomia, Via Piero Gobetti 101, 40129 Bologna, Italy\label{aff97}
\and
Astronomical Observatory of the Autonomous Region of the Aosta Valley (OAVdA), Loc. Lignan 39, I-11020, Nus (Aosta Valley), Italy\label{aff98}
\and
ICL, Junia, Universit\'e Catholique de Lille, LITL, 59000 Lille, France\label{aff99}
\and
ICSC - Centro Nazionale di Ricerca in High Performance Computing, Big Data e Quantum Computing, Via Magnanelli 2, Bologna, Italy\label{aff100}
\and
Instituto de F\'isica Te\'orica UAM-CSIC, Campus de Cantoblanco, 28049 Madrid, Spain\label{aff101}
\and
CERCA/ISO, Department of Physics, Case Western Reserve University, 10900 Euclid Avenue, Cleveland, OH 44106, USA\label{aff102}
\and
Laboratoire Univers et Th\'eorie, Observatoire de Paris, Universit\'e PSL, Universit\'e Paris Cit\'e, CNRS, 92190 Meudon, France\label{aff103}
\and
Departamento de F{\'\i}sica Fundamental. Universidad de Salamanca. Plaza de la Merced s/n. 37008 Salamanca, Spain\label{aff104}
\and
Dipartimento di Fisica e Scienze della Terra, Universit\`a degli Studi di Ferrara, Via Giuseppe Saragat 1, 44122 Ferrara, Italy\label{aff105}
\and
Istituto Nazionale di Fisica Nucleare, Sezione di Ferrara, Via Giuseppe Saragat 1, 44122 Ferrara, Italy\label{aff106}
\and
Center for Data-Driven Discovery, Kavli IPMU (WPI), UTIAS, The University of Tokyo, Kashiwa, Chiba 277-8583, Japan\label{aff107}
\and
Dipartimento di Fisica - Sezione di Astronomia, Universit\`a di Trieste, Via Tiepolo 11, 34131 Trieste, Italy\label{aff108}
\and
Minnesota Institute for Astrophysics, University of Minnesota, 116 Church St SE, Minneapolis, MN 55455, USA\label{aff109}
\and
Institute Lorentz, Leiden University, Niels Bohrweg 2, 2333 CA Leiden, The Netherlands\label{aff110}
\and
Universit\'e C\^{o}te d'Azur, Observatoire de la C\^{o}te d'Azur, CNRS, Laboratoire Lagrange, Bd de l'Observatoire, CS 34229, 06304 Nice cedex 4, France\label{aff111}
\and
Institute for Astronomy, University of Hawaii, 2680 Woodlawn Drive, Honolulu, HI 96822, USA\label{aff112}
\and
Department of Physics \& Astronomy, University of California Irvine, Irvine CA 92697, USA\label{aff113}
\and
Department of Astronomy \& Physics and Institute for Computational Astrophysics, Saint Mary's University, 923 Robie Street, Halifax, Nova Scotia, B3H 3C3, Canada\label{aff114}
\and
Departamento F\'isica Aplicada, Universidad Polit\'ecnica de Cartagena, Campus Muralla del Mar, 30202 Cartagena, Murcia, Spain\label{aff115}
\and
Instituto de Astrof\'isica de Canarias (IAC); Departamento de Astrof\'isica, Universidad de La Laguna (ULL), 38200, La Laguna, Tenerife, Spain\label{aff116}
\and
Department of Physics, Oxford University, Keble Road, Oxford OX1 3RH, UK\label{aff117}
\and
Institute of Cosmology and Gravitation, University of Portsmouth, Portsmouth PO1 3FX, UK\label{aff118}
\and
Department of Computer Science, Aalto University, PO Box 15400, Espoo, FI-00 076, Finland\label{aff119}
\and
Instituto de Astrof\'\i sica de Canarias, c/ Via Lactea s/n, La Laguna 38200, Spain. Departamento de Astrof\'\i sica de la Universidad de La Laguna, Avda. Francisco Sanchez, La Laguna, 38200, Spain\label{aff120}
\and
Ruhr University Bochum, Faculty of Physics and Astronomy, Astronomical Institute (AIRUB), German Centre for Cosmological Lensing (GCCL), 44780 Bochum, Germany\label{aff121}
\and
Univ. Grenoble Alpes, CNRS, Grenoble INP, LPSC-IN2P3, 53, Avenue des Martyrs, 38000, Grenoble, France\label{aff122}
\and
Department of Physics and Astronomy, Vesilinnantie 5, 20014 University of Turku, Finland\label{aff123}
\and
Serco for European Space Agency (ESA), Camino bajo del Castillo, s/n, Urbanizacion Villafranca del Castillo, Villanueva de la Ca\~nada, 28692 Madrid, Spain\label{aff124}
\and
ARC Centre of Excellence for Dark Matter Particle Physics, Melbourne, Australia\label{aff125}
\and
Centre for Astrophysics \& Supercomputing, Swinburne University of Technology,  Hawthorn, Victoria 3122, Australia\label{aff126}
\and
School of Physics and Astronomy, Queen Mary University of London, Mile End Road, London E1 4NS, UK\label{aff127}
\and
Department of Physics and Astronomy, University of the Western Cape, Bellville, Cape Town, 7535, South Africa\label{aff128}
\and
ICTP South American Institute for Fundamental Research, Instituto de F\'{\i}sica Te\'orica, Universidade Estadual Paulista, S\~ao Paulo, Brazil\label{aff129}
\and
IRFU, CEA, Universit\'e Paris-Saclay 91191 Gif-sur-Yvette Cedex, France\label{aff130}
\and
Oskar Klein Centre for Cosmoparticle Physics, Department of Physics, Stockholm University, Stockholm, SE-106 91, Sweden\label{aff131}
\and
Astrophysics Group, Blackett Laboratory, Imperial College London, London SW7 2AZ, UK\label{aff132}
\and
INAF-Osservatorio Astrofisico di Arcetri, Largo E. Fermi 5, 50125, Firenze, Italy\label{aff133}
\and
Dipartimento di Fisica, Sapienza Universit\`a di Roma, Piazzale Aldo Moro 2, 00185 Roma, Italy\label{aff134}
\and
Aurora Technology for European Space Agency (ESA), Camino bajo del Castillo, s/n, Urbanizacion Villafranca del Castillo, Villanueva de la Ca\~nada, 28692 Madrid, Spain\label{aff135}
\and
Centro de Astrof\'{\i}sica da Universidade do Porto, Rua das Estrelas, 4150-762 Porto, Portugal\label{aff136}
\and
HE Space for European Space Agency (ESA), Camino bajo del Castillo, s/n, Urbanizacion Villafranca del Castillo, Villanueva de la Ca\~nada, 28692 Madrid, Spain\label{aff137}
\and
Institute of Astronomy, University of Cambridge, Madingley Road, Cambridge CB3 0HA, UK\label{aff138}
\and
Department of Astrophysics, University of Zurich, Winterthurerstrasse 190, 8057 Zurich, Switzerland\label{aff139}
\and
INAF-Osservatorio Astronomico di Brera, Via Brera 28, 20122 Milano, Italy, and INFN-Sezione di Genova, Via Dodecaneso 33, 16146, Genova, Italy\label{aff140}
\and
Theoretical astrophysics, Department of Physics and Astronomy, Uppsala University, Box 515, 751 20 Uppsala, Sweden\label{aff141}
\and
Department of Physics, Royal Holloway, University of London, TW20 0EX, UK\label{aff142}
\and
Department of Astrophysical Sciences, Peyton Hall, Princeton University, Princeton, NJ 08544, USA\label{aff143}
\and
Niels Bohr Institute, University of Copenhagen, Jagtvej 128, 2200 Copenhagen, Denmark\label{aff145}
\and
Center for Cosmology and Particle Physics, Department of Physics, New York University, New York, NY 10003, USA\label{aff146}
\and
Center for Computational Astrophysics, Flatiron Institute, 162 5th Avenue, 10010, New York, NY, USA\label{aff147}}

%
\abstract{
To date, galaxy image simulations for weak lensing surveys usually approximate the light profiles of all galaxies as a single or double Sérsic profile, neglecting the influence of galaxy substructures and morphologies deviating from such a simplified parametric characterisation. While this approximation may be sufficient for previous data sets, the stringent cosmic shear calibration requirements and the high quality of the data in the upcoming \Euclid survey demand a consideration of the effects that realistic galaxy substructures and irregular shapes have on shear measurement biases. Here we present a novel deep learning-based method to create such simulated galaxies directly from \textit{Hubble} Space Telescope (HST) data. We first build and validate a convolutional neural network based on the wavelet scattering transform to learn noise-free representations independent of the point-spread function (PSF) of HST galaxy images. These can be injected into simulations of images from \Euclid's optical instrument VIS without introducing noise correlations during PSF convolution or shearing. Then, we demonstrate the generation of new galaxy images by sampling from the model randomly as well as conditionally. In the latter case, we fine-tune the interpolation between latent space vectors of sample galaxies to directly obtain new realistic objects following a specific Sérsic index and half-light radius distribution. Furthermore, we show that the distribution of galaxy structural and morphological parameters of our generative model matches the distribution of the input HST training data, proving the capability of the model to produce realistic shapes. Next, we quantify the cosmic shear bias from complex galaxy shapes in \Euclid-like simulations by comparing the shear measurement biases between a sample of model objects and their best-fit double-Sérsic counterparts, thereby creating two separate branches that only differ in the complexity of their shapes. Using the Kaiser, Squires, and Broadhurst shape measurement algorithm, we find a multiplicative bias difference between these branches with realistic morphologies and parametric profiles on the order of $(6.9\pm 0.6)\times 10^{-3}$ for a realistic magnitude-Sérsic index distribution. Moreover, we find clear detection bias differences between full image scenes simulated with parametric and realistic galaxies, leading to a bias difference of $(4.0\pm 0.9)\times 10^{-3}$ independent of the shape measurement method. This makes complex morphology relevant for stage IV weak lensing surveys, exceeding the full error budget of the Euclid Wide Survey ($\Delta\mu_{1,2} < 2 \times 10^{3}$). }
%

    \titlerunning{\Euclid preparation. Galaxy morphology bias calibration}
    \authorrunning{B. Csizi et al.}
    \maketitle
%
%
%
%
   
\section{\label{sc:Intro} Introduction}
Identifying the origin of the accelerated expansion of the Universe by constraining the dark energy equation of state parameter $w$ is one of the most 
challenging and pressing open questions in cosmology. To tackle the task of unravelling the characteristics of dark energy, several next-generation surveys such as
\Euclid \citep{Laureijs11, EuclidSkyOverview}, the \textit{Nancy Grace Roman} Telescope \citep{Spergel2015}, and the Legacy Survey of Space and Time at the \textit{Vera C. Rubin} Observatory 
\citep{Ivezic2019} will need to measure weak lensing (WL) image distortions at extremely high accuracy. These distortions have been imprinted on the observed shapes of distant galaxies by the gravitational fields of the foreground cosmic large-scale structure. 
Such measurements call for precise calibrations to meet the tight requirements. Detailed and realistic image simulations are hence a key ingredient for the 
latest generation of weak lensing surveys,  as they allow for the cosmic shear measurement methods that will be applied to be calibrated, and thus the full predictive power of the data for the 
inference of cosmological parameters can be leveraged. 

There have been many efforts to quantify the effect of the properties of image simulations on shape measurements \citep[see for example][]{Hoekstra2017, HernandezMartin2020} and to subsequently improve
the simulation quality to more closely match the real observations concerning, for example, galaxy number densities, morphological properties, redshifts and magnitudes, and
instrumental or atmospheric effects \citep{Mandelbaum2018a, Kannawadi2019, MacCrann2022, Li2022, Li2023}. Until now, however, the galaxy morphologies included in these simulations have lacked the 
complexity and irregularity of real galaxies, or they have relied on injecting real high-resolution \textit{Hubble} Space Telescope (HST) observations into image simulations by adjusting the imaging properties to match the target survey's instrument and observing properties \citep{Li2022}. The latter method is applicable to ground-based stage III experiments but has been shown to cause strong issues in the shear estimation due to noise correlations \citep{Scognamiglio2025}. For the former simulation framework, the light distribution of an object is simulated as a single analytic Sérsic profile or as a two-component model consisting 
of the sum of a Sérsic bulge and an either Sérsic or exponential disc. While stage III surveys, such as the Dark Energy Survey \citep{DES2016}, the Hyper-Suprime-Cam Survey 
\citep{Aihara2018}, and the Kilo-Degree Survey \citep{deJong2013}, were able to rely on this 
simplification due to the lower shape measurement bias requirements, novel stage IV projects such as \Euclid will have to account for the influence of galaxy
substructures on the cosmic shear analysis. In weak lensing, cosmic shear is measured from spatial correlations in galaxy ellipticities using large source samples, thus requiring simulations that accurately reproduce the shapes of real objects to calibrate the measurement. Previous attempts at creating more realistic simulations include the emulation of lensing data from HST images \citep{Mandelbaum2012, Mandelbaum2018a}, the generation of galaxies via shapelet functions \citep{Massey2004}, and the sampling of different deep learning models trained on real data \citep{Lanusse2021}. Results from the GREAT3 challenge \citep{Mandelbaum2015} showed a percent-level bias difference with respect to parametric galaxy morphologies for most shape measurement methods using HST emulation, which was more prominent for simulated space-based data due to the high resolution and small pixel scales, while the Shear Testing Programme 2 \citep{Massey2007b} revealed a sub-percent bias difference using the shapelet galaxies, albeit at ground-based pixel scales and point-spread functions (PSFs). The analysis of the full Euclid Wide Survey requires a shear calibration accuracy to better than 0.2\,\% \citep{Cropper2013}, making it imperative to account for the impact of galaxy morphologies. The effect stems from the fact that second-order moments of galaxy light profiles, which are commonly used by shape measurement algorithms, are coupled
with higher-order moments by the shear \citep{Massey2007, Zhang2011, Bernstein2010}. These higher-order moments are, however, dominated by morphologies deviating from simple parametrisations. Previous estimates determined that the Euclid Wide Survey will be able to resolve substructures down to a surface brightness of $22.5\,\mathrm{mag\,arcsec}^{-2}$ and down to $24.9\,\mathrm{mag\,arcsec}^{-2}$ in the  Deep Fields \citep{Bretonniere-EP13}. This results in approximately 250 million galaxies with resolved morphologies over the entire mission lifetime.
As there does not exist a well-established parametric model of galaxy morphologies that is more realistic than a two-component description, aside from simulating a substructure using parametric shapes that are disturbed with knots in the light profile \citep{Sheldon2017}, a different path is needed to simulate galaxy images. 

With deep learning techniques on the rise for tasks of computer vision, such as image generation or classification, capitalising on this growing research field for 
the aforementioned goal is a promising approach. \cite{Spindler2021}, \citet{Lanusse2021}, \citet{Smith2022}, and \cite{Holzschuh2022} have shown how variational autoencoders \citep[VAEs;][]{Kingma2013} and generative adversarial networks \citep[GANs;][]{Goodfellow2014} can be applied to
generate galaxy images with high-resolution training data, a method that has since also been applied for forecasts on galaxy morphologies with \Euclid 
\citep{Bretonniere-EP13}. Aside from VAEs and GANs, diffusion models have gained popularity as a powerful image generation method \citep{Ho2020}, but they are accompanied by increased complexity. Aside from image generation, latent space machine learning models have also found other applications, for example, galaxy classification \citep{ChengTY2021}, discovery of strong lenses \citep{ChengTY2020}, modelling of galactic dust extinction \citep{Thorne2021}, and dimensionality reduction of galaxy spectra \citep{Portillo2020}.

In this work, we propose a new generative model that allows for noise-free and PSF-independent reconstruction of real galaxy images and is able to generate a distribution of new objects following an 
input distribution of morphological parameters. While it is mostly interesting to generate new images due to the necessity of including $10^7$--$10^9$ galaxies in order to reach the precision of the \Euclid shear calibration requirements
\citep{Martinet-EP4}, the noise-free reconstruction of inputs with minimal residuals can also be relevant for the \textit{Euclidisation} procedure proposed by \cite{Scognamiglio2025}, where the authors
presented a method to convert HST images of isolated galaxies (hereafter, postage stamps) into \Euclid observations. Such a pipeline requires PSF (de)convolution, shearing, and down-sampling steps on a noisy image, which gives rise to
noise correlations that impact the shape measurement \citep{Gurvich2016}. Moreover, our model can be applied to future high signal-to-noise ratio observations in the Euclid Deep Fields in order to
obtain a larger sample of observed but noise-free galaxies that can be injected into simulations without introducing correlated noise or relying on the output of a black-box machine learning model. 

In this paper, we perform a proof-of-concept study on the calibration of biases from a complex galaxy morphology for the \Euclid mission. First, we present the architecture and training data of our deep learning model based on the wavelet scattering transform after summarising the weak lensing formalism
and the theoretical background on galaxy morphological statistics in Sect. \ref{sc:WL} and Sect. \ref{sc:GalMorph}. We then compare our reconstructed images with their inputs in terms of their 
structural parameters and a set of morphological statistics in Sect. \ref{sc:NoiseRemoval}. Afterwards, we describe our method for generating new galaxies according to either an input distribution of Sérsic indices or their overall visual characteristics. Finally, in Sect. \ref{sc:ShearMeas}, we quantify the shear bias introduced by galaxy substructures through the comparison of simulations of \Euclid VIS-like postage stamps with samples from our model and their respective best-fit parametric models before concluding in Sect. \ref{sc:conclusion}.

\section{\label{sc:WL} Weak lensing formalism}
\subsection{Definitions}
Weak lensing describes the coherent, statistical distortion of objects in the Universe due to gravitational deflection by mass density fluctuations along the line of sight, see \citet{Massey2010}, \citet{Kilbinger2015}, and \citet{Mandelbaum2018b} for reviews. The local differential mapping between lensed and unlensed coordinates can be described via the Jacobian matrix, whose elements are
\begin{equation}
    \mathcal{A}_{ij} = \delta_{ij} - \partial_i \partial_j \Psi\, ,
\end{equation}
where $\Psi$ denotes the lensing potential, $\delta_{ij}$ is the Kronecker delta, and $\partial_{i,j}$ are the derivatives along the respective coordinate of the lens plane. This matrix can be parametrised by the introduction of a convergence $\kappa$ and a two-component shear $\gamma = (\gamma_1, \gamma_2)$, 
to simplify the Jacobian to 
\begin{equation}
    \mathcal{A}_{ij} = 
    \begin{pmatrix}
    1 - \kappa - \gamma_1 & -\gamma_2 \\
    -\gamma_2 & 1 - \kappa + \gamma_1 
    \end{pmatrix}\, .
\end{equation}
Usually, the shear is expressed as a complex number $\gamma = \gamma_1 + \mathrm{i}\gamma_2$ and in terms of the reduced shear 
\begin{equation}
    g = g_1 + \mathrm{i}g_2 = \frac{\gamma}{1 - \kappa}\, , 
\end{equation}
which is advantageous due to this quantity's connection to the ellipticity $\epsilon$, making it directly measurable. In the absence of preferential intrinsic galaxy orientations, the expectation value of the intrinsic ellipticity $\langle\epsilon_\mathrm{s}\rangle$ vanishes, leading to 
\begin{equation}
    g = \langle\epsilon_\mathrm{o}\rangle\, ,
\end{equation}
where $ \langle\epsilon_\mathrm{o}\rangle$ is the expectation value of the observed ellipticity. This simple relation shows that a measurement of the ellipticity of galaxies allows measurement of the shear when averaging over a sufficiently large number of sources.

\subsection{Shear measurement}
There exist several methods for measuring the shear from a galaxy image. It can be done directly from galaxy brightness moments, for instance with the Kaiser, Squires, and Broadhurst (KSB) shape measurement algorithm \citep{Kaiser1995, Hoekstra1998}, or using forward-modelling of a light 
distribution and then fitting it to the data by maximising a likelihood function, such as \textsc{im3shape} \citep{Zuntz2013}, \textit{lens}fit \citep{Miller2007}, or \lensmc \citep{Congedo2024}. Moreover, \textsc{Metacalibration} \citep{Sheldon2017} has been applied with both of these types of methods to get, in principle, model bias-corrected estimates of the shear signal. We measure galaxy shapes in this work using the KSB implementation from the HSM module of the \texttt{GalSim} package \citep{Hirata2003, Rowe2015}. Therein, second-order brightness moments 
\begin{equation}
    Q_{ij} = \frac{\int I(\vec{\theta})\, W(\vec{\theta})\,\theta_i\,
    \theta_j\,\mathrm{d}^2\vec{\theta}}
    {\int I(\vec{\theta})\, W(\vec{\theta})
    \, \mathrm{d}^2 \vec{\theta}}
\end{equation}
are used to infer the complex ellipticity 
\begin{equation}
    \epsilon_\mathrm{o} = \epsilon_1 + \mathrm{i}\epsilon_2 = \frac{Q_{11} - Q_{22} + 2\mathrm{i}Q_{12}}{Q_{11} + Q_{22}}\, .
\end{equation}
Here, $\vec{\theta} = (\theta_1, \theta_2)^{\rm T}$ is the position vector relative to the object centre, which is defined such that the weighted first-order brightness moment vanishes, $I(\vec{\theta})$ is the image light distribution, and $W(\vec{\theta})$ is an arbitrary weight function, which is usually a Gaussian. Within the KSB formalism the ellipticity is then corrected for the impact of the point-spread function using further brightness moments and measurements of stars.

Model fitting algorithms such as \lensmc on the other hand employ a Bayesian approach to forward-model the pixel data and then estimating the ellipticity as the mean of a posterior distribution 
\begin{equation}
    \hat{\epsilon} = \int \epsilon\; p(\epsilon | \vec{D})
    \;\text{d}\epsilon
\end{equation}
by sampling the galaxy model parameter space, for instance with Markov-Chain Monte-Carlo (MCMC), and marginalising over nuisance parameters. Here, $p(\epsilon | \vec{D})$ is the ellipticity marginal posterior on the pixel data $\vec{D} = I(\vec{\theta})$ given by
\begin{equation}
    p(\epsilon | \vec{D}) = \frac{1}{p(\vec{D})}
    \int p(\vec{D}|\epsilon,\xi,\phi)\; p(\epsilon,\xi,\phi)
    \;\text{d}\xi\,\text{d}\phi\, ,
\end{equation}
with $p(\vec{D})$ being the marginal likelihood and $\xi$ and $\phi$ as the intrinsic and linear nuisance parameters \citep{Congedo2024}.

\subsection{Bias calibration}
Shear estimators are affected by a range of different bias sources, for example from selection biases or PSF correction errors \citep{Bernstein2002, Hirata2003, Conti2017}. In linear
approximation, the shear bias, defined as the difference between an observed reduced shear $g_i^\text{obs}$ and the true reduced shear $g_i^\text{true}$ with $i=1, 2$ (assuming no mixing between the two components), can be written as 
\begin{equation}
    g_i^\text{obs} -g_i^\text{true} = \mu_i g_i^\text{true} + c_i + n_i\, ,
    \label{eq:bias-def}
\end{equation}
with $\mu_i$ and $c_i$ being the multiplicative and additive shear biases, respectively, and $n_i$ as a statistical noise component. The indices $i=1, 2$ denote the shear component along the Cartesian axes and along the $\pi/4$ diagonals, respectively. Alternatively, the bias can be defined via a spin-2 equation with spin-0 and spin-4 multiplicative biases, which then facilitates an inclusion of non-linear shear bias terms, if required \citep{Kitching2022}. The magnitude of the multiplicative bias is generally a function of galaxy morphological parameters \citep[see, e.g.,][]{HernandezMartin2020}, galaxy signal-to-noise ratio \citep{Schrabback2010, Hoekstra2015}, redshift \citep{Kannawadi2019}, and blending \citep{MacCrann2022}, therefore requiring a redshift tomography-dependent calibration.  

To mitigate the effect of biased galaxy shape estimation on shear analysis (and in consequence cosmological inference), several techniques can be exploited. On one side, methods such as shape and pixel noise cancellation \citep{Massey2007, Jansen2023} have been shown to be able to efficiently scale down the necessary simulation volume for shear calibration, which in return is used to correct the survey measurements for the determined bias. Other methods such as \textsc{Metacalibration} have been employed successfully to remove noise biases, model biases and selection effects from the shape measurement directly during the measurement process \citep{Sheldon2017, Sheldon2020}. Nevertheless, detailed simulated 
images are needed due to blending, detection \citep{Hoekstra2021}, and redshift blending \citep{MacCrann2022, Li2023}. In this work, we will focus on the influence of complex galaxy morphologies on the shear measurement bias. This is a model bias that enters the shape measurement due to inaccuracies of the underlying model (either the fitted profiles or the estimation of ellipticities via moments), and thus it depends on the applied method. While this specific model bias has been previously ignored for ground-based surveys with lower resolution imaging, where substructure is not well resolved, this assumption does not necessarily hold for a space-based mission such as \Euclid, given its unprecedented precision and survey area.

\section{\label{sc:GalMorph} Galaxy morphologies}
\subsection{Sérsic profiles}
The overall complex structure of galaxies is a product of their complicated evolution history, see \cite{Conselice2014} for a review. Nevertheless, the observations of a majority of galaxies at currently feasible resolutions for large ground-based WL surveys can be well approximated using a simple analytic prescription. The most common parametric form of a galaxy-like light distribution is the 
Sérsic profile
\begin{equation}
    I(r) \propto \exp\left[ -b_n \left(\frac{r}{r_\mathrm{e}}\right) ^{1/n} \right]\, ,
\end{equation}
where $n$ is the Sérsic index, $r_\mathrm{e}$ is the half-light radius, and $b_n$ is a scaling factor that depends on $n$ \citep{Sersic1963}. With this model, galaxy light profiles can be simulated as either a single Sérsic or a double Sérsic profile consisting of separate bulge 
and disc components. Simulating galaxies in this way is advantageous due to the simplicity and existing measurements of the model parameters across previous survey areas, which 
enables a simple yet realistic generation of simulated footprints.

\subsection{Morphological statistics}

\Euclid's VIS instrument \citep{Cropper16, EuclidSkyVIS} will observe billions of galaxies over 14\,000\,deg$^2$, with many of them covering only a few pixels given the $\ang{;;0.1}$ pixel scale. A large subset will have shapes that 
can be well approximated by a Sérsic parametrisation, but a non-negligible portion of the sample will also display structural features such as 
irregularities, clumps, spiral arms, or tidal streams. Moreover, the fraction of such peculiar galaxies changes gradually with redshift, with high-$z$ objects ($z>1.2$) showing irregularities more commonly, especially since optical observations show rest-frame UV structures for high-$z$ galaxies, which are dominated by star-forming regions \citep{Abraham1996, Conselice2005, Bundy2005}. Even though these objects will have mostly low signal-to-noise ratios, parametric fits can still result in large residuals for a substantial number of objects. Given that the \Euclid WL sample will include galaxies with redshifts up to $z\approx 2$ \citep{Ilbert-EP11}, accurate galaxy shapes will accordingly be even more relevant.  

Statistical proxies for disturbed morphologies can be evaluated on input HST data and the deep learning model output to validate its capability to capture modes that deviate from smooth structures. \cite{Hackstein2023} summarised a set 
of such proxies to estimate the power of galaxy image generators. For instance, the MID statistics (multi-mode, intensity, deviation)
by \cite{Freeman2013} provide an estimate of peculiar features of a galaxy morphology by tracing the existence and intensity ratio of multiple nuclei as well as the deviation from simple 
elliptical representations. Similarly, the Concentration, Asymmetry \& Smoothness (CAS) morphology indicators \citep{Conselice2000, Conselice2003} trace irregular shapes by defining the following set of estimators:
\begin{align}
    C &= 5\, \log_{10} \left(\frac{r_{80}}{r_{20}}\right)\, ; \\
    A &= \sum_{i,j} \frac{|I_0(\theta_i, \theta_j) - I_{\pi}(\theta_i, \theta_j)|}{|I_0(\theta_i, \theta_j)|} - |B_{\pi}|\, ;\\
    S &= \sum_{i,j} \frac{|I_0(\theta_i, \theta_j) - I_\mathrm{S}(\theta_i, \theta_j)|}{|I_0(\theta_i, \theta_j)|} - |B_{\mathrm{S}}|\, .
\end{align}
\begin{figure*}
    \centering
    \includegraphics[width=1.8\columnwidth]{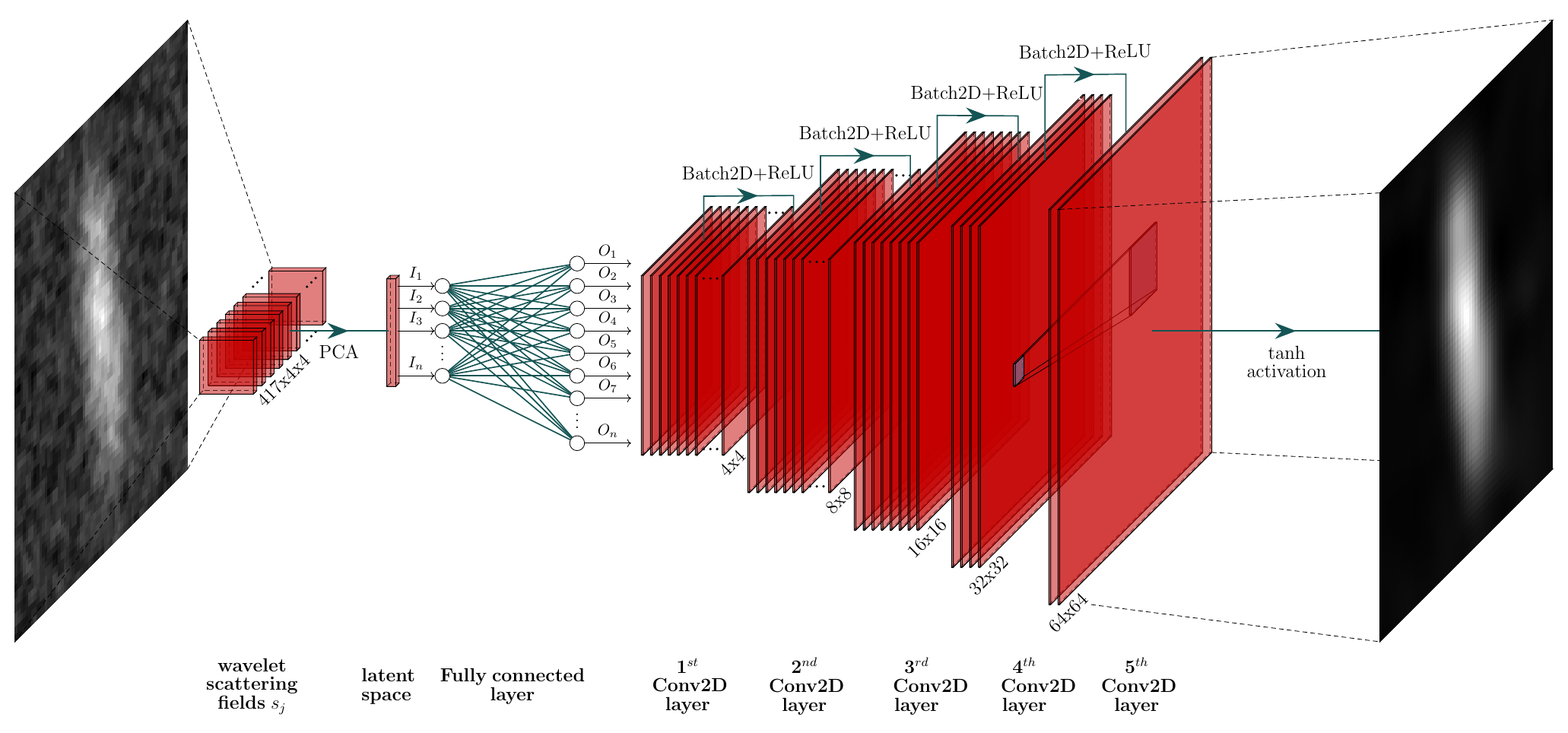}
    \caption{Architecture of the CNN. Noisy input images are embedded into a latent space vector by performing a PCA 
    on the wavelet scattering fields $s_2^{j_1, l_1, j_2, l_2}$ with $J=4, L=8$, which is then propagated through one fully connected layer and five convolutional layers, 
    each with a $5\times 5$ kernel, batch normalisation and ReLU. The final output of the generative model is produced by a $\tanh$ activation function.}
    \label{fig:neural-net-architecture}
\end{figure*}
The concentration $C$ measures the bulge concentration by relating the radii $r_{80}, r_{20}$ of apertures within which 80\% and 20\% of the total flux are located. Additionally, the asymmetry parameter $A$ and the smoothness $S$ quantify the rotational symmetry with respect to the flux, and the magnitude of small-scale structures, respectively. Here, $I_0(\theta_i, \theta_j)$ is the galaxy image intensity at pixels $\vec{\theta}=(\theta_i, \theta_j)^{\rm T}$, $I_{\pi}(\theta_i, \theta_j)$ is the same image rotated by $\pi$ around the image centre, $|B_{\pi}|$ is the average asymmetry of the rotated image background, $I_\mathrm{S}(\theta_i, \theta_j)$ is the image smoothed by a boxcar filter, and $|B_{\rm S}|$ is the average smoothness of the background. 

Furthermore, another such statistic, the Gini coefficient, is sensitive to the intensity concentration in a compact component of the light profile and can be calculated with
\begin{equation}
    G = \frac{1}{k(k-1)\,\bar{I_0}} \sum_{i}^{k}
    (2i-k-1)\,I_0(\vec{\theta}_i)\, .
\end{equation}
The value $I_0(\vec{\theta}_i)$ is the $i$-th pixel value of the individual galaxy, here with pixels sorted by increasing intensity, and $\bar{I_0}$ is the mean 
over all $k$ pixels \citep{Abraham2003}. We also calculated the $M_{20}$ coefficient 
\begin{equation}
    M_{20}= \log_{10} \left( \frac{\sum_i Q_i}{Q_\text{tot}}
    \right)\, ,
\end{equation}
where $Q_\mathrm{tot}$ is the second-order moment of the total galaxy light distribution (sum over all pixel fluxes multiplied by their squared distance to the image centre) and $\sum_i Q_i$ are the second-order moments summed over only the brightest 20\% of pixels, so $\sum_i I_0(\vec{\theta}_i) < 0.2\,I_0(\vec{\theta})$. This parameter is thereby able to indicate merger signatures and 
clumpiness \citep{Lotz2004}.

\section{\label{sc:NoiseRemoval} Reconstruction of HST data}

\subsection{The wavelet scattering transform}

One main attribute of deep learning models is a process of dimensionality reduction. To 
learn the distribution of training images, these models usually compress the 2D array
of image pixels into a latent vector $\boldsymbol{z}$, which is then 
later expanded to image size using convolutional layers. While these latent representations
within VAEs and GANs effectively constitute a black-box, where usually no physical meaning can be 
attributed to the latent variables, we employ an image compression method that is based on the
wavelet scattering transform \citep[WST;][]{Mallat2012} as an encoder for the network. Such convolutional neural networks (CNNs) have 
previously been proposed by \cite{Bruna2013} and applied to common deep learning test data sets 
by \cite{Angles2018}. This operation is useful, due to the transform's ability to capture morphological information. With this mathematically motivated latent space, we can later on also sample 
galaxies by clustering them according to their wavelet scattering coefficients. Thus we avoid learning a latent 
variable model or encountering the typical limitations of other generative models, such as
mismatches between aggregate posterior and prior in VAEs \citep{Tomczak2017} or mode collapse for
GANs \citep{Salimans2016}.

The wavelet scattering transform is an operation that applies a set of convolutions by dilated 
and rotated wavelet filters $\psi_\lambda$ to a 2D array. A family $\{\psi_\lambda\}_{j,l}$ of such
filters is specified by a dyadic sequence of scales $2^j$ with $j\in\mathbb{Z}, J\geq j > 0$, a number of rotations with 
angles $l, l\in\mathbb{Z}, L \geq l > 0$, and a rotation operation $r_l$ on the data $\vec{x}$:

\begin{equation}
   \{\psi_\lambda\}_{j,l} = \frac{1}{2^j}\psi_\lambda
   \left(\frac{r_l^{-1}\vec{x}}{2^j}\right)\, .
\end{equation}

Given an input image $I_0(\vec{\theta})$, the zeroth-order scattering transform is defined simply as the mean
of the input. The first-order coefficients are calculated by convolving the image with the family of wavelet filters
$\{\psi_\lambda\}_{j_1,l_1}$ and then by taking the mean of the modulus of the obtained scattering fields. Similarly, the second-order 
scattering coefficients are given by the convolution of the first-order fields by another set of wavelets $\{\psi_\lambda\}_{j_2,l_2}$:
\begin{align}
    s_0 &= \langle I_0(\vec{\theta}) \rangle\, ; \\
    \vec{s_1}^{j_1, l_1} &= \left\langle\left|  I_0(\vec{\theta}) \star \{\psi_\lambda\}_{j_1,l_1}  \right|\right\rangle\, ; \\
    \vec{s_2}^{j_1, l_1, j_2, l_2} &= \left\langle\left|  \left|I_0(\vec{\theta}) \star  \{\psi_\lambda\}_{j_1,l_1}\right| 
    \star  \{\psi_\lambda\}_{j_2,l_2}  \right|\right\rangle \,.
\end{align}
Here, $\star$ designates a convolution. This results in $J^nL^n+1$ scattering coefficients, where $n$ is the maximum order of the scattering transform, which exceeds the target latent space dimension for typical values of $J=4, L=8$. By
averaging over all orientations $(l_1, l_2)$, one can reduce this number to $J+J^2+1$ \citep{Greig2023}. This, however, does not preserve the angular dependence of the wavelet filtering and thus only probes the size scales of the image, which thus reduces the morphological information. We only average over the orientations $l_1$, which preserves shape information but discards information on the orientation. Additionally, one can further limit the dimensionality by a factor of $2^{n-1}$, 
by discarding coefficients with $j_1 \geq j_2$, as they only contain high-frequency information \citep{Cheng2021}. This is almost exclusively noisy pixel-level information, which will not be highly relevant for \Euclid with respect to HST, given the resolution difference. 
For the reduced scattering coefficients of $I_0(\vec{\theta})$, one thus
ultimately obtains
\begin{align}
     \vec{s_1}^{j_1} &= \left\langle\left\langle\left|  I_0(\vec{\theta}) \star \{\psi_\lambda\}_{j_1,l_1}  \right|\right\rangle\right\rangle_{l_1}\, ; \\
    \vec{s_2}^{j_1, j_2, l_2} &= \left\langle\left\langle\left|  \left|I_0(\vec{\theta}) \star  \{\psi_\lambda\}_{j_1,l_1}\right|\star  \{\psi_\lambda\}_{j_2,l_2}  
    \right|\right\rangle\right\rangle_{l_1}, \;\;\; j_2>j_1\, .
\end{align}
Here, $\langle\dots\rangle_{l_1}$ denotes an average over all $l_1$ indices. The choice of $J$ and $L$ depends on the size of the images and the scales and angles that need to be probed. In principle, the wavelet scattering transform can be extended to higher orders; this is, however, computationally expensive. \cite{Bruna2013} have furthermore shown that the information content is extremely small beyond the second order. 
We compute wavelet scattering coefficients with the \texttt{kymatio} package with \texttt{pytorch} backend \citep{Andreux2018, Paszke2019}. 

\subsection{Training data}
Our galaxy image training data set consists of HST images observed in the F814W filter as part of the COSMOS program \citep{Scoville2007}. The \texttt{GalSim}
package supplies a catalogue of postage stamps of deblended, PSF deconvolved galaxies with magnitudes down to $m_\mathrm{AB}^{\text{F814W}}\leq 23.5$ from this survey
\citep{Leautaud2007,Mandelbaumdataset,Mandelbaum2014}. For the sample selection of this data set, some additional cuts were imposed on the COSMOS data to reject objects with contamination from stars or image defects, as well as objects lying in masked regions of the ground-based $BVIz$ imaging, to ensure good photometric redshift estimates. The exact cuts can be found in the appendix of \cite{Mandelbaum2014}. 

We draw these 56\,062 galaxies on $64\times 64$ images at a pixel scale of $\ang{;;0.05}$, that is half of the nominal pixel size of the \Euclid VIS instrument, 
and convolve them by a simple Gaussian PSF with $\sigma=\ang{;;0.07}$ for training. We can later easily deconvolve the noise-free reconstructions by the same PSF 
without introducing noise correlations. Next, we discard galaxies with a low signal-to-noise ratio (S/N $\leq 10$), as well as large galaxies that exceed the image size of the
postage stamps to avoid truncation. This is done by creating a $3\sigma$ binary segmentation map and removing objects whose edges do not lie within the stamp. Alternatively, we could increase our postage stamp size, although only to the detriment of much longer training times. Moreover, 
galaxies that exceed 64 pixels at $\ang{;;0.05}$ are not relevant for the cosmic shear analysis due to their angular size of $\ge\ang{;;3}$. Such objects do not carry a significant shear signal and can thus be excluded from the analysis. These cuts leave us with
46\,720 galaxies, which we divide into 43\,520 training images and 3\,200 test images. 

\begin{figure}
    \centering
    \includegraphics[width=1.0\columnwidth]{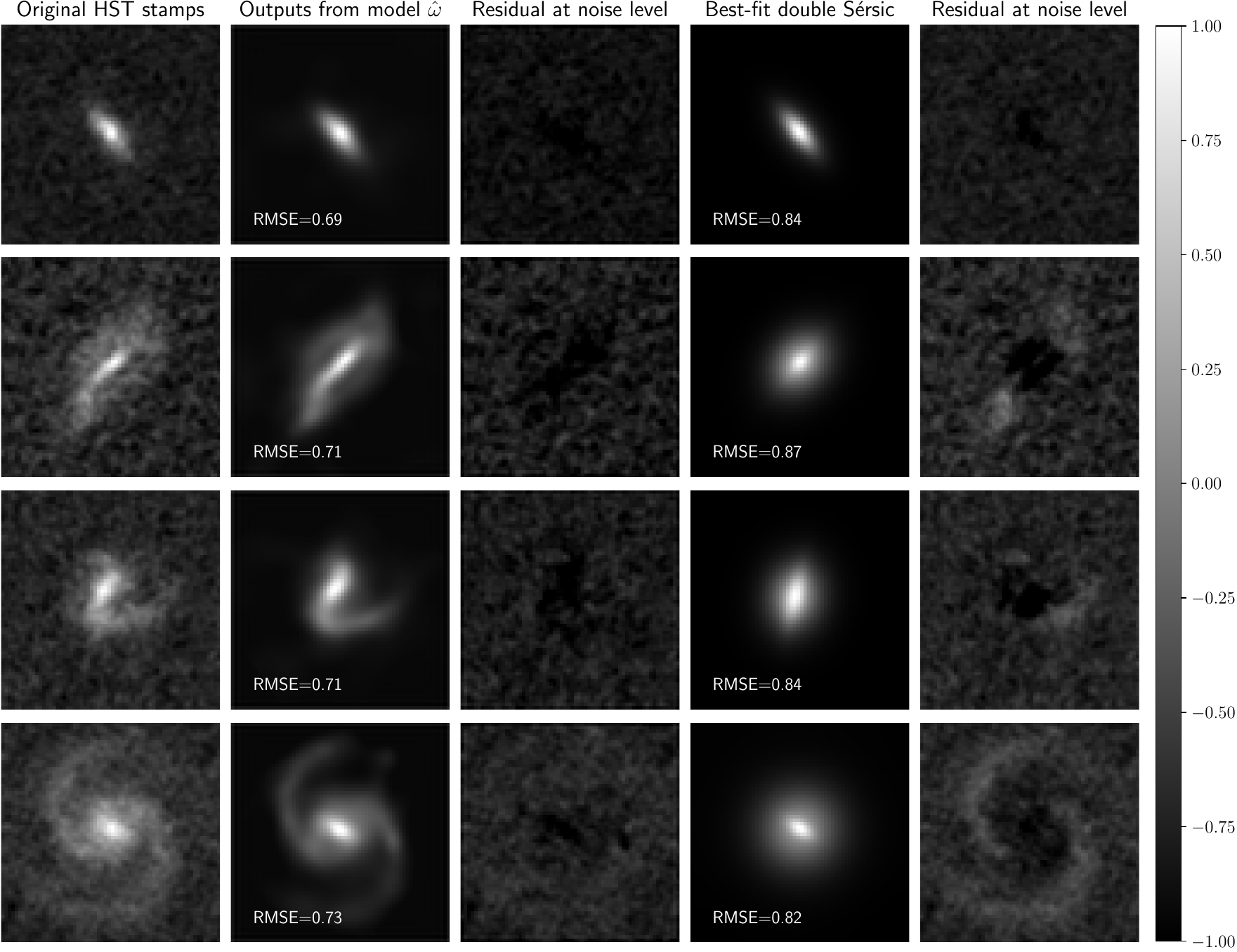}
    \caption{Qualitative check of the HST reconstruction from the generator $\hat{\omega}$. The left column shows original input galaxies, while the second column displays
    the output of the generator after training. In the middle we plot the residual at noise level between the first two columns, followed by the best-fit double Sérsic from \texttt{pysersic} and the residuals between the original HST stamps and the Sérsic fits. We also display the root mean squared error (RMSE) values between the original image and the reconstruction, for both the parametric and realistic columns.}
    \label{fig:neural-comparision_hst_wst}
\end{figure}

\subsection{Network architecture and training}

Using the aforementioned WST, we first calculate the scattering fields of our training images up to second order with $J=4, L=8$, 
resulting in a 3D vector of size $(1+LJ+L^2J(J-1)/2, w/2^J, h/2^J)$ for an image with size $(w,h)$. To further reduce the dimensionality we 
perform a principal component analysis \citep[PCA, see][]{Shlens2014} to compress the data into a latent space embedding $\{\boldsymbol{z}_s(I_{0,i})\}$ of input 
images $I_{0,i}$ with 64 components, which has been tested before by \cite{Angles2018} for deep learning generative models. This proved to be more robust on noisy inputs 
than training the network directly on the scattering coefficients, as the scattering coefficients first capture morphological information at fixed scales, which mostly do not contain the high-frequency noise), which is then further de-noised by the PCA. This two-step process facilitates the selection of morphological modes only, and produces a latent space that is not arbitrary, but carries significant, correlated information. Later on, however, we also compute the reduced scattering coefficients of the reconstructed 
noise-free HST data, to classify the objects and facilitate conditional sampling of galaxies.

Overall, the model architecture resembles an autoencoder, with a CNN decoder, but a manual encoder to create latent vectors. The CNN itself consists of a linear fully connected layer followed by 5 transposed convolutional layers with batch normalisation and ReLU activation functions \citep{Agarap2018}.
These iteratively expand the compressed latent space data via convolutions with $5\times 5$ kernels until a final $\tanh$ activation to get the generator output. 
Figure \ref{fig:neural-net-architecture} depicts an overview of the CNN architecture. We then also calculate the second-order reduced scattering coefficients 
$\vec{s_2}^{j_1, j_2, l_1}$ (hereafter, $\vec{s_2}$) of the generated reconstructions, which have previously been shown to be able to trace morphologies in galaxy images \citep{Cheng2021}.  

We trained our model $\hat{\omega}$ on the embeddings $\{\boldsymbol{z}_s(I_{0,i})\}$ by minimising a $L^1$ loss function such that
\begin{equation}
    \hat{\omega}\left(\{\boldsymbol{z}_s(I_{0,i})\}\right) = \underset{\omega \in\mathcal{G}}{\mathrm{argmin}}\sum_{i=1}^n |
    I_{0,i} - \omega(\{\boldsymbol{z}_s(I_{0,i})\})|\, ,
\end{equation}
where $\mathcal{G}$ represents the class of CNNs with the specified architecture. The generator is trained in batches of size 128 and its hyperparameters are iteratively 
improved by an ADAM optimiser \citep{Kingma2014}. 

To compare the outputs of our model with the original images, we first perform a qualitative inspection by plotting the samples from the generative model on a few test 
images next to the corresponding HST galaxy. Additionally, we estimate the best-fit double-Sérsic profile of each galaxy to visualise the gain of our model with respect to 
common parametric methods. To obtain this fit, we use the \texttt{pysersic} package, which employs Bayesian inference methods for this task \citep{Pasha2023}. After 
estimating a prior on the fit parameters from the input image, the code finds a posterior distribution by either full MCMC or using stochastic 
variational inference \citep[SVI,][]{Hoffman2012}. We use the latter (mainly due to its speed), which initially finds the maximum a posteriori (MAP) parameters with SVI and then 
samples from a narrow Gaussian distribution around these values to obtain the best-fit model. 

In Fig. \ref{fig:neural-comparision_hst_wst}, we show this comparison between the input HST data and the realisations from $\hat{\omega}$ for a selection of galaxies. As one can observe, 
the CNN is able to easily capture details that deviate from the parametric representation, resulting in an overall reduced residual and a detection of features in the surface brightness 
that consistently surpasses the capabilities of the employed parametric models. This gain is naturally not as pronounced for galaxies that closely match the regular disc-bulge or elliptical morphology,
as for example visible in the first row of the plot. Nevertheless, our model can introduce additional substructure, which is relevant for shear bias calibration. We also display the pixel-wise root mean square error (RMSE) values between the original HST image and either the generative model reconstruction or the parametric fit. As all images are normalised to the same interval and the reconstructions do not contain noise, the values are not qualitative comparisons, but rather a way to show that the best-fit Sérsic profile is consistently less accurate.

\begin{figure}
    \centering
    \includegraphics[width=1.0\columnwidth]{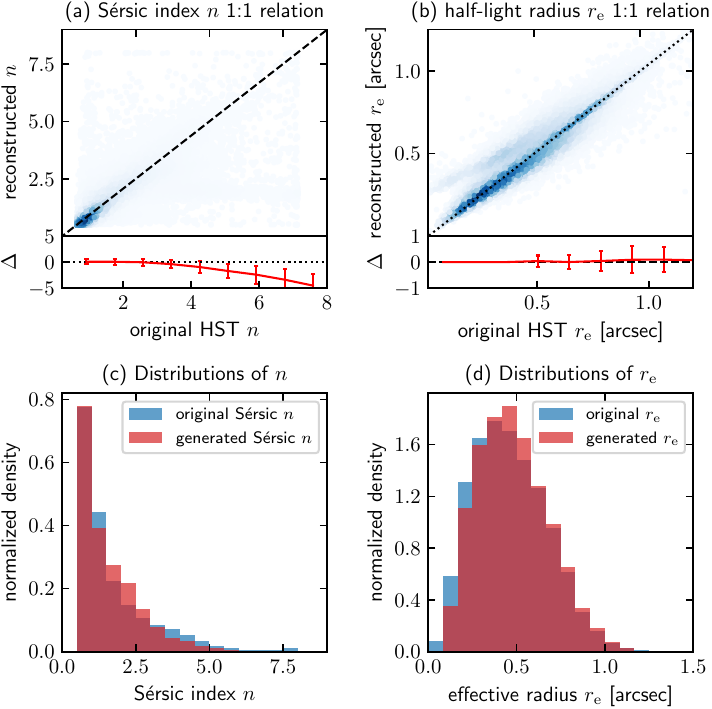}
    \caption{Comparison of Sérsic index $n$ and half-light radius $r_\mathrm{e}$ recovery on original and reconstructed images. Panel (a) shows the 1:1 relation for the Sérsic index, and panel
    (b) shows the 1:1 relation for the half-light radius. Panels (c) and (d) display the overall distributions of the two parameters in both samples. The residual $\Delta$-plots show the means and standard deviations of the relative difference between the original and reconstructed subsets over equi-spaced bins.}
    \label{fig:hst-wst-sersic-distribution}
\end{figure}

While there are of course residuals besides pure noise, extremely small-scale deviations from the original images are not concerning for the shear calibration simulations. As the 
\Euclid VIS instrument operates at half the pixel resolution of our output, the generated sample has to be processed by a \textit{Euclidisation} pipeline which adds a correct PSF and
re-samples the image with roughly $2\times 2$ binning, thus reducing the overall resolution. Such differences might affect the possibility of training the CNN directly on Euclid Deep Field data, which is an option that could be investigated in the future. 

\begin{figure}
    \centering
    \includegraphics[width=1.0\columnwidth]{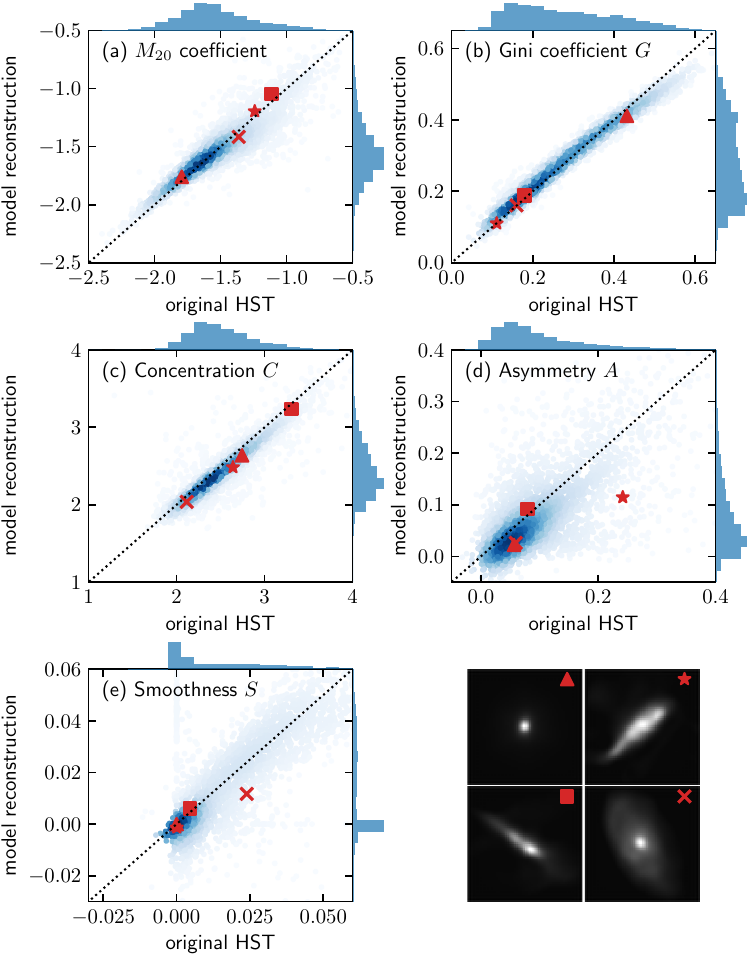}
    \caption{Comparison of morphological proxies between original HST images ($x$-axis) and reconstructions by the generative model ($y$-axis). Panel (a) shows the $M_{20}$ index, and panel (b) shows the Gini coefficient. Panels (c-e) display the CAS statistics, and panel (f) depicts four example images from the test data set with their values for the respective parameters shown according to the coloured markers.}
    \label{fig:hst-wst-morph-distribution}
\end{figure}

The proposed model is not the first generative neural network for galaxy morphologies developed recently, as previously another architecture has been used within the context of \Euclid \citep{Bretonniere-EP13, Bretonniere-EP26}. There are, however, some key differences. The presented architecture facilitates extremely fast training and sampling, the latter of which is necessary for the large simulation volume required for a successful \Euclid shear calibration campaign. Moreover, it allows for an easy extension to multi-channel inputs and outputs in order to learn and simulate morphologies in multiple filter bands jointly. This improvement will be part of future work and is an important step in obtaining true morphologies across both \Euclid instruments and for calibration of colour gradient biases \citep{Semboloni2013}.

\subsection{Recovery of galaxy structural parameters}
As an additional step for the validation of the reconstructions, we look at the distribution of the common galaxy structural parameters in the test data set, namely the Sérsic
index $n$ and the half-light radius $r_\mathrm{e}$. This allows us to check if the generative model generalises well upon application to data outside of the training set, which is paramount for subsequent sampling stability and reliability. Moreover, as shape measurement biases depend on the $n$ and $r_\mathrm{e}$ distributions, an accurate recovery is necessary for the next steps of the main goal of the work. 

We again estimate these Sérsic model properties by performing fits of single Sérsic profiles on the original HST image, as well as on the generated output. We mimic the inputs with our new sample by matching the flux and noise properties between the input and the generated image. Additionally, we restrict our model fitting by fixing the priors on both data sets to assure identical centroid positions and fluxes for each galaxy pair. 

\begin{figure*}
    \centering
    \includegraphics[width=2.0\columnwidth]{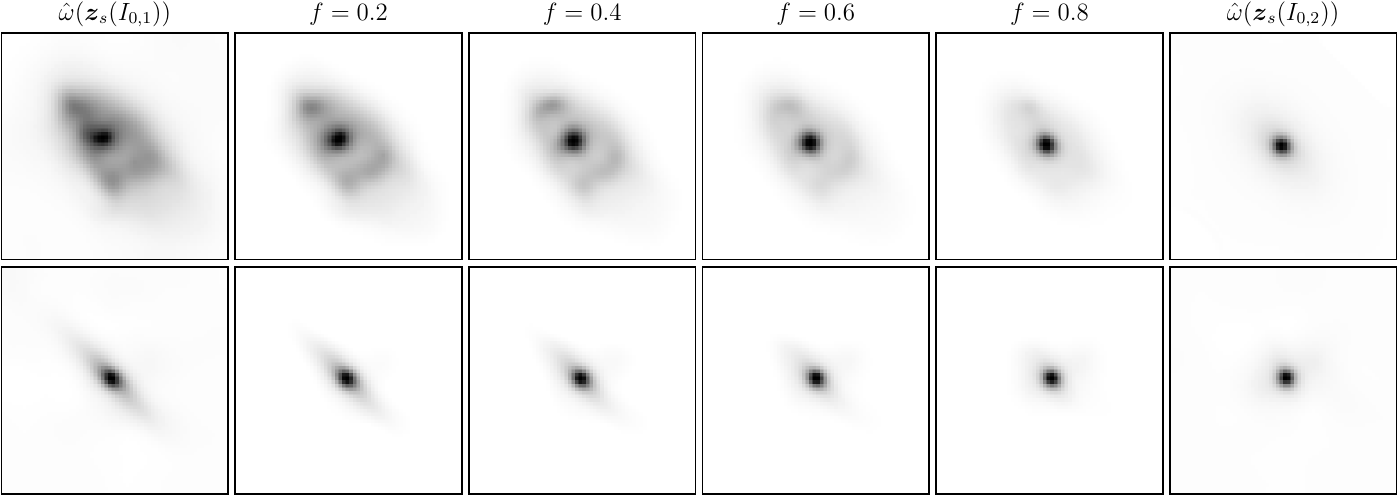}
    \caption{Interpolation between two input galaxies with embeddings $\{\boldsymbol{z}_s(I_{0, 1})\}$ and 
    $\{\boldsymbol{z}_s(I_{0, 2})\}$. The columns in the middle display the intermittent results by feeding interpolated embeddings in 0.2 increments into the generative model $\hat{\omega}$. The images are displayed inverted to shown faint details.}
    \label{fig:generator-interpolation}
\end{figure*}

Recovering the general input distributions is an additional indicator for success of the generative model and important for sampling of new galaxy images, as these parameters are necessary for realistic generations, due to the existing 
knowledge on spatial distribution and number densities of these structural parameters for true galactic populations \citep{Shen2003}. Moreover, a correlation between shear measurement bias and Sérsic index 
has previously been shown by \citet{Pujol2020} and  \citet{HernandezMartin2020}. Therefore, samples with accurate Sérsic index distributions in the tomographic redshift bins are required for calibration of the shape measurement. 

Figure \ref{fig:hst-wst-sersic-distribution} shows the 1:1 relations and histograms between the best-fit parameters calculated on the original HST image and the corresponding reconstruction output by the generator. The latter is thereby matched by flux to its original counterpart and then given a noise level that resembles the one on the original HST input image. We see that our generative model $\hat{\omega}$ generalises well on the test data, as the measured structural parameters are recovered well for the
majority of galaxies. The mean difference of Sérsic indices is $\langle \Delta n \rangle = 0.13\;(-0.005)$ with a standard deviation of $\Delta\sigma_n = 0.68\;(0.46)$, where the numbers in parentheses are the values when only considering objects up to $n\leq 4.0$. For the half-light radii, the mean scatter and its standard deviation are similarly small, with $\langle \Delta r_\mathrm{e} \rangle = -0.02\;(-0.01)$ and $\Delta\sigma_{r_\mathrm{e}} = 0.08\;(0.07)$. Towards high Sérsic indices $n\geq 4$, the fit accuracy breaks down, although it should be noted that the sample size is small at these values. Still, the overall distribution remains precise, with some excess at intermediate Sérsic indices in the generated sample. The scatter towards higher values of $r_\mathrm{e}$ on the reconstructed sample can be mostly attributed to the fitting procedure, where the models for strongly peaked galaxies with high Sérsic indices are not 
easily distinguishable using the SVI posterior estimation and thus often produce offset half-light radii. To check the recovery of these galaxies, we show in App. \ref{apdx:sfdiff} a sub-sample of such galaxies with the strongest fit offsets between the HST image and the generator output. Looking at the actual images of the galaxies with the strongest offset from the 1:1 relation, we see that the CNN recovered their overall shape similarly well, meaning that the large difference is a product of the degrading fit accuracy. This mostly happens for very concentrated galaxies (high-$n$ objects) with high S/N.

\subsection{Recovery of morphological statistics}

Next, we check the statistics on disturbed morphologies for both galaxy image samples. Again, an accurate reconstruction with subsequent noise and flux matching should be able to 
recover the values of the input for the various morphology parameters introduced in Sect. \ref{sc:GalMorph}. In Fig. \ref{fig:hst-wst-morph-distribution}, we compare a set of morphological proxies, namely $M_{20}$, the Gini coefficient $G$, and the CAS tracers. Concentration $C$, $M_{20}$ and $G$ are recovered well for the majority of galaxies. The parameters $A$ and $S$ follow the 1:1 relation as well, although with more scatter compared to the other parameters, especially towards the higher end of their respective ranges. We also depict four example galaxies with varying shapes to indicate where
the main galaxy populations reside within the plots. The estimation of these statistical parameters is again dependent on the noise matching, which is presumably the strongest source of scatter, as shown by \cite{Conselice2000},  \cite{Conselice2003}, and \cite{Lotz2004}. Furthermore, offsets can be introduced by the segmentation algorithm applied to separate the galaxy from the background, where redshift-dependent biases may arise due to surface brightness dimming \citep{Freeman2013}. We implement the segmentation method from the \texttt{photutils} package \citep{Bradley2022} for our analysis. Overall, we find a good and robust recovery of most of the morphological proxies, which proves the capabilities of the reconstructive power of our generative model. 

In general, there is only limited knowledge so far on the distribution of these properties in observed samples of galaxies across redshift bins, as not a large amount of data exists, which has a high enough resolution to reliably determine such proxies. We do however expect a dependency of the shear bias from complex morphologies on some of the parameters, as shapes with more disturbance from smooth profiles should lead to larger overall deviations for an ellipticity estimator. In Sect. \ref{sc:ShearMeas}, we will check this dependency for \Euclid-like simulations.

\section{\label{sc:GalGen} Generation of new galaxies}

\subsection{Galaxy-galaxy interpolation}

To generate new galaxies from our trained model $\hat{\omega}$, several options are at hand. The common approach for VAEs or GANs is to sample 
directly from the distribution of latent space variables. This, however, is accompanied by the risk of also generating images whose latent
values originate from the multivariate distribution spanned by the training set, but do not possess shapes that fit into the pool of observed
galaxies. This can occur for example when the latent space is not compact, resulting in a possibility of arbitrary output shapes. Additionally, this makes conditional sampling difficult without training a secondary latent space model \citep[see, e.g.,][]{Lanusse2021} or using 
a larger set of input galaxies that can be binned without restricting the generalisation power of the generative model. 

Another possible path for the conditional generation of galaxies is linear interpolation, by leveraging the linearity of the scattering fields \citep{Angles2018}, which extends 
onto our PCA components, as the PCA is itself a linear operation. Such latent space linear arithmetic calculations have also been shown to allow for
informative sampling in GANs \citep{Bojanowski2017} and prescribe a common test for generative model performance. Assuming that the \texttt{GalSim} COSMOS data set is representative of the general plethora of possible galaxy shapes, every additional galaxy can be
interpreted as an intermediate shape and ergo as an interpolation between two different galaxies from the whole sample. The only galaxy population that presumably does not fit into
this space are highly irregular galaxies, for them, the shapes are not conformable with any known physical model anyway, as their name suggests. Thus, a potential irregular 
object created from a generator cannot be definitively confirmed or refuted as a realistic representation. Still, caution is required also for common shapes, due to the fact that the latent space between two objects is not necessarily fully covered, so interpolating between any arbitrary galaxy pair might not lead to realistic shapes. Interpolation between two edge-on galaxies that are rotated by 90$^{\circ}$ with respect to each other could ensue intermediate realisations of, for example, cross-like shapes and therefore produce overall more irregular morphologies, which we found when just randomly interpolating between galaxies from the training set. Hence,
a fine-tuning of the operation is necessary to ensure realistic shape distributions. Another caveat of training data set size is the small area of the COSMOS field, with potentially large cosmic variance, meaning that the assumption that the full multidimensional parameter space is covered can be wrong. In the future, the training set needs to be expanded towards a more diverse galaxy sample.

Given the embeddings  $\boldsymbol{z}_s(I_{0, 1})$ and $\boldsymbol{z}_s(I_{0, 2})$ of two original HST galaxies calculated with the formalism described 
in Sect. \ref{sc:NoiseRemoval}, we can obtain a linearly interpolated latent space vector $\boldsymbol{z}_s(I_{0, f})$  with
\begin{equation}
    \boldsymbol{z}_s(I_{0, f}) = \left(1 - f\right) \, \boldsymbol{z}_s(I_{0, 1}) + f \, \boldsymbol{z}_s(I_{0, 2})\, ,
    \label{eq:interpolation}
\end{equation}
where $f\in[0, 1]$ is the fraction of interpolation between the individual latent space components of both initial galaxies. Feeding such new realisations
into the trained generative model allows for a transformation from a linear operation in the latent space into a non-linear interpolation in image space. In principle, this may be extended to extrapolations, with $f$ values that lie outside of the mentioned interval. Alternatively, one could calculate the interpolation directly on the scattering fields, as the WST is, however, not invertible, and a gradient descent method is thus needed, which requires the training of a secondary CNN for a regularised inversion. 

\begin{figure*}
    \centering
    \includegraphics[width=1.8\columnwidth]{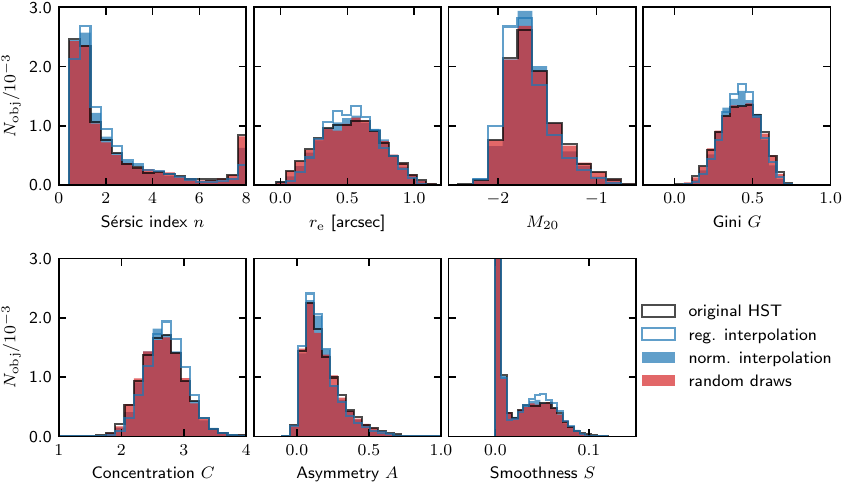}
    \caption{Comparison of structural parameter distributions between original HST images reconstructed by the generator (black), samples composed by interpolating between galaxies (blue line), normalised interpolation (blue), and samples formed by random draws from the latent distribution (red). The term $N_\mathrm{obj}$ is the number of objects found per bin.}
    \label{fig:random-vs-interpolation}
\end{figure*}

Latent space interpolation has previously been shown to be prone to distribution mismatch, where the latent priors are narrowed by sampling in this manner, leading to possibly incomplete coverage of the posterior distribution of images \citep{Kilcher2017}. To alleviate this issue, several methods can be incorporated, as for example normalised interpolation \citep{Agustsson2017}, where the 
intermediate embedding vectors are given by 
\begin{equation}
    \boldsymbol{z}_s(I_{0, f}) = \frac{\left(1 - f\right) \, \boldsymbol{z}_s(I_{0, 1}) + f \, \boldsymbol{z}_s(I_{0, 2})}{\sqrt{(1-f)^2} + f^2}\, .
    \label{eq:norm-interpolation}
\end{equation}

Randomly loading pairs of embeddings from the joint training+test data set and interpolating between them with an arbitrary value for $f$ thus constitutes another simple way, aside from random draws of the latent space distribution, to generate new objects. The number of possible novel galaxy instances hereby by far exceeds the requirements for the \Euclid shear calibration,
as a 0.1 spacing for $f$ already produces visibly varying morphologies and allows for combinations of the order of $\mathcal{O}(10^{10})$ with $\sim 50\,000$ training objects, which can be further increased in the future by incorporating a larger postage stamp sample. 

Figure \ref{fig:generator-interpolation} shows how this procedure translates into the image space. Depicted are original galaxies on the left- and rightmost subplots, with four intermediate realisation obtained via 0.2 increments for $f$ in Eq. \eqref{eq:interpolation}. It is apparent how the overall shape is shifted in a continuous way amidst the two sample objects, demonstrating the capabilities of the CNN and the interpolation procedure. 

To preserve the input shape distribution and avoid unrealistic morphologies due to non-compactness of the embeddings space or orientation-related sampling issues, a more restrictive interpolation may thus need to be incorporated. While the standard random interpolation might produce realistic shapes for ellipticals, this is not generally the case for all pairs of galaxies, thus requiring a fine-tuning of the interpolation. 

\subsection{Interpolation fine-tuning}

To circumvent such concerns, several solutions can be realised. For once, we can limit the interpolation fraction. This, however, reduces the amount of clearly discernible galaxies from our data set and could result in too many similar looking galaxies that do not necessarily cover the range of realistic shapes that will be observed by \Euclid. Another option is to disturb existing galaxy images not in a specific direction of the latent space, but by diffusion or random walks in the neighbourhood of their embedding vectors. This though could again give rise to an overall unrealistic distribution of numerous too similar objects or risk the generation of non-physical objects, as we do not have knowledge of the latent space topology mapped by the optimised generative model. 

Therefore, we test a different method that should allow for more variable generation and simplify the realisation of conditional sampling by Sérsic index or $r_\mathrm{e}$ distributions. To be able to use the full range of interpolation fractions $f$ and diminish the likelihood of unrealistic light profiles, we first rotate all original HST galaxies to match along their major axis, which we set as the axis of the best-fit Sérsic model. In consequence, we avoid the possibility of generating objects with for example cross-like shapes, as the interpolation then always takes place along the same axis. Galaxies for which a clear symmetry cannot be reasonably assigned, that is irregulars, do not pose a threat to this framework, as they have intrinsically peculiar shapes and will therefore naturally lead to generations of new irregular instances, irrespective of their rotation. Then, we recreate the embeddings for these new images and re-train the generator $\hat{\omega}$ on this data set. We choose the 45$^\circ$ diagonal with respect to the $x$-axis as the designated direction, as we can thereby steer clear of rotating large galaxies previously residing along this direction out of the image bounds. It should be noted that rotation in the image space can results in information loss due to interpolation onto a new pixel grid with \texttt{GalSim}. This should, however, be irrelevant towards the application for \Euclid, as the pixel scales differ by a factor of two and fine details will be smeared out by the re-binning and noise application. With this new data set of generated galaxies, we can obtain new instances over the full range of $f$ by drawing pairs of objects and interpolating between them along the diagonal axis.

\subsection{Random draws versus interpolations}

Next, we check whether one of both interpolation methods, regular and normalised, provides an advantage with respect to the common random multivariate sampling approach for galaxy generation if no conditionality is needed. For this, we employ all three techniques to create 10$^4$ random galaxies, respectively, and also randomly choose 10$^4$ galaxies from the reconstructed HST training data. Afterwards, we compare the distributions of structural parameters and $CAS$+$GM_{20}$ statistics between all sets of objects and selection from the input training set. 

Figure \ref{fig:random-vs-interpolation} displays the histograms of the measured properties for all subsets. Clearly, random latent space drawing can  more consistently trace the parameters distributions of the original HST data set sample, with almost identical histograms. The regular interpolation method on the other hand is similarly reliable on the Sérsic index recovery, but fails to capture the distribution tails for the effective radius and the $CAS$+$GM_{20}$ morphology proxies, even though the means are captured correctly. 

This is a logical consequence of the method, due to the aforementioned distribution mismatch. The probability of drawing a galaxy from one of the tails is low to begin with, and the likelihood of generating such an object will be decreased upon interpolation with a sample that most likely does not reside in the same regime. This leads to a narrowing of the latent distribution, which translates to the image space and hence to the measured properties. As can be seen in the plot, normalised interpolation reduces this effect, but is still not capable of achieving the quality of the results from random draws.

Still, we note that linear interpolating along a predefined axis did not produce objects with non-physical structural parameters or morphology statistics outside the input distribution. This proves the capability of the technique for the task at hand and indicates a tightly packed latent space. While the random draw method delineates a powerful tool for galaxy generation if only the input distribution shall be recovered, we will further explore the interpolation approach for conditional sampling of galaxy properties.

We note here that while the $CAS$+$GM_{20}$ parameters provide a well-established set of morphological estimators, there is no clear prior information on which values or distributions would describe non-physical shapes. Moreover, irregular galaxies can most likely reside anywhere in this parameter space, as their origins lie in turbulent process that can create a wide range of complex structures. Thus, a distribution match between samples and observation does not necessarily prove the realism of the generations, but is still a valuable indicator of the model performance. 

\subsection{Galaxy classification}

One step towards conditional sampling can be to group the galaxy data set roughly into main categories of for example ellipticals, spirals, and irregular galaxies. Overall, there exist a multitude of methods for
galaxy classification along the Hubble sequence. This can be achieved using intrinsic galaxy properties such as star-formation rate (SFR) or colour \citep{Kennicutt1998}, joint analysis of morphological tracers (e.g. Gini$-M_{20}$), or even citizen science projects such as Galaxy Zoo \citep{Darg2010}. While classification with Galaxy Zoo is able to make more distinct classifications, it cannot handle the amount of data in stage IV surveys such as \Euclid. Machine learning techniques trained on Galaxy Zoo results, however, have recently been shown to be able to classify \Euclid morphologies directly from the images \citep{Aussel2024}. This, as well as using morphological proxies, requires pixel data and relies on rather arbitrary thresholds for classification. Leveraging galaxy population properties, which on the one hand only uses photometry, is usually only able to robustly separate the bimodal distribution of early-type ellipticals and late-type spiral galaxies \citep{Baldry2004}. For deep generative models, a classification can also be obtained via the learned embeddings. Here, studies for instance on GANs have previously shown the manifold clustering capabilities of their low-dimensional latent space distributions \citep{Mukherjee2018}.

\begin{figure}
    \centering
    \includegraphics[width=0.8\columnwidth]{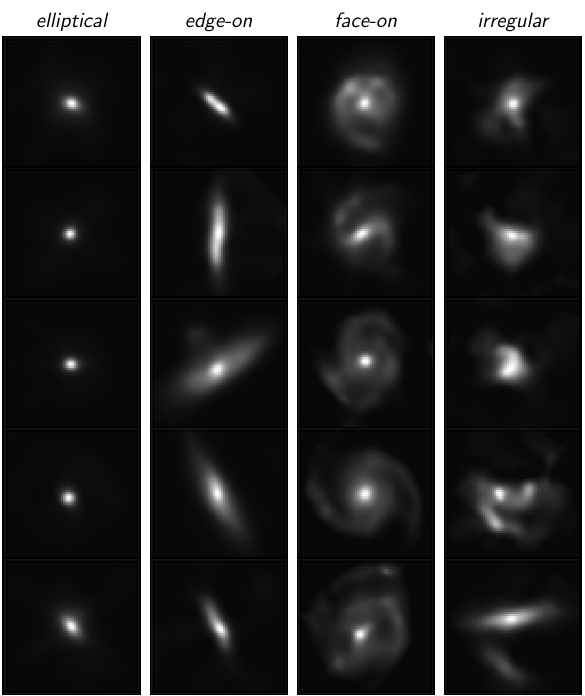}
    \caption{Example galaxies for the four classes \textit{elliptical, edge-on, face-on}, and \textit{irregular}, classified by fitting a bGMM to their second-order reduced scattering coefficients $\vec{s_2}$ and assigning labels to each component.}
    \label{fig:galaxy-classification}
\end{figure}

We here employ a different method by performing a clustering analysis on our galaxies via the wavelet scattering transform. Given the scattering coefficients of each image, we can determine if and how the multi-dimensional parameter set correlates with each galaxy's intrinsic light profile and accordingly its overall class affiliation. For this, we calculate the second-order reduced scattering coefficients $\vec{s_2}$ with $L=4, J=6$ for each galaxy, as they have been shown to correlate with galaxy morphology \citep{Cheng2021}. Next, we fit a Bayesian Gaussian mixture model (bGMM) to their distribution. In general, a Gaussian mixture model is a probabilistic model that describes a weighted sum of $k$ multivariate normal distributions $\mathcal{N}$ given by
\begin{equation}
    p(\boldsymbol{x}) = \sum_n^k \pi_n 
    \mathcal{N}(\boldsymbol{x}\; |\; \mu_n, \Sigma_n)\, ,
    \;\;\;\;\; \sum_n^k \pi_n = 1\, , 
\end{equation}
where $\pi_n$ is the weight of the $n$-th component and $\mu_n,\Sigma_n$ are the respective means and covariance matrices, $p(\boldsymbol{x})$ is the distribution of input vectors, in our case with $\boldsymbol{x}=\vec{s_2}$. In a bGMM, the parameters of the model are not found with an expectation maximisation algorithm, as is the case for common GMMs, but by variational inference of an approximate posterior using a Dirichlet prior on the parameters and then maximising the log-likelihood $\ln\mathcal{L}(p(\boldsymbol{x}))$. We use the bGMM implementation from the \texttt{scikit-learn} package \citep{Pedregosa2011}. Afterwards, we attribute a keyword to each component, based on the general visual appearance of the items within the respective cluster. These keywords are \textit{elliptical, edge-on, face-on, irregular}. 

Figure \ref{fig:galaxy-classification} visualises a random sample of galaxies for each class. It should be acknowledged that the class \textit{edge-on} hereby does not necessarily only consist of disc plus bulge galaxies with an inclination angle of 90$^\circ$ relative to the line of sight, but only groups objects which appear elongated and could thus also include, for example, lenticular galaxies with a large axis ratio. Moreover, the differentiation between face-on spirals and a subset of the irregular class is not trivial, resulting in possible contamination of the peculiar sample by regular spiral galaxies. Besides, the exact differentiation between face-on and edge-on is overall rather arbitrary, as naturally galaxies exist over the full range of inclination angles. While this method is by no means a classifier that can currently compete with machine learning models such as \texttt{Zoobot} \citep{Walmsley2022}, it still showcases the morphological information carried by the wavelet scattering transform, as also shown by \cite{Cheng2021}. Due to the fact that we do not necessarily need to sample from overall morphological galaxy populations for the quantification of the shear bias, a more complex model is not needed. Still, these general morphology flags provide a first method towards conditional sampling of the COSMOS data set by drawing objects from the latent distribution of each cluster.

\subsection{Conditional sampling}

\begin{figure}
    \centering
    \includegraphics[width=\columnwidth]{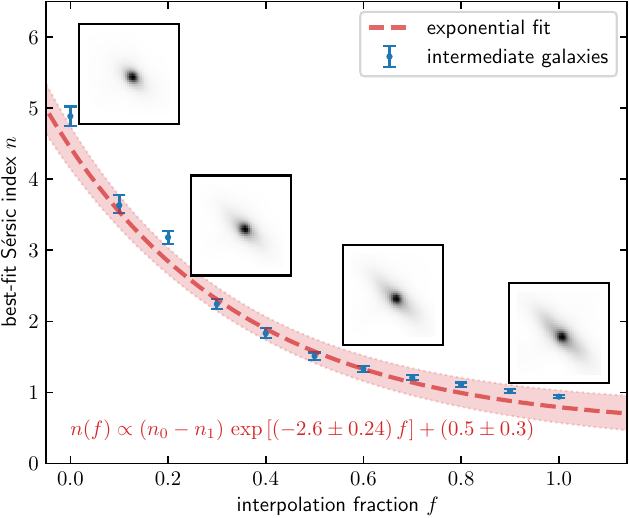}
    \caption{Dependence of the Sérsic index on the interpolation fraction $f$. The blue points and error bars are the respective best-fit Sérsic index and the estimated error from \texttt{pysersic}. The red dashed curve shows our best-fit parametric model, with the dotted region designating the $1\sigma$ confidence interval. The image insets are inverted stamps at intermediate interpolation points. }
    \label{fig:exp_fct_interpolation}
\end{figure}

Image simulations for shear calibration are needed not only for random fields, but may need specific structural property distributions such as for cluster fields, therefore requiring conditional sampling methods. To sample new galaxies by their Sérsic indices, we can leverage the latent space interpolation technique. To draw an object at a specific Sérsic index $n_i$, we draw a galaxy with $n_0 > n_i$ and another object with $n_1 < n_i$ from the joint reconstructed training/test data set. Then, conditional sampling only requires finding a robust functional correlation between the interpolation fraction $f$ and the Sérsic index. To obtain such an empirical description of this dependence, we draw random pairs of galaxies with $n_0 > n_1$ and calculate their intermediate realisations by propagating the interpolated latent representations in 0.1 steps for $f$ through the generator, then measure their structural parameters and finally parametrise the functional form of the Sérsic index development. 

We find that the correlation of $f$ and $n$ can be approximated well by an exponential law following
\begin{equation}
    n(f) \propto (n_0 - n_1)\, \mathrm{e}^{-\beta\, f} + \eta, 
    \label{eq:int-frac-n}
\end{equation}
where the parameters $\beta = 2.6\pm 0.24, \eta= 0.5\pm 0.30$ were determined empirically as the mean values from an exponential fit to the $n(f)$ relation for multiple galaxy pair draws. Figure \ref{fig:exp_fct_interpolation} shows the functional form of $n(f)$ with the 1$\sigma$ confidence interval and example images of a galaxy pair for the interpolation along the exponential law. A similar parametrisation can also be constructed for the half-light radius, to sample galaxies according to this property. 

\begin{figure}
    \centering
    \includegraphics[width=\columnwidth]{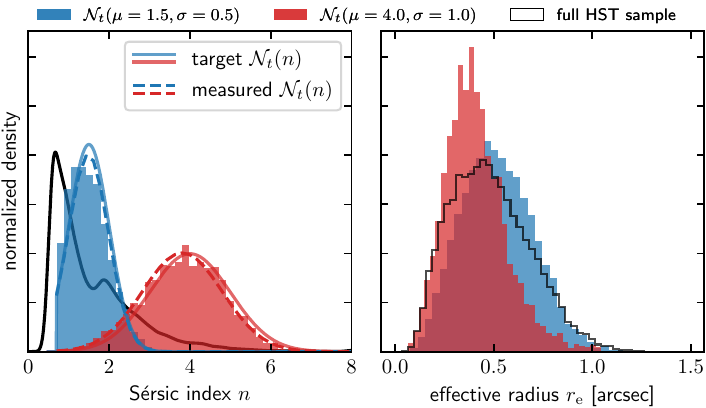}
    \caption{Results of conditional sampling of galaxies with a normal Sérsic index distribution. The left plot shows the obtained Sérsic index distributions, on the right we show the corresponding effective radii. The black curve depicts the overall distributions from the HST training data.}
    \label{fig:sersic_sampling}
\end{figure}

\begin{figure*}
    \centering
    \includegraphics[width=2.0\columnwidth]{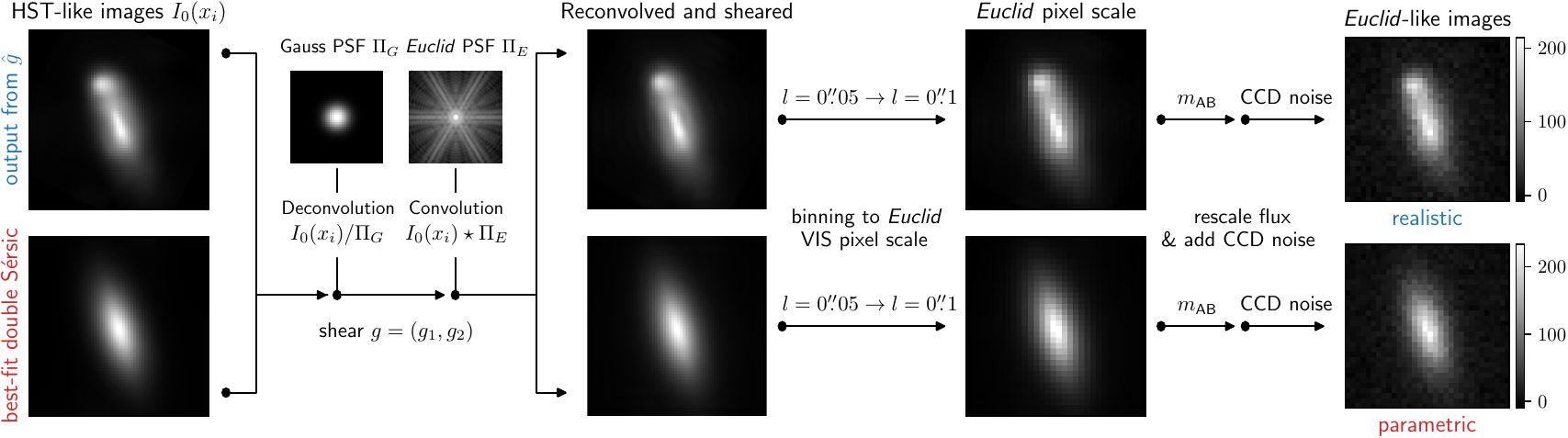}
    \caption{Procedure for creating \Euclid VIS-simulated galaxies from the generator output and their parametric double-Sérsic counterparts. First, the generated images are deconvolved by their Gaussian training PSF, and then a shear is applied, after which they are convolved with the \Euclid PSF from App. \ref{apdx:PSF}. Next, the galaxies are re-binned to the correct pixel scale of $l=\ang{;;0.1}$. Finally, the total image flux is rescaled according to an apparent magnitude following Eq. \eqref{eq:flux-mag} and CCD noise is added using the \texttt{GalSim.CCDNoise} tool.}
    \label{fig:euclidization}
\end{figure*}

The main drawback of this approach is the fit accuracy at high Sérsic indices. As precise fits of Sérsic profiles are difficult for strongly peaked galaxies without time-intensive sampling, the correct interpolation might suffer under galaxy draws that are mismatched from the target index, thus leading also to differing intermediate representations. To alleviate contamination of the target sample, we reject every galaxy that deviates by more than $\Delta n = 0.5$ from its target value. We observe that the rejection fraction increases with $n$, which confirms the hypothesis of mismatches due to high-Sérsic fitting difficulties. 

Given this correlation, we can sample either specific single objects or distributions of Sérsic indices from the generator $\hat{\omega}$. To achieve this, we draw values for $n$ randomly from the target prior. Then, we choose two galaxies with $n_0>n>n_1$ from the train/test images and calculate the required interpolation fraction according to Eq. (\ref{eq:int-frac-n}) and generate the corresponding image from the interpolated embedding vector. Afterwards, we measure the Sérsic index of the obtained object to validate the result. 

To present the capability of this sampling method, we apply this routine to generate two arbitrarily chosen samples of 10$^4$ objects following a truncated normal distribution $\mathcal{N}_t(\mu, \sigma, [\min, \max])$ for the Sérsic index over the interval $[0.5, 8.0]$, peaked at means of $\mu=2$ and $\mu=4$ and with respective standard deviations of $\sigma=0.5$ and $\sigma=1.0$. Figure \ref{fig:sersic_sampling} shows a comparison of the target Sérsic index distributions and the obtained samples, as well as the distributions of the corresponding half-light radii. We find that we are able to accurately construct galaxies given a target Sérsic index distribution. Moreover, the technique also allows for sampling of individual objects at specific $n$ values over the range of the training data distribution. Additionally, the effective radius of the generated sample does not strongly depart from the range of the original HST distribution, which confirms that the interpolation approach does not create galaxies outside of the input range that might thus exhibit unrealistic shapes. 

While the rejection fraction of generated galaxies with offsets larger than 0.5 from their target value increases strongly with $n$ (up to $\sim 0.8$ for $7<n<8$), the majority of galaxies, especially at low Sérsic indices, where the observed number density is higher anyway, are created accurately and with morphological proxies that trace the input distribution, as shown in Fig. \ref{fig:random-vs-interpolation}. Furthermore, even high rejection fractions are not concerning given the speed of the generation, as $\sim10^4$ objects can be propagated through the generator in approximately one minute using only one CPU core. This can easily be accelerated even more with GPUs, given that the model architecture is written entirely with \texttt{pytorch} and already optimised for usage with \texttt{cuda} devices.

This conditional method allows for the direct and fast simulation of a large number of galaxies for insertion into full \texttt{GalSim} scenes and the \Euclid VIS pipeline. Not only can one sample using a specific value or distributions for Sérsic indices, but also by galaxy class (\textit{elliptical, face-on, edge-on, irregular}) via interpolation between objects from the respective label. Additionally, the two methods can be united, to for instance create a population of edge-on galaxies at Sérsic index $n=2$. Still, the diversity of such generated data is strongly susceptible on the size of the training data. This can be improved in the future by larger training data sets from different HST fields or Euclid Deep Field observations. Right now, this is not needed for the main goal of this work, which is the calibration of the shear measurement bias due to galaxy substructures.

\section{\label{sc:ShearMeas} Quantification of shear bias from complex galaxy morphologies}
\subsection{Measurement setup}

To reach the scientific requirements of \Euclid, the bias needs to be calibrated to an accuracy of $|\delta\mu| < 2\times 10^{-3}$, $|\delta c| < 1\times 10^{-4}$ \citep{Cropper2013}. However, most of this error budget is needed for PSF modelling errors. Hence, even sub-percent level biases have to be accounted for, as they exceed the target precision by more than one order of magnitude. Now that we have presented our method for generating galaxies with realistic, noise-free morphologies from the \texttt{GalSim} COSMOS sample, we will apply the technique for the estimation of the shape measurement bias by complex galaxy structures. The difference in shear biases between realistic and parametric objects can be leveraged for this task and determine if this difference is relevant for Euclid Wide Survey cosmic shear measurements. We perform the estimates here using KSB due to its speed, future work will apply more modern shape measurement methods such as \lensmc and \textsc{Metacalibration}. 

First, we need a sample of generated galaxies and their parametric counterparts which are both rendered identically (aside from the complexity of their shapes) and possess the same noise and PSF properties. For this, we sample randomly from the latent space of the trained CNN and create $5\times 10^5$ new images. We mainly want to quantify the bias with respect to the whole range of possible shapes and as a function of structural parameters and morphological proxies, a conditional sampling approach is therefore not needed. Afterwards, we perform double-Sérsic fits for each generated object using \texttt{pysersic}. We note that we here do not account for redshift--magnitude correlations yet, which can change the distribution of shapes due to increased irregularity at higher redshifts. For tomographic bias difference estimates this has to be accounted for. This can for example be done by adding the redshift as an additional latent dimension and learning the joint distribution of redshifts and shapes. 

As we try to determine the bias for \Euclid VIS, the generated galaxies have to be modified towards a \Euclid-like emulation. Such an \textit{Euclidisation} procedure has previously been developed by \cite{Scognamiglio2025}, although we can here abstain from noise whitening and symmetrisation. Instead, we just add CCD noise (consisting of Poisson shot noise from source and background, as well as Gaussian read-out noise) to the image using the corresponding \texttt{GalSim} function. Accounting for both of these noise sources is important for \Euclid, given its lower sky background in comparison to previous ground-based weak lensing surveys. First, however, we deconvolve the generated images by their original Gaussian PSF, apply a shear and convolve them with a fixed \Euclid PSF, as presented in \citet{Tewes2019} and  \citet{Jansen2023}. Next, we draw the galaxies at the correct pixel scale of $\ang{;;0.1}$, producing $32 \times 32$ pixel postage stamps. Then the galaxies need correct fluxes and noise levels. For this, we assigned to each galaxy a magnitude $m_\text{gal}$ following the distribution from the test+training data set but allowed magnitudes of up to $m_{AB} \leq 24.5$, which approximately corresponds to the VIS target-limiting magnitude. Finally, the noise was added as described above. For this emulation, we used the assumptions on the \Euclid VIS detector and observing conditions from \cite{Tewes2019}, namely the nominal VIS exposure time $t_\text{exp}$, gain $G_\text{VIS}$, read-out noise $R$, zero point $Z_\mathrm{p}$, and sky brightness $m_\text{sky}$ \citep{Refregier2010, Laureijs11, Niemi2015, Cropper16}. The sky level and galaxy flux at given object and sky magnitudes were thereby calculated following \cite{Tewes2019} with
\begin{align}
    F_\text{sky}  &= \frac{t_\text{exp} l^2}{G_\text{VIS}}
     10^{-0.4(m_\text{sky} - Z_\mathrm{p})}\, ; \\
    F_\text{gal} &= \frac{t_\text{exp}}{G_\text{VIS}}
     10^{-0.4(m_\text{gal} - Z_\mathrm{p})}\, , 
    \label{eq:flux-mag}
\end{align}
where $l$ is the pixel scale in arcsec. These parameters will have to be adjusted in the future once accurate measurements have been performed following the performance verification of the \Euclid mission. Figure \ref{fig:euclidization} exhibits an overview of this process to obtain pairs of realistic and parametric galaxies that emulate \Euclid VIS images. To perform shape noise cancellation, we additionally create duplicates of the original HST-like images rotated by 90 degrees for both sets of simulated galaxies, but with an identical noise field (the noise is not rotated), which increases the effective number of galaxies by a factor of two to $10^6$. 

We separately save an oversampled image of the applied \Euclid PSF, which is needed by the KSB method for a shear estimate. This PSF image can be found in App. \ref{apdx:PSF}. On top of that, we assign each object to one of the galaxy classes introduced in Sect. \ref{sc:GalGen} by calculating their second-order wavelet scattering coefficients and also measure the S/N  following the definition by \cite{Tewes2019} and the $CAS$+$GM_{20}$ statistics of the galaxies to later quantify the bias as a function of these properties. 

\begin{figure}
    \centering
    \includegraphics[width=1.0\columnwidth]{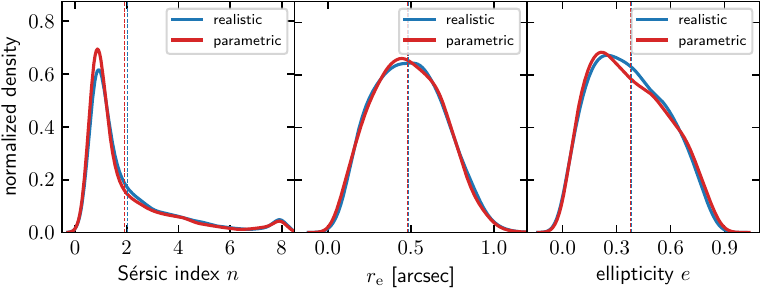}
    \caption{Sérsic index $n$, half-light-radius $r_\mathrm{e}$ and ellipticity $e$ distributions measured on the realistic (blue), and parametric (double-Sérsic, red) sample galaxies. The data are shown as a kernel density estimate on the histograms. The vertical dashed lines indicate the mean of the respective distribution.}
    \label{fig:n-re-e-dist}
\end{figure}

\subsection{Realistic morphologies versus double-Sérsic}

As a first measurement, we take the sample of 10$^6$ noise-less galaxies from the generator and their parametric double-Sérsic fits and emulate \Euclid VIS images following the described prescription, creating two measurement branches. Initially, we do this at fixed magnitudes over the range $m_\text{AB}\in[20.0, 24.0]$. With this, we intend to quantify the overall effect of the bias as a function of magnitude up to approximately the limit of the Euclid Wide Survey. As the fine details of galaxy substructures get increasingly washed out with a decreasing S/N, the galaxy morphology bias may exhibit a dependence on the S/N. While there is an additional dependency of galaxy complexity on redshift and magnitude correlations, we do not yet account for this here. Still, in the future this has to be addressed, as two independent effects can affect the bias difference: The evolution of the galaxy population with more irregular objects at higher redshift and thus lower S/N, as well as the improved resolution of substructures at higher S/N, with a possibly stronger impact on the shape measurement. As an initial estimate of this effect, we will later on show the dependence of the bias on the $CAS$+$GM_{20}$ statistics at a fixed magnitude. As aforementioned, this can in the future be implemented by adding latent space dimensions consisting of the relevant conditional parameters, which are for example magnitude and redshift.  

\begin{figure*}
    \centering
    \includegraphics[width=1.8\columnwidth]{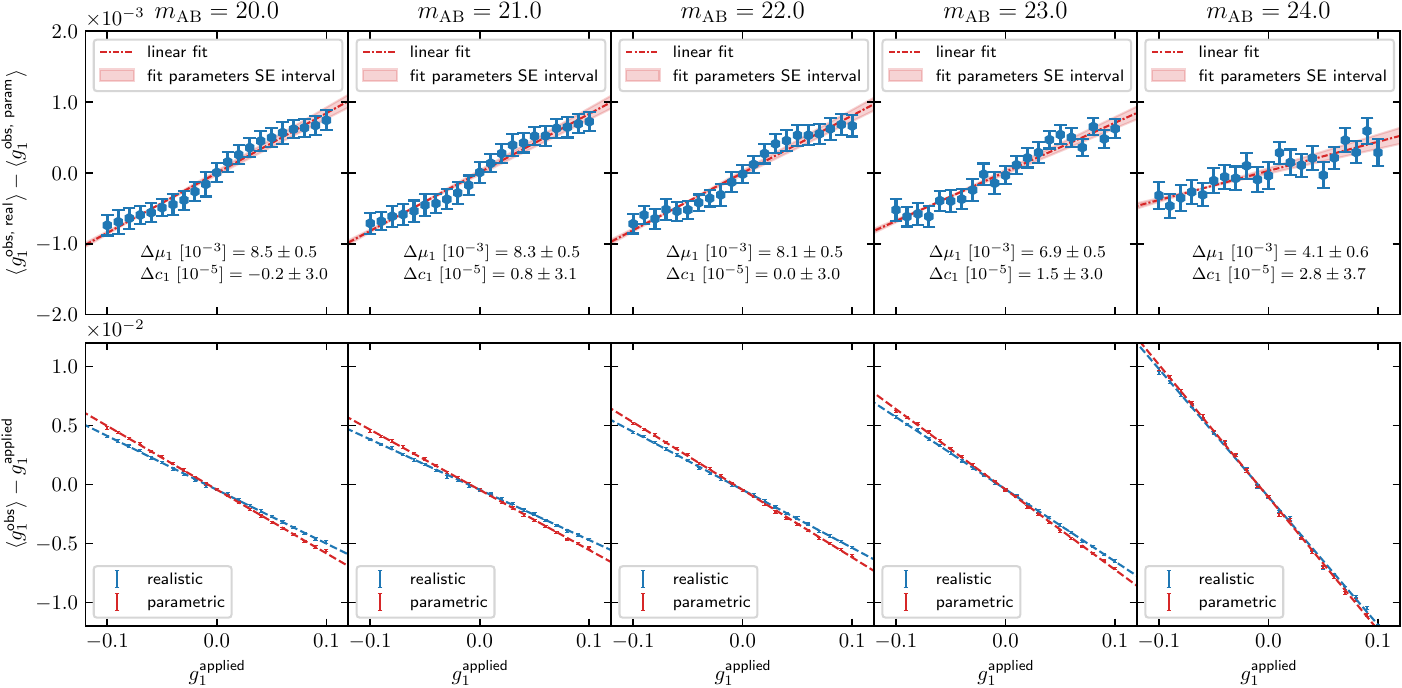}
    \caption{Results of the cosmic shear measurement with KSB on $1\times 10^6$ simulated galaxies at a fixed magnitude. The simulation was performed once with realistic galaxies from the deep learning model and once with their double-Sérsic counterparts. The bottom row depicts the individual shear biases measured on both branches, the upper row shows the difference of the mean shear at each increment with a linear fit and the standard error confidence interval for the fitted parameters $\Delta\mu_1, \Delta c_1$.}
    \label{fig:bias-mag-dep}
\end{figure*}

To ensure that both branches have identical Sérsic indices, half-light radii, and ellipticity distributions, we measured these properties with \texttt{pysersic} on the HST-like generator sample and the fitted double-Sérsic profiles. This is an important validation in order to warrant that the measured bias difference does not originate from deviations in said distributions, which has been shown to have a direct influence on the bias \citep{Hoekstra2015, HernandezMartin2020, Pujol2020}. In Fig. \ref{fig:n-re-e-dist} we show the distributions of these properties in both branches for the full set of 10$^6$ galaxies. As one can see, the distributions match over the full data set, which therefore guarantees that no significant influence from these parameters can be expected on the shear bias difference from complex morphology. Slight differences can also be attributed to the fast fitting procedure that might not always find the true best-fit Sérsic model on either of the two branches. Moreover, no real Sérsic fit can be obtained for complex objects with multiple brightness peaks for instance, naturally leading to differences between branches. One can argue that such differences can lead to a bias itself. This, however, would also be there in relation to real galaxies, where similar issues arise from the simplicity of a parametric light distribution. Thus, such discrepancies are a part of the morphology bias that should not be removed by using only objects with identical structural parameter distributions.  

During the emulation, we apply shears over a range of values for $g_1$, while we fix $g_2=0$. We choose a shear interval of $g_1^\text{applied}\in [-0.1, 0.1]$ and sample it in increments of 0.01. Then, we measure the shear components with KSB for each generated pair of galaxies. A key difference to the usual parametric image simulations is the fact that our training galaxies, and thus also the CNN output, already contain a non-zero cosmic shear, as they originate from real HST data. This can neither be removed nor accounted for, as we do not know the shear of the generated galaxies. However, that only adds an additional source of shape noise and also affects parametric models where usually an observed ellipticity distribution is used as the input.

After measurement of the shear on all galaxies over the specified shear interval, we calculate the mean $\langle g_1^\mathrm{obs}\rangle$ and standard error $\sigma/\sqrt{N}$ of the observed shear over the entire sample, where $\sigma$ is the standard deviation of the measured shear distribution over the full sample and $N$ designates the number of galaxies in the sample, that is $N=10^6$. Thereby, we employ the shape noise cancellation, where the mean and errors are calculated over the pre-computed means of each galaxy pair. If KSB fails for one of the two objects, or both, we simply reject the full galaxy pair to avoid selection bias effects for now. This way, we obtain the multiplicative and additive biases for both measurement branches, namely the realistic and the parametric samples, using Eq. (\ref{eq:bias-def}). We determine the errors on the linear fits of the bias via bootstrapping of random samples from the shear measurement and obtaining mean and standard deviations of the fit parameters from the bootstrap sample. In the absence of other bias sources that differ between the branches, the difference 
\begin{equation}
    \left\langle g_i^\text{obs, real} \right\rangle - 
    \left\langle g_i^\text{obs, param}\right\rangle 
    = \Delta\mu_i\, g_i^\text{applied} + \Delta c_i
    \label{eq:bias-morphology}
\end{equation}
can be used as a metric to determine the bias introduced on the shape measurement from realistic galaxy morphologies as opposed to simple Sérsic profiles. While there exist other sources of bias, as for example from pixel noise or a cut-off of the light profile at the edges of the postage stamp (which should be neglectable as our model was only trained on images that do not exceed the specific stamp size), these effects should be apparent in both data sets. Accordingly, these effects should be irrelevant for the main aim of this work, as we do not expect these bias sources to vary between both measurement branches due to their same distributions in S/N, $n$, $r_\mathrm{e}$, and ellipticity. 

While the aforementioned magnitude limit designates the target for the inclusion into the cosmic shear analysis for \Euclid, low surface brightness galaxies with high half-light radii might exhibit extremely low S/N despite lying in the overall magnitude range. Therefore, we applied a cut at $\mathrm{S/N}=10$ in the parametric branch and did not include galaxies from both branches into the bias estimation that fall within this sample.

In Fig. \ref{fig:bias-mag-dep}, we show the results of measuring the shear bias following Eq. \eqref{eq:bias-morphology}, where each column shows the measurements at a specific fixed magnitude that was applied to all sample objects. This allows one to trace how the bias changes with decreasing magnitude. While in general morphology and magnitude are tightly correlated, we neglected this correlation on purpose in this case to be able to trace the development of the bias over matching shape distributions. Overall, it is apparent that there is a significant offset between the linear biases of the two branches over the entire magnitude range. This offset does not appear to strongly change with increased magnitude and thus decreased S/N up to $m_\mathrm{AB}=22.0$, with values at a $\sim 0.8\,\%$ level for $\Delta\mu_1$ and a $\sim 10^{-5}$ level for $\Delta c_1$ for each respective measurement with the KSB algorithm. At a fainter magnitude of $m_\mathrm{AB}=24.0$ though, the bias is significantly reduced below the $0.5\,\%$ level for $\mu_1$. We note here also the increased scatter at lower S/N, due to the degradation of the KSB shear estimator for faint galaxies \citep{Hirata2003}. Likewise the absolute value of the shape measurement bias increases at this magnitude (see the steep slopes in the lower panel plots). \cite{Schrabback2010} proposed corrections for noise-related multiplicative bias components for this algorithm, though we deem these here not necessary, as KSB only serves as an initial fast test. Noise-related differences occur in both branches, cancelling out in the relative comparison of $\mu$. We also notice and confirm the non-linearity in the \texttt{GalSim} KSB shear estimator previously detected by \cite{Jansen2023} for high shears. Ignoring higher-orders of the shear bias also biases the linear term. We show antisymmetric quadratic fits to the shear bias difference in App. \ref{apdx:nonlin} and list the corresponding parameters for comparison. 

Overall, the results show that realistic galaxies with complex morphologies and substructures can bias the shape measurement at a relevant level given \Euclid's tight requirements for the cosmic shear analysis. Our results confirm the order of magnitude of this effect as found in the space-based PSF branch of the GREAT3 challenge \citep{Mandelbaum2015}. While the effect is reduced at lower S/N, where the majority of the galaxy population in the Wide Survey lies, the bias can still not be neglected there and thus has to be accounted for. Still, our sample for now assumes the presence of galaxies of all types within the COSMOS sample at each magnitude, which is not the case in reality. In true observations, disturbed and irregular structures will preferentially occupy the high-redshift and faint-magnitude regime, which could thus reduce or increase the amount of bias in specific tomographic redshift bins. In the future, we will extend this analysis to provide a shear bias calibration that incorporates the distinct target redshift tomography of the Euclid Wide Survey that will be applied for the cosmological weak lensing analysis. The presented work is mostly a proof-of-concept study, which shows that complex morphologies should be added to the \Euclid simulation pipeline for accurate shear calibration. However, even if KSB and the missing morphology-redshift correlation were to overestimate the bias by a full order of magnitude, it would still be relevant for the Survey. For instance, looking at the $m_\mathrm{AB}=24.0$ estimates, it is likely that the bias difference is underestimated. The sample includes more elliptical galaxies than the typical faint and high-$z$ sample, which would mean even tomographic bins with mostly faint objects would be biased significantly due to increased complexity.  

Moreover, different shape measurement methods need to be tested as well, especially \lensmc \citep{Congedo2024}, as it is the designated code for the \Euclid data release 1 (DR1). While KSB is fast and thus advantageous for initial testing, it will not be used for stage IV surveys and hence does not provide an estimate that is fully useful for future cosmological analyses. Nevertheless, these initial findings show the importance of the accurate calibration of this effect, because multiplicative biases at the $\sim 0.5\,\%$ level exceed the necessary calibration precision by a factor of 2.5 \citep{Cropper2013}. As the applied PSF is not elliptical though, the additive bias should be consistent with zero, which is the case within the errors. More galaxies and a more realistic PSF are required in the future to check if PSF anisotropy has a significant impact on the bias difference from complex morphologies. 
 
These initial estimates provide a rough quantification of the order of magnitude of the influence by complex galaxy morphologies on the shape measurement, where the magnitude and the S/N of the image correlate with the bias level. This can be expected, as the details of galaxy substructures are washed out with decreasing S/N, which gives rise to the expectation of a reduced bias level at faint magnitudes and hence a low S/N. Our results support this assumption. As mentioned, the amount and degree of complexity of galaxy shapes is a function of redshift, and it might thus still affect the bias for true distributions. If a population of galaxies at low $z$ and bright magnitude has a higher percentage of elliptical objects than galaxies at higher redshifts, the bias can be reduced in comparison to a sample with increased peculiarity. Hence, it is a necessary step to extend the model towards realistic correlations between morphological properties and redshift; this, however, is beyond the scope of this work. 

\begin{figure}
    \centering
    \includegraphics[width=1.0\columnwidth]{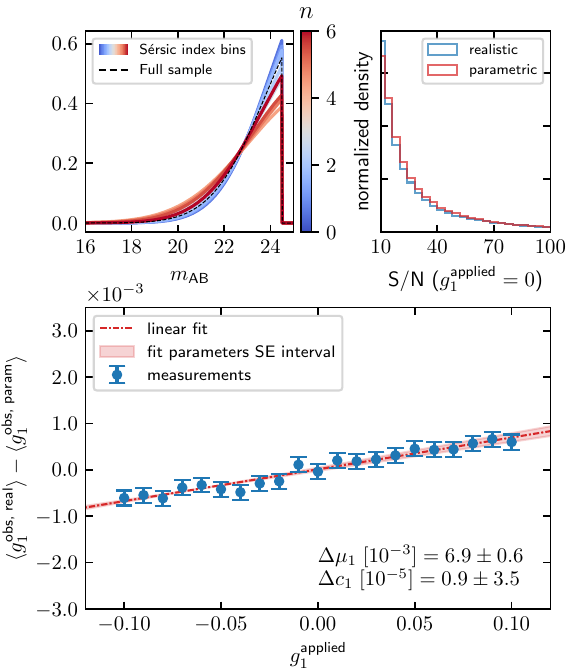}
    \caption{Shear bias difference estimate on a realistic magnitude distribution. The first subplot shows the magnitude distribution modelled with truncated Gaussians over a set of Sérsic index bins and the second subplot shows the measured S/N distribution after the \Euclid-like conversion of the generated images. The large panel shows the estimate on the shear bias between the realistic and parametric branches calculated according to Eq. \eqref{eq:bias-morphology}.}
    \label{fig:bias-real-dist}
\end{figure}

\begin{figure*}
    \centering
    \includegraphics[width=1.8\columnwidth]{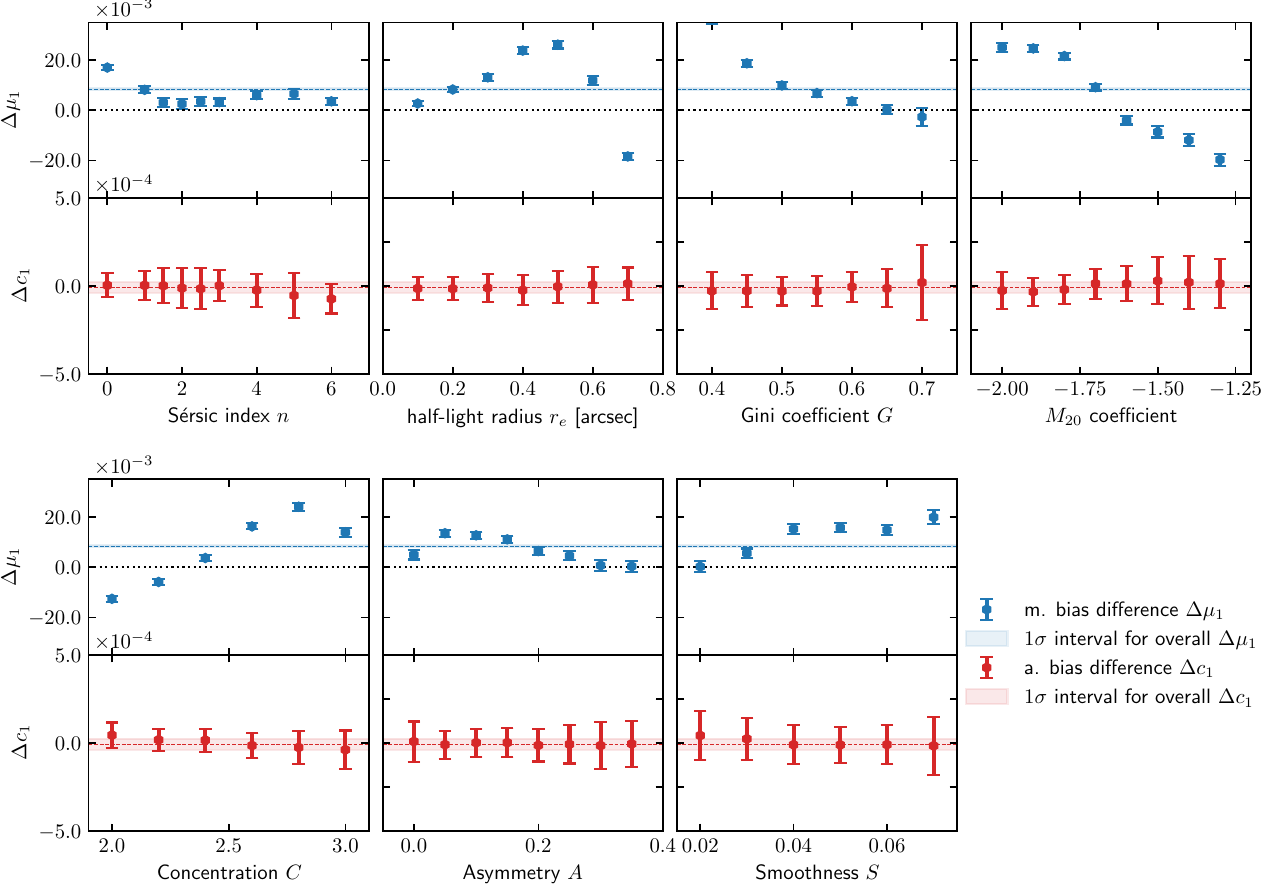}
    \caption{Results of shear bias difference measurements as a function of structural parameters and morphological proxies. The points with error bars depict the multiplicative and additive biases in the respective bin, the dotted lines with the blue or red areas show the means and errors for $\Delta\mu_1, \Delta c_1$ when measuring over the whole sample of $1\times 10^6$ galaxies. The data points are placed at the lower bound of the respective bin, the right-most points include all objects above this threshold.}
    \label{fig:bias-morph-dep}
\end{figure*}

\subsection{\label{subsec:biasrealmag}Bias for a realistic magnitude distribution}

Next we also estimate the mean morphology related bias for a galaxy population that follows the expected magnitude distribution of the Euclid Wide Survey \citep{Scaramella-EP1}. Given that we use the COSMOS data as a training sample, we can use its magnitude and Sérsic index distributions to model the corresponding properties of our emulated \Euclid data set. To ensure closely matching characteristics, we separate the magnitude distribution into Sérsic index bins and then fit a truncated Gaussian with an upper limit set at $m_\text{AB}=24.5$ on each component. Then, we sample magnitudes for each generated object according to the magnitude distribution in the respective bin. Thereby, we intend to mimic the true observations as closely as possible with a basic magnitude-morphology relation. However, this does not account for possible correlations between magnitude and tracers of more complex morphologies. This will be achieved in future work as we expand the framework to account for such dependencies and create fully realistic galaxy populations within each tomographic redshift bin. Finally, we again convert the generated sample into VIS-like postage stamps with additional shears over the same interval and then measure the S/N and the estimated $g_1$ from KSB.

Figure \ref{fig:bias-real-dist} shows the initial magnitude and measured S/N distributions, as well as the bias estimate on $10^6$ galaxies in the lower subplot. We again observe an overall shear bias difference, that is $\Delta \mu_1 = (6.9\pm 0.6)\times 10^{-3}$ for the multiplicative bias and $\Delta c_1 = (0.9\pm 3.4)\times 10^{-5}$ as the additive component. The $c$-bias is again consistent with zero within its errors, as expected for a circular PSF. Overall though, it is apparent that either the \Euclid cosmic shear analysis needs to be aware of the value of the $\mu$ and $c$ biases across the tomographic bins or the simulation pipeline has to be extended to include galaxies with realistic, complex morphologies. While the exact amount of bias still has to be determined using \lensmc and, optionally, other shape measurement codes, a bias at the $|\delta\mu| \gtrsim 7.0\times 10^{-3}$ level cannot be neglected, as it lies above the target calibration accuracy determined for the \Euclid science goals. While it is of course clear that the bias might be reduced for \lensmc, we note that the order of magnitude found using KSB lies in the range of previous estimates from \cite{Mandelbaum2015}. Therein, also other shape measurement methods relying on for example model fitting were tested, with similar bias differences of at least $10^{-3}$ found. This suggests that we we may expect a similar order of magnitude for \lensmc and assume relevance for the \Euclid analysis. The aforementioned \Euclid requirements were, however, determined for the full Wide Survey footprint and will hence be lower for DR1. 


\subsection{Bias dependence on morphological parameters}

The contributions to the bias originate from peculiar shapes themselves, such as spiral arms, merger remnants, tidal streams, or other possible sources of complexity. Moreover, substructures also exist to some degree in galaxies whose shape can be extremely well modelled by a single or double Sérsic profile. Such features correlate with the morphological proxies and the Sérsic profile parameters presented in Sect. \ref{sc:GalMorph}. Using the $CAS$+$GM_{20}$ statistics as well as Sérsic index $n$, and half-light-radius $r_\mathrm{e}$, we can determine the bias as a function of these properties. 

To do so, we take the measurement at fixed magnitude $m_\mathrm{AB}=20.0$ (to ignore S/N effects on the shear bias difference), bin the measured distributions of $g_1, g_2=0$ for each parameter and then average over the shears of all objects in the respective bin. As the overall number of sample galaxies is not exceptionally high, the  errors in bins with low number counts can thus again be potentially large. Nevertheless, this gives a rough estimate on how the occurrence of specific disturbed morphologies can effect the shape measurement. 

In Fig. \ref{fig:bias-morph-dep}, we show the results of this analysis. Looking at the additive bias difference $\Delta c_1$, we show that it is consistent with zero for all bins. This is expected due to the isotropy of the PSF. In contrast, the multiplicative bias difference displays clear dependencies on all the metrics. First of all, it decreases slightly with Sérsic index, which can be anticipated, as higher Sérsic indices indicate highly peaked galaxies with usually smooth profiles at the given pixel scale. Looking at the half-light radius, we observe an initial increase of the bias difference, followed by a sharp decline after $r_\mathrm{e} \gtrsim \ang{;;0.5}$. Within the Gini-$M_{20}$ space, both parameters induce a decrease in the multiplicative bias with increased value. High $G$ values indicate strongly peaked galaxies, where the bias difference here drops to zero as both branches are almost identical. Furthermore, $M_{20}$ is related to merger signatures such as for instance multiple nuclei as it traces the overall moments of the brightest image pixels and does not rely on circular apertures such as $C$ \citep{Lotz2004}. 

Additionally, we detect a correlation between the concentration $C$ and $\Delta\mu$. While initially higher $C$ relates to an increased bias, very concentrated objects again lead to a reduction of the bias difference. The concentration parameter $C$ has furthermore been shown to have a strong positive connection to intrinsic galaxy properties such as stellar mass and velocity dispersion \citep{Graham2001}. Given that the more massive galaxies, as well as high Sérsic index galaxies, are almost exclusively elliptical and thus do not possess spiral arms or large active star-forming regions, these result can be anticipated, as images of such objects resemble their Sérsic-fits more closely and thereby do not alter the estimated shear to the same degree as early-type galaxies. For Sérsic indices $n>4$ this results in a remaining bias of $\Delta\mu_1\sim 0.3\,\%$. One could argue that for such objects the bias should reduce even lower, given the absence of any substructure. However, the sub-sample of such galaxies also contains cases where the best-fist Sérsic profile might be highly peaked, but the underlying true shape can exhibit weak tails or even multiple nuclei that cannot be modelled by a Sérsic fit. Thus, the set of galaxies which closely match in both branches is contaminated, resulting in a small residual bias. 

The trends for the asymmetry and the smoothness or clumpiness also indicate interesting behaviour. Asymmetries of galaxies are mainly caused by merger signatures, spiral structures, and star-forming regions \citep{Conselice2003, Hambleton2011}. 
The low-$A$ population includes ellipticals and high-Sérsic index galaxies, where the previous morphological tracers showed a clearly reduced $\Delta\mu$. Similarly, the symmetry, which traces the clumpiness, shows an increase of the bias with $S$. This means that high-$S$ objects, which often possess multiple nuclei that are not captured by Sérsic models, lead to strong shape differences in the branches and thus increase the relative bias difference. For the asymmetry on the other hand, the bias decreases with $A$, although with an initially low value for the lowest-$A$ bin. The correlation is therefore not purely linear, similar to the concentration. 

Aside from the intrinsic increase of asymmetry and clumpiness with increasing redshift, an additional effect amplifies this effect. Observing all galaxies in the same filter, in this case the \Euclid VIS band, means that higher redshift objects will be seen at restframe wavelengths further in the UV regime, which then leads to even more complex and irregular observed morphologies. Especially for \Euclid, where very faint objects will represent the majority of the sample, this could increase the shear bias from complex morphology even more at low S/N, which is counteracted by the washing out of structures at faint magnitudes. Thus, simply using the F814W band to mimic \Euclid VIS images can also influence the bias difference, as the two bands do not fully coincide, with VIS being much broader. Future improvements of the generative model will alleviate this issue by training on multi-band data which can be used to then combine to the full VIS baseline.  

Overall, these results show that the shear bias measurement does not depend on only structural parameters such as Sérsic index or effective radius, but can also be traced by statistics for complex morphologies. As neglecting substructure of galaxies in shear calibration naturally introduces an estimator-dependent bias that is relevant for stage IV surveys, the bias is clearly dependent on the analysed galaxy population and, the degree of peculiarity in their shapes. Especially spirals, merger signatures and multiple components increase the bias substantially. On the other hand, its magnitude is irrelevant for \Euclid's accuracy requirements for ellipticals and in general symmetric or strongly peaked galaxies. 

\subsection{Detection and blending}

Cosmic shear results are not only dependent on the intrinsic biases of the shape measurement methods themselves, but also on selection effects and the detection process \citep{Hirata2003, Hoekstra2017, Conti2017, Martinet-EP4, Kannawadi2019, Hoekstra2021}. This can enter the results in two ways: Either by actual detection differences in calibration simulations and real observations or by failure of the shape measurement method to determine the ellipticities of individual galaxies. The detection significance of a faint object is usually dependent on its alignment relative to the shear or the PSF \citep{Hirata2003, Bernstein2002}. The detection process happens with a circular kernel, meaning that the images of galaxies with shear perpendicular to their major axis will appear rounder because of the shear, leading to an increased detection probability. This can lead to a multiplicative bias. Moreover, the detection bias estimate is correlated with the shape noise, as objects might be only detected in one of the rotated versions if shape noise cancellation is applied. In simulations, detection bias can thus be removed by requiring the detection of both rotated counterparts, which is, however, not feasible for real observations \citep{Pujol2020, Hoekstra2021}. We note that objects with failed shear estimates for one of the pair galaxies were previously omitted for the analysis of $\Delta\mu_1$, meaning that the results presented in the previous subsections do not include selection bias from KSB. 

Given our two measurement branches, it is a necessary step to check if there is a difference in the detection with \textsc{SExtractor} \citep{Bertin1996} if we do not look at isolated galaxies, but full image scenes with potential blending effects. This is relevant for \textsc{Metacalibration} too \citep{Sheldon2017}, which will also be applied for \Euclid WL. This method is generally unbiased for isolated galaxies, and can account for selection effects if the selection is performed on a measured quantity of the artificially sheared images \citep{Sheldon2017}, but will suffer from detection biases or other possible selection effects \citep{Sheldon2020}. 

To first of all check if there is a relevant difference from detection itself, we create image scenes per measurement branch, realistic and parametric, with zero applied shear. We use galaxies from our model drawn with the aforementioned magnitude-morphology correlation and paste them at random positions in scenes of size $4000\times 4000$ pixels, roughly matching the size of one \Euclid VIS CCD. We furthermore include additional galaxies up to  $m_\mathrm{AB} = 29.0$, as \citet{Hoekstra2017}, \cite{Martinet-EP4} previously showed the significance of including faint galaxies beyond the \Euclid WL magnitude limit for the shear measurement bias. Simultaneously, matching scenes with the 90$^\circ$-rotated galaxies are created, to perform shape noise cancellation and increase the sample size. Afterwards, Gaussian read noise and Poisson shot noise are added to the image, with matching noise fields in rotated scene pairs. While our simulations for now ignore blending with large and bright objects, as well as stars and diffraction spikes, we expect this to influence both branches in the same manner. In the current analysis we only investigate the impact of realistic and parametric light distributions as such. Moreover, galaxies are not randomly positioned in reality, leading to increasing blending fractions due to clustering, especially by faint objects, which we here ignore for now. We assume an average galaxy number density of $\rho_n\sim 250$\,arcmin$^{-2}$ up to $m_\mathrm{AB} = 29.0$, but also repeat the process with doubled and tripled number densities to increase the amount of blended objects.

\begin{figure}
    \centering
    \includegraphics[width=1.0\columnwidth]{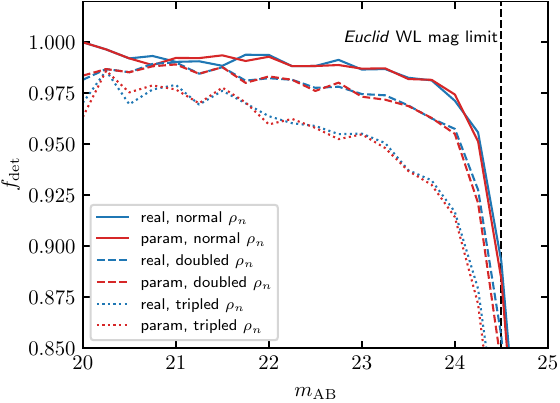}
    \caption{\textsc{SExtractor} detection fraction in the realistic (blue) and parametric (red) branch. The solid, dashed, and dotted lines show the results for a regular, doubled, and tripled source number density $\rho_n$.} 
    \label{fig:detection_bias}
\end{figure}

After rendering of the simulations, we run \textsc{SExtractor} on the images scenes and first compare the blending and detection fractions of both measurement branches. Looking at the blending fraction $f_\mathrm{blend}$, we find that \textsc{SExtractor} flags more objects as blended in the realistic branch, and also detects overall more galaxies therein. The overall blending fraction is $2.8\,\%, $ $4.2\,\%$, or $9.2\,\%$ for realistic objects with normal, doubled, or tripled number density, respectively, or $2.1\,\%$, $4.9\,\%$, and $8.5\,\%$ with double-Sérsic profiles (calculated from the number of detected galaxies and the number of flagged de-blends). We note that these numbers strongly depend on the \textsc{SExtractor} parameters, which we list in App. \ref{apdx:C}.

The increased blending fraction arises from the complex structure and most likely irregular galaxies, where objects with multiple nuclei might be de-blended into separate objects. While the overall difference is not large, with around $2\,\%$ more detections in the realistic branch over the full magnitude range (only 0.2\,\% below $m_\mathrm{AB}= 24.5$), this still has to be kept in mind. \textsc{SExtractor} sometimes struggles to distinguish irregular objects as single galaxies and thus detects some of them inaccurately. Overall, some discrepancy between the detections is expected, especially at fainter magnitudes, where most of the detection differences can be observed. If the best-fit double-Sérsic has a lower surface brightness, for example for asymmetric objects with faint tails that are only fitted with a very broad profile with large $r_\mathrm{e}$, a portion will fall below the \textsc{SExtractor} detection limit, while the respective corresponding realistic object might still be selected. We show a cutout of a simulated frame with examples of detection differences in App. \ref{apdx:C}.

To show how the detection differences depend on magnitude, we depict in Fig. \ref{fig:detection_bias} the detection fraction $f_\mathrm{det}$ for all measurements. The sample is almost complete up to a magnitude of around 24.0 with the standard galaxy number density, with some loss of objects for increased $\rho_n$, which can be expected due to blending. The detection fraction for both types of simulations is almost identical, although with slight differences in the number of detected galaxies, that is $\sim0.2$\,\% more in the realistic branch at standard $\rho_n$ for $m_\mathrm{AB}\leq 24.0$. Which exact galaxies are detected varies slightly over the magnitude bins, as seen by the solid lines. Furthermore, there is more scatter for bright objects at higher blending fractions, which can be relevant for simulations where sources are not randomly positioned. Therefore, estimates have to be made on how such slight selection effects apparent even at bright magnitudes for both measurement branches are relevant and how selections effects between real observations and parametric simulations in general affect \Euclid. 

\subsection{Detection bias}

In order to determine the actual shear biases from detection-related selection, we create the simulations with standard $\rho_n$ for each value of the constant shears $g_1\in [-0.1, 0.1]$ used for the previous estimates on the morphology bias and run \textsc{SExtractor} on the full frames. We also produce identical images with the 90$^\circ$ rotated versions to determine detection differences due to orientation with respect to the shear and analyse them in the same manner. Then we select only the ellipticities of galaxies that were actually detected by searching the nearest-neighbour galaxies for each input in the output catalogue. If the centroid offset is larger than three pixels \citep{Hoekstra2021}, we count this as a mismatch and thus a non-detection by \textsc{SExtractor}, as it is then most likely not a detection of an input shear catalogue object but from the additionally added faint galaxies with $24.5<m_\mathrm{AB}\leq 29$. Afterwards, we determine the multiplicative biases in both branches for three different cases:
\begin{enumerate}[i]
    \item Removing objects that are non-detections by \textsc{SExtractor}, separately in each branch: $\mu^\mathrm{D}$ 
    \item Removing non-detections and blended objects as identified by \textsc{SExtractor} with $0< \texttt{FLAGS} \leq 3$, separately in each branch:
    $\mu^\mathrm{DB}$
    \item Removing object with failed KSB measurements, without removing rotated counterparts if they are measured by KSB, ignoring detection differences, separately in each branch: $\mu^\mathrm{KSB}$.
\end{enumerate}

We determine the detection and selection biases separated by the contributions from \textsc{SExtractor} and KSB, including from blending. To mimic the detection bias results from \cite{Hoekstra2021} for a \Euclid-like setup, we here do not perform a S/N cut, but simply select galaxies with $20\leq m_\mathrm{AB}\leq 24.5$. 

To furthermore avoid model-dependent influences on the detection bias and isolate the \textsc{SExtractor} contribution, the bias is hereby not calculated with the KSB estimates, but with the true input ellipticities after shearing of the galaxies, following  \citet{Conti2017}, \cite{Kannawadi2019}, and \citet{Hoekstra2021}. Here, however, we do not have intrinsic ellipticites of the objects, at least not in the realistic branch, as they are not parametric profiles, but just sampled shapes from the generative model. As the parametric galaxies are by construction modelled on the realistic branch and we have shown in Fig. \ref{fig:n-re-e-dist} that their ellipticity distributions match, we use the input ellipticites from the Sérsic objects for both branches. The complex observed ellipticites, consisting of the intrinsic shape $\epsilon_\mathrm{s}$ and the added reduced shear $g=g_1+\mathrm{i}g_2$ with $g_2=0$ and $|g|<1$ are given by
\begin{equation}
    \epsilon_\mathrm{o} = \frac{\epsilon_\mathrm{s} + g}{1+g^*\epsilon_\mathrm{s}}\, ,
\end{equation}
where $^*$ denotes a complex conjugate \citep{Seitz1997, Hoekstra2021}. The complex intrinsic ellipticites are calculated from the absolute ellipticity values of the Sérsic fits and their rotation angles. The bias is then again determined via a linear fit of the means of the ellipticities $\epsilon_\mathrm{o}$ with the applied cuts i--iii as a function of $g_1$.  

We first determine the biases from all three cases separately for realistic and parametric objects, and then compute the bias differences between the branches $\Delta\mu^\mathrm{D}, \Delta\mu^\mathrm{DB}, \Delta\mu^\mathrm{KSB}$ to quantify how the selection process affects the simulation sets differently. There, the difference is given by $\Delta\mu^\mathrm{x}=\mu^\mathrm{x}_\mathrm{real} - \mu^\mathrm{x}_\mathrm{param}$, with $\mathrm{x}\in \{\mathrm{D}, \mathrm{DB}, \mathrm{KSB}\}$. We summarise the results in Table \ref{tab:det-bias-res} for all three cases. 

\begin{table}
\centering
\caption{Detection and selection biases.}
\smallskip
\label{tab:det-bias-res}
\smallskip
    \begin{tabular}{lccc}
    \hline
    & & & \\[-9pt]
    & \hspace{2pt}$\mu^\mathrm{D}$ $[10^{-3}]$ & \hspace{2pt}$\mu^\mathrm{DB}$ $[10^{-3}]$ & $\mu^\mathrm{KSB}$ $[10^{-3}]$ \\
    \hline\hline
    & & &  \\[-9pt]
    without S/N cut: & & & \\
    realistic & $-7.0\pm 0.6$ & $-5.6\pm 0.7$ & $-28.3\pm 2.2$ \\
    parametric & $-9.2\pm 0.6$ & $-11.9\pm 0.6$  & $-25.3\pm 2.4$ \\
    $\vec{\Delta\mu^\mathrm{x}}$ &  $\vec{2.2\pm 0.8}$ & $\vec{6.3\pm 0.9}$ & $\vec{-3.3\pm 3.1}$ \\
    with S/N cut: & & & \\
    realistic & $-3.5\pm 0.7$ & $-2.8\pm 0.7$ & $-15.1\pm 1.0$ \\
    parametric & $-4.6\pm 0.6$ & $-6.8\pm 0.7$  & $-11.3\pm 0.9$ \\
    $\vec{\Delta\mu^\mathrm{x}}$ &  $\vec{1.1\pm 0.8}$ & $\vec{4.0\pm 0.9}$ & $\vec{-3.8\pm 1.2}$ \\
    \hline
    \end{tabular}
\tablefoot{Summary of detection/selection bias contributions from \textsc{SExtractor} and KSB for both branches, and the bias differences $\Delta\mu^\mathrm{x}$ between both branches, with $\mathrm{x}\in \{\mathrm{D}, \mathrm{DB}, \mathrm{KSB}\}$, for the influences from detection, detection plus blending, and KSB measurement selection.}
\end{table}

A strong source of selection bias comes from KSB itself, where keeping objects where one from each rotated galaxy pair is not measured produces a strong bias that differs between the branches, leading to a residual of $\Delta\mu^\mathrm{KSB} = -(3.8\pm 3.1)\times 10^{-3}$. This is higher than the effect from KSB selection determined by \cite{Hoekstra2021}, but also very noisy due to the large size of the error. We note that there exist differences in the respective simulation and measurement setups that can account for such variations. Different KSB implementations and morphologies were applied, which naturally leads to variations. The \texttt{GalSim} KSB implementation fails for a substantial number of objects ($\sim$4\,\%), with more failures in the realistic branch. This can be expected due to the existence of irregular shapes that make an ellipticity estimate more difficult. 

More importantly, the \textsc{SExtractor} results significantly differ in both branches due to detection and blending. We find that the detection and detection-plus-blending bias estimates lie close to the values from \cite{Hoekstra2021} for parametric objects. The slight differences can be accounted for by varying PSF and morphology models (Sérsic vs. double-Sérsic), as well as the choices of detector, background, and noise characteristics. The biases for realistic galaxies, however, strongly deviate from these results. Overall, this leads to a bias difference $\Delta\mu^\mathrm{DB} = (6.3\pm0.9)\times 10^{-3}$ associated with bias from the detection or removal of objects flagged as a blend. This increase compared to the only detection-related bias difference $\Delta\mu^\mathrm{D}=(2.2\pm0.8)\times 10^{-3}$ is especially interesting, as it suggests that blending leads to a positive bias in the realistic branch, but to a negative bias for parametric shapes. This can be explained by the aforementioned blending differences for irregular shapes, where objects modelled with realistic morphologies are more likely to be flagged. Thus, not only are galaxies that are actually blended omitted but also isolated objects that look similar to blends. We find that the realistic objects which are additionally flagged as blended in comparison to the parametric simulations have almost exclusively large effective radii. Such galaxies often show for example spiral arms or other complex substructure in the realistic branch, which can sometimes lead to \textsc{SExtractor} flagging despite the absences of actual blending. Flagging due to the shape itself will, however, most likely appear in a galaxy pair for both versions, leading to no selection effects. While it then may seem counter-intuitive that the bias decreases compared to the double-Sérsic simulations, it can be explained by looking at the galaxies that are flagged in both branches: Some of them may be irregular and thus also have a flagged rotated counterpart. This effectively reduces the number of objects with differently flagged rotations, which in return reduces the number of objects that contribute to the blending bias on the ellipticity estimate. 

We note that using the true input ellipticies of the profiles in both branches can be an additional source of systematic error in the estimate of the bias in the realistic branch. This can though not be corrected for, as no true ellipticites exist for non-parametric shapes. An alternative option could be to determine the true ellipticities with a shear response on the noise-free shapes, similar to the \textsc{Metacalibration} method. This though exceeds the scope of this work. Using the measured KSB $g_1$ values on the other hand would also potentially lead to difference due to the morphology bias, although this might be not significant, as mostly faint galaxies induce a detection bias, where the difference due to morphology is low anyway. Overall, we expect that using the identical ellipticities in both branches is the most robust method to avoid model bias influences.

Another effect that will affect both the shear measurement and detection biases are correlated ellipticities. For example in dense regions, such as galaxy clusters, the intrinsic shapes may be correlated and more heavily blended, while also exhibiting overall higher shears outside the 10\,\% interval where the shears are currently measured. This will be addressed in future work by simulating scenes with realistic positions and correlated shapes. 

As we applied an additional cut of $\mathrm{S/N}>10$ during the morphology bias estimate, the detection and blending bias that is relevant for the results from Sec. \ref{subsec:biasrealmag} will be lower, as there will be fewer objects without a detection. With this additional S/N cut, the bias differences are reduced to $\Delta\mu^\mathrm{D} = (1.1\pm0.8)\times 10^{-3}$, $\Delta\mu^\mathrm{DB} = (4.0\pm0.9)\times 10^{-3}$, and $\Delta\mu^\mathrm{KSB} = -(3.8\pm1.2)\times 10^{-3}$, respectively. While the absolute values in the respective branches increase when low signal-to-noise galaxies are included, the relative difference between the branches stays stable within the error bars. 

The results show that realistic morphologies not only bias the shape estimator itself, but can account for detection and selection differences between parametric image simulations and observed data. Their different blending characteristics can induce strong biases in the ensemble shape measurement, independently of the algorithm applied to measure the shear signal. Even when neglecting these additional blending effects, the effect from the detection alone leads to a bias that is significant for the \Euclid DR1 and independent of the applied shape measurement method. This underlines the need to derive accurate bias corrections from simulations that include complex morphologies for the shape measurement methods that are used for the \Euclid science analysis.

\section{\label{sc:conclusion} Conclusions}
In this work, we have performed a proof-of-concept study about the calibration of model biases from complex galaxy morphologies for the \Euclid mission. We have presented a new deep generative CNN for high-resolution, noise-free galaxy postage stamps that uses a physically motivated latent space via the wavelet scattering transform. In addition to fast training and a latent space model that encodes morphological information, which can also be independently applied in unsupervised learning for galaxy classification, the model generalises well over the plethora of galaxy shapes. It is able to recover structural parameters and morphological proxies such as the $CAS$+$GM_{20}$ statistics between input and output distributions, which are also correlated with each other. We showed that various sampling techniques can be leveraged to obtain new objects from the generative model, for example, also via conditional sampling by the Sérsic index $n$. In subsequent work, we will extend the model to incorporate correlated sampling via redshift, magnitude, and structural parameter distributions, as for now we have only assumed a connection between the magnitude and the Sérsic index. 

Next, we estimated the bias difference introduced by realistic galaxy morphologies in shape measurements. For this, we simulated galaxies from the model together with their parametric counterparts, applying additional shear. With the KSB method, we found a multiplicative bias difference at the $\Delta\mu_1\sim 0.7\,\%$ level for a realistic magnitude and S/N distribution with an empirical dependency on the Sérsic index. This bias therefore lies above the target shear accuracy for \Euclid by a factor of three. While this method will not be applied to \Euclid data, it confirms the findings from \cite{Mandelbaum2015} for a \Euclid-like setup and suggests that the bias will have to be calibrated for \Euclid given the results with other shape measurement methods from this previous work. We detected a correlation between $\Delta\mu_1$ and the S/N, the Sérsic index $n$, the half-light radius, and the $CAS$+$GM_{20}$ statistics of the galaxies. Furthermore, we showed that the bias reduces heavily to below a $\sim 0.3\,\%$ level for objects whose shapes closely resemble each other, for example smooth profiles at high S/N or low-S/N objects. This proves that the origin of the measured model biases really lies in the complex structures and is not an artefact of the image rendering process. Still, this first measurement assumes a random distribution of galaxy complexity across magnitudes. While we implemented a relation between $m_\text{AB}$ and $n$ based on observations, there is still not much quantitative information about the dependency on other morphological proxies. In reality, where the fraction of irregular galaxies is highest at high redshifts, the bias might be reduced, as the complex substructures that arise from various mechanisms and thereby emerge in specific regions of the $CAS$+$GM_{20}$ space could be washed out at a low S/N. This would lead to an overall lower effect on the shear estimation. However, if that is not the case, high-$z$ samples could be more strongly biased due to their increased irregularity. Either way, the bias difference from galaxy morphology is most likely a function of the source redshift. Therefore, ignoring the effect will influence the tomographic shear calibration and thus also bias the cosmological results.

We also showed that the detection by \textsc{SExtractor} is not completely identical between both measurement branches when creating matching images scenes. The detection process recovers almost the same galaxies, but more galaxies are detected below the $m_\mathrm{AB}=24.5$ magnitude limit of the Euclid Wide Survey in the realistic image scenes, which we attribute mostly to the fact that in some cases \textsc{SExtractor} de-blends complex shapes with multiple brightness peaks into separate galaxies. Thus, selection effects caused by detection and blending discrepancies lead to a bias difference of $\Delta\mu \sim 4.0\times 10^{-3}$ between realistic and parametric objects over the full sample with an applied $\mathrm{S/N}>10$ cut. This implies that realistic morphologies must be included in weak lensing image simulations in order to reach the \Euclid requirements
for the estimation of detection biases alone, disregarding model biases, which are independent of the employed shape measurement method. 

Furthermore, KSB is not the designated shape measurement algorithm for \Euclid; hence future calibrations will need to be performed with \lensmc or \textsc{Metacalibration}, where the shape measurement bias could be decreased due to the forward modelling plus fitting nature of \lensmc and the generally unbiased approach of the shear response measurement by \textsc{Metacalibration}, where detection and selection effects are still relevant. Fitting codes, for example, have previously been shown to exhibit a lower morphology bias in comparison to KSB \citep{Massey2007b, Mandelbaum2015, Hoekstra2021}, though they 
 were only tested with simpler galaxy models via shapelets or emulation from HST data and still at a level relevant for \Euclid. While this does not necessarily mean that a measurement with \lensmc will be less biased due to complex morphology, the procedure of the code indicates that the bias will be reduced, as \lensmc fits more realistic double-Sérsic models to the data. Thus, a wide variety of shapes can be modelled with such model fitting codes, likely leading to a reduction in the relative biases between realistic and parametric galaxies. This will be determined quantitatively in future work. Once the redshift dependency has been embedded into the model framework, the cosmic shear morphology bias of \Euclid can be calibrated as a function of tomographic redshift bins for both $g_1$ and $g_2$ using \lensmc and \textsc{Metacalibration}. We note that this project is a proof-of-concept study whose aim is to obtain a general estimate of the relevance of the bias and to present the deep learning model that can be applied for the full calibration of this bias within the \Euclid simulation pipeline. Nevertheless, even without the caveats of an analysis solely with KSB, the detection bias results show that there will be residual biases in the shear estimate due to detection differences in simulations and observations if complex morphologies are not accounted for. This alone justifies the usage of realistic shapes for shear calibration. 

Eventually, the CNN can be even further extended towards multiple colour channels using HST training data from multi-band observations of, for example, CANDELS data in the GOODS, COSMOS, UDS, and AEGIS fields \citep{Koekemoer2011, Grogin2011, Stefanon2017}. This will enable calibration of the \Euclid cosmic shear analysis for the colour gradient bias, which is caused by the wavelength dependence of the \Euclid PSF and the spatial colour gradients within galaxies, allowing future calibrations of the colour gradient bias to go beyond the simplified bulge and disc analyses employed in \citet{Semboloni2013}, \citet{Er2018}. This bias cannot be directly calibrated by \textsc{MetaCalibration} or \textsc{MetaDetection} \citep{Sheldon2020}, which underlines the necessity of this analysis for next-generation cosmological surveys. With realistic multi-band morphologies from the generator, we intend to determine a colour gradient bias that depends on the actual local SED distribution of each galaxy and thus permits more accurate calibration.

\begin{acknowledgements}
The Innsbruck authors acknowledge support provided by the Austrian Research Promotion Agency (FFG) 
and the Federal Ministry of the Republic of Austria for Climate Action, Environment, Energy, Mobility, 
Innovation and Technology (BMK) via the Austrian Space Applications Programme with grant numbers 899537, 900565, and 911971,
as well as support provided by the Deutsche Forschungsgemeinschaft (DFG, German Research Foundation) under grant
415537506. HHo ackowledges support by the European Union (ERC-AdG OCULIS, project number 101053992). LL is supported by the Austrian Science Fund (FWF) [ESP 357-N]. GC acknowledges support provided by the United Kingdom Space Agency with grant numbers ST/X00189X/1,
ST/W002655/1, ST/V001701/1, and ST/N001761/1. ANT acknowledges the UK Space Agency for its support.\\
\AckEC  
\end{acknowledgements}

\bibliographystyle{aa}
\bibliography{Bibliography/EROplus,Bibliography/Paper,Bibliography/Euclid}

\begin{appendix}
\onecolumn

\section{\label{apdx:sfdiff} Sérsic-fit difference examples}

\begin{figure}[!ht]
    \centering
    \includegraphics[width=1.0\columnwidth]{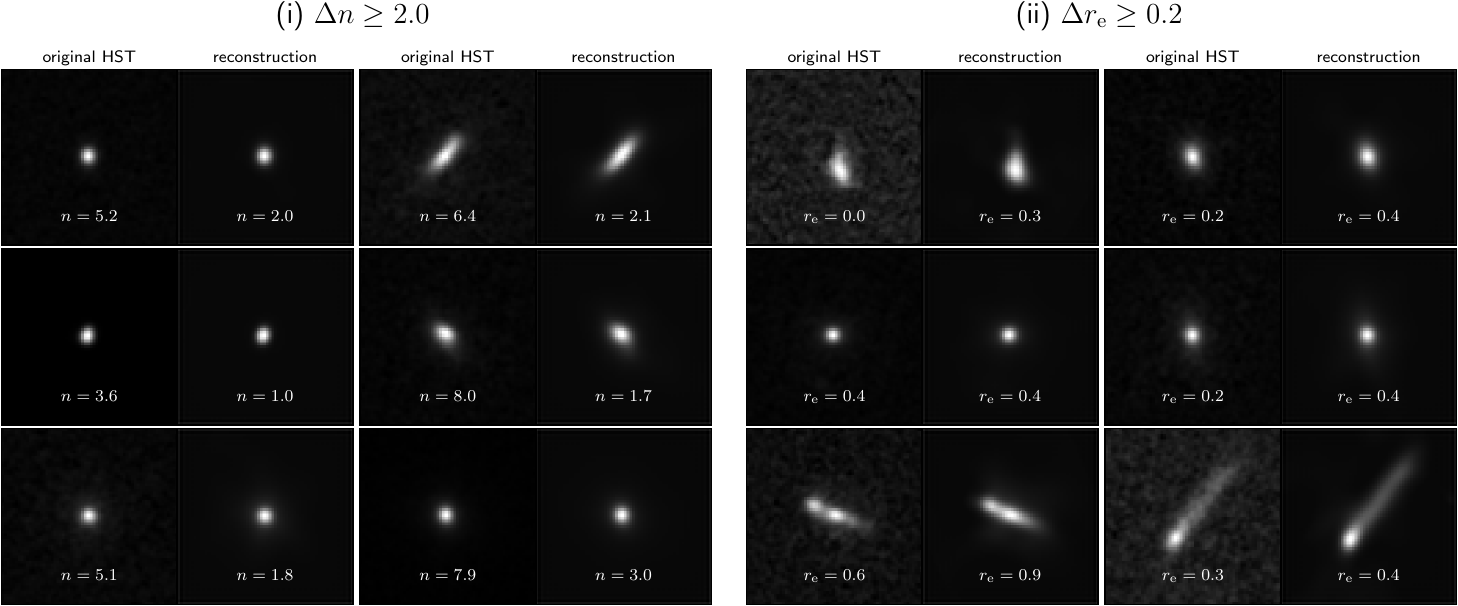}
    \caption{Example of galaxies where the difference between the original and generated $n$ or $r_{\rm e}$ exceeds $\Delta n=2$ or $\Delta r_\mathrm{e}=0.2$.}
    \label{fig:badfits}
\end{figure}

Figure \ref{fig:badfits} shows a sample of galaxies with fit results that strongly deviate between the original image and the estimate of the reconstruction with matched noise, as described in 
Sect. \ref{sc:NoiseRemoval}.

It is clear that exclusively strongly peaked, elliptical galaxies with high S/N are affected by low-quality fits obtained using \texttt{pysersic} concerning the Sérsic index, while the galaxies with inaccurate half-light radius estimates also include objects with faint features or clear bulge plus disc visibility. For such cases, the SVI method finds inaccurate posterior estimates for at least one of both images. Either full MCMC sampling or improved noise matching (which is not very relevant at high S/N) could be able to reduce this mismatch, albeit only with much longer fitting times ($\sim 10\,$s with Laplacian SVI vs. $> 60\,$s with MCMC). Given the objective of this work, this is not necessary, as we are for now mainly interested in the overall reconstructive power of the generative model for the estimation of the shear bias. In the future, however, accurate Sérsic fits are needed if realistic galaxies of a distinct distribution of structural parameters need to be injected into full-size \Euclid VIS simulations. Different ways of generating structural parameter fits for billions of galaxies have been investigated within the \Euclid Morphology Challenge \citep{Bretonniere-EP26}.

In principle, the monotonic decrease of the reconstruction accuracy seen in Fig. \ref{fig:hst-wst-sersic-distribution} could also be related to the architecture of the model. The generator tries to minimise the loss during training, for which especially the most abundant populations of morphologies have to be reconstructed accurately. Thus, the overall sparsely populated region of $n\geq 4.0$ could be neglected. However, looking at the reconstructed images, we do not think that this is the case here. 

\section{\label{apdx:2Dparam} Two-dimensional morphological parameter distributions}
\begin{figure}
    \centering
    \includegraphics[width=0.7\columnwidth]{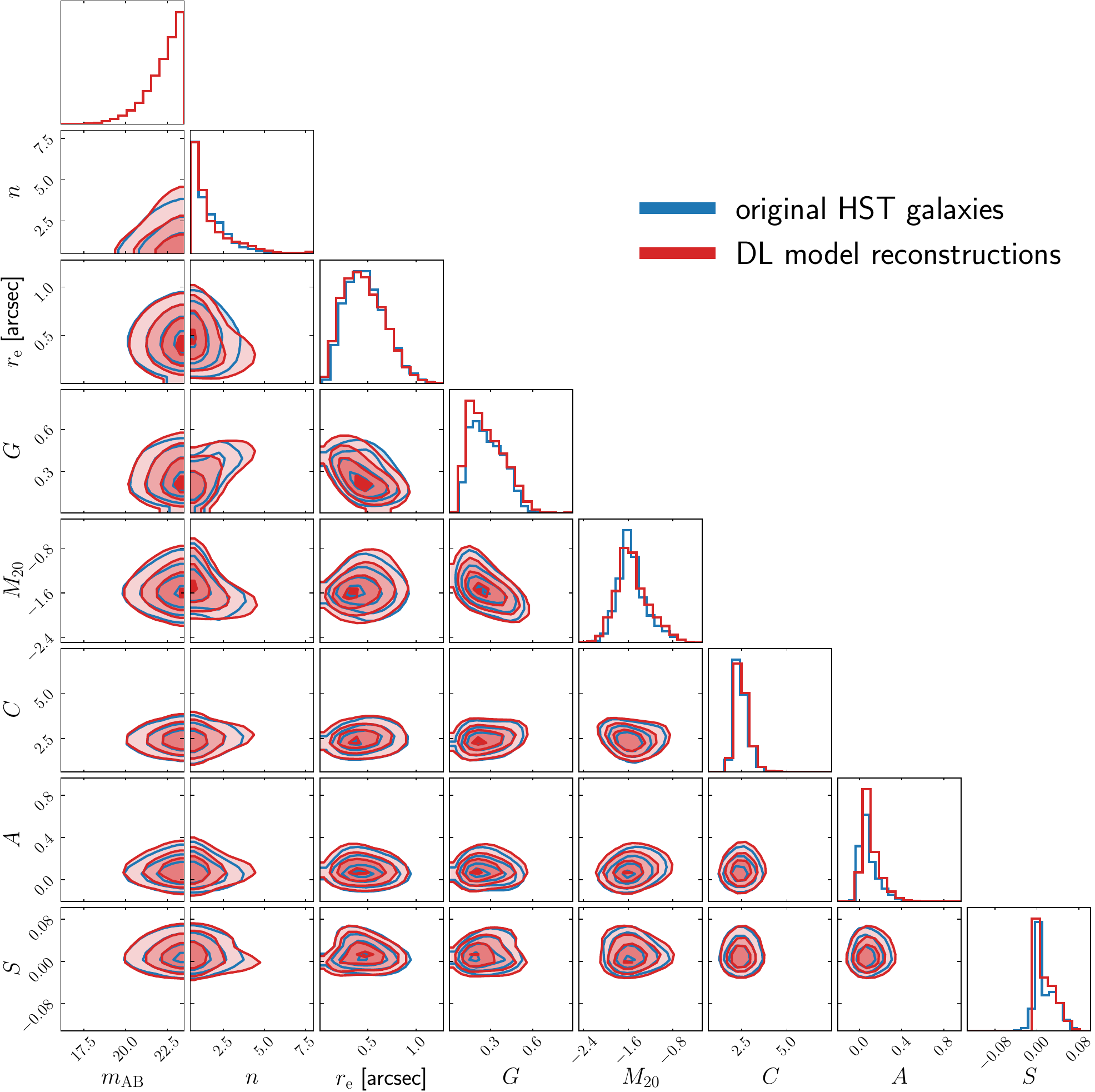}
    \caption{Two-dimensional parameter distributions of magnitude, structural parameters, and morphological proxies. The blue contours show the measurements done directly on the HST images, while the red contours depict the results of measuring on the reconstructed deep learning model outputs. }
    \label{fig:comp2D}
\end{figure}

In Fig. \ref{fig:comp2D} we show a comparison between the 2D correlated distributions of the training input sample and the measurements on the reconstructed outputs from the deep learning model. We can see that the histograms and 2D contours match overall well, with also correlated parameter distributions recovered well by the model.
\clearpage

\section{\label{apdx:PSF} \Euclid-like point-spread function}

\begin{figure}[h!]
    \centering
    \includegraphics[width=0.50\textwidth]{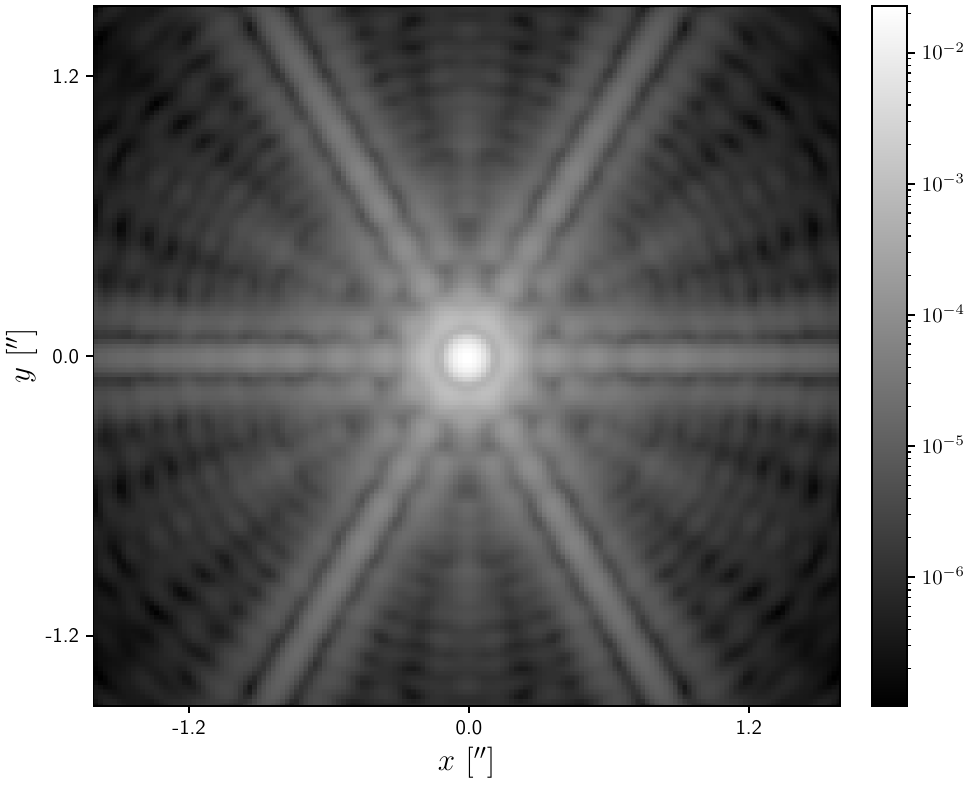}
    \caption{Oversampled version of the \Euclid-like PSF on a logarithmic grey scale. Axes are scaled in arcseconds.}
    \label{fig:psf}
\end{figure}

Figure \ref{fig:psf} shows the \Euclid-like PSF used within this work and mentioned in Sect. \ref{sc:ShearMeas}. The generated galaxies are deconvolved with their training PSF and then convolved with this PSF afterwards to create \Euclid-like simulations, as illustrated in Fig. \ref{fig:euclidization}. For more accurate PSF correction, the PSF model is usually oversampled. Figure \ref{fig:psf} displays a version of the PSF that is oversampled by a factor of 5 on a logarithmic grey scale.

The PSF model is created from a stack of monochromatic PSFs over the wavelength range of the VIS bandpass from 500\,nm to 900\,nm, with weighting of each component by a stellar spectrum, that is, a Vega spectrum. Additionally, it assumes the \Euclid telescope's optical characteristics such as mirror size and obscuration. A more detailed description on how this PSF was modelled can be found in \cite{Tewes2019} and \cite{Jansen2023}.

\section{\label{apdx:nonlin} Non-linearity of the shear bias}
The results of the shear bias measurement at bright magnitudes shown in Fig. \ref{fig:bias-mag-dep} exhibit a strong non-linear component, which has previously also been found by \cite{Jansen2023}. To show how second-order effects in the shear bias affect the linear term, we also fit an antisymmetric quadratic function to the shear bias difference. This function is defined as 
\begin{equation}
    \label{eq:nonlin}
    g_i^\text{obs} -g_i^\text{true} = \alpha \,|g_i^\text{true}|\,g_i^\text{true}
    +\mu_i g_i^\text{true} + c_i + n_i\, ,
\end{equation}
where $\alpha_i$ is an additional fit parameter. Following \cite{Kitching2022}, the linear shear bias can also be written as a spin-2 equation 
\begin{equation}
    g^\text{obs} = (1+\mu_{i,0})g^\text{true} + 
    \mu_{i,4}\left(g^\text{true}\right)^* + c + n\, .
\end{equation}
There, $\mu_{i,0}$ and $\mu_{i,4}$ are spin-0 and spin-4 quantities, respectively, and $^*$ denotes a complex conjugate. This simplifies the extension to higher-order terms of the shear bias with integer and half-integer power of the shear bias. We refer to \cite{Kitching2022} for a detailed derivation and description of this process. Here, we simply fit the quadratic multiplicative biases with \eqref{eq:nonlin}. 

\begin{figure}[h!]
    \centering
    \includegraphics[width=1.0\columnwidth]{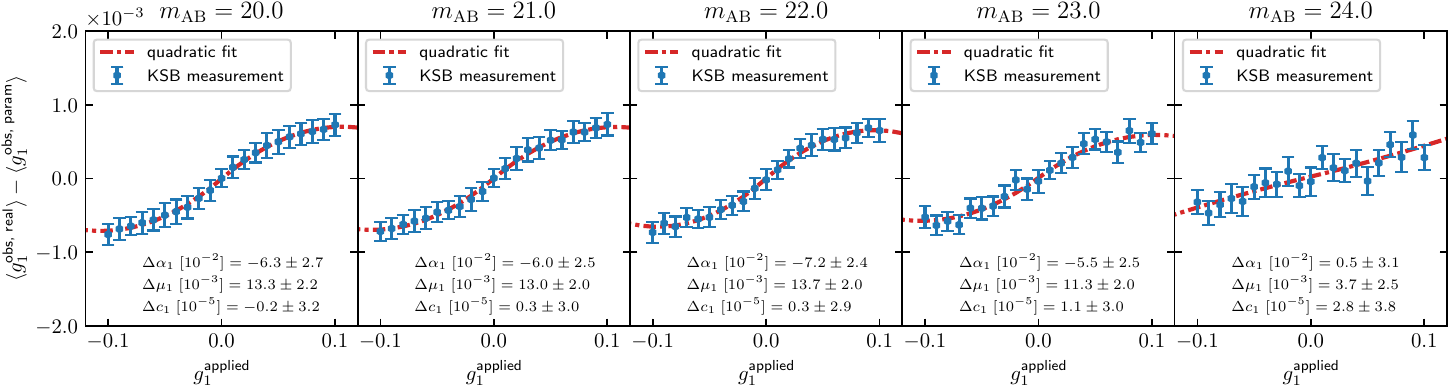}
    \caption{Results of the cosmic shear measurement with KSB on $10^6$ simulated galaxies at fixed magnitude, once with realistic galaxies from the deep learning model, and once with their double-Sérsic counterparts. The plots show the relative difference between separate bias estimates on both branches, realistic and parametric galaxies. The red curve shows the best-fit antisymmetric quadratic model according to Eq. (\ref{eq:nonlin}). Inside the plot we summarise the difference $\Delta$ of the fit parameters $\alpha_1, \mu_1$ and $c_1$.}
    \label{fig:nonlinbias}
\end{figure}

We show in Fig. \ref{fig:nonlinbias} the same KSB measurements of the shear bias difference as in Fig. \ref{fig:bias-mag-dep}. However, in  Fig. \ref{fig:nonlinbias}, the measurements are shown with the antisymmetric quadratic fit from Eq. (\ref{eq:nonlin}). It is apparent that this function fits the data better for bright samples, but no clear higher-order contribution can be seen in the faint galaxy sample, resulting in an overall low term $\Delta\alpha_1$ there. Compared to the fully linear estimates, the linear term $\Delta\mu$ is strongly increased at low magnitudes, but fits the estimate within the error bars for the measurement at $m_\mathrm{AB}=23.0$. 

While the non-linearity seems to be relevant for bright objects, we do not observe strong higher-order terms for faint samples or for a realistic magnitude and S/N distribution as shown Fig. \ref{fig:bias-real-dist}, which is why we do not explore similar comparisons for this measurement here. 

\section{\label{apdx:C} Scene example for detection comparison}

\begin{figure}[h!]
    \centering
    \includegraphics[width=1.0\columnwidth]{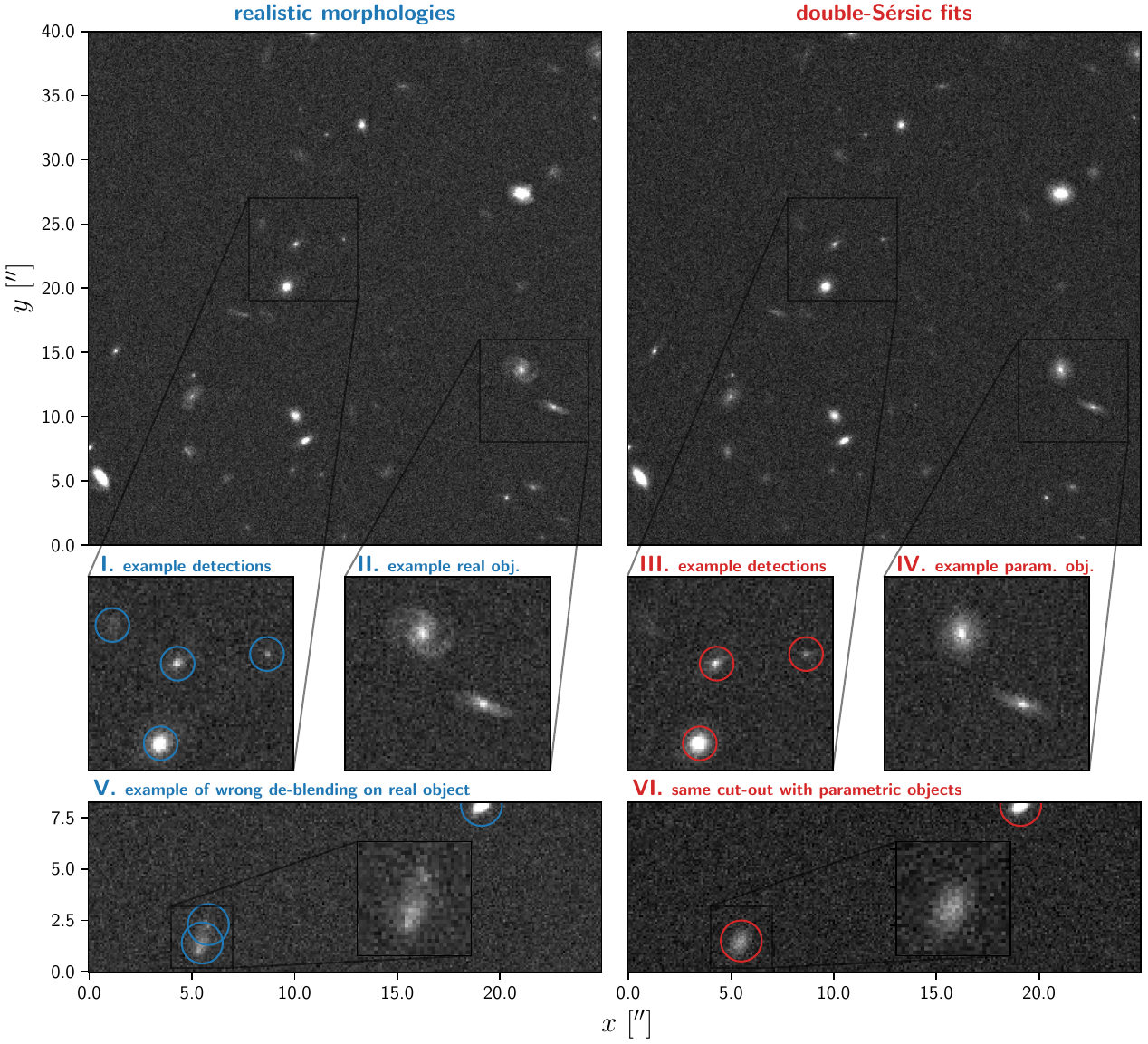}
    \caption{Example from the full image simulations in both branches used for the \textsc{SExtractor} detection comparison. The top panels show $400\times 400$ pixel cutouts from an image scene, with doubled number density $\rho_n$, with pixel intensities clipped between the 10$^\mathrm{th}$ and 90$^\mathrm{th}$ percentiles. Below are zoom-ins into interesting regions: Examples I. and III. illustrate a detection difference, where a realistic object lies just above the threshold, while its best-fit double-Sérsic counterpart is not detected. The zoom-ins II. and IV. show an example of a clearly more complex spiral galaxy and how this is represented in the parametric branch, highlighting the ability of the model to produce more realistic image simulations. The lower panels V. and VI. show another example cutout from the same full scene where a single object is de-blended into two components by \textsc{SExtractor}.}
    \label{fig:detect-comp}
\end{figure}

We here show in Fig. \ref{fig:detect-comp} a comparison of image scenes generated from the two branches to illustrate differences in the \textsc{SExtractor} detection, as described in Sect. \ref{sc:ShearMeas}. We show identical cutouts from simulations with a doubled number density $\rho_n = 500\,$arcmin$^{-2}$ to showcase the potential differences in the detections. We clearly see in the zoom-ins II. and IV. that there are objects with for example spirals in the realistic simulations that lack adequate modelling of their morphologies, leading to potential additional deblending into multiple components also at bright magnitudes from the faint structures that are omitted when using Sérsic profiles. Moreover, stamps I. and III. show how detection differences might arise for faint objects, that is, when the surface brightness is not fully identical in the branches (the faint object in the upper-left corner is not above the detection threshold in the parametric branch). The lower panel zoom-ins depict an example where the de-blending produces more objects and a higher blending fraction in the realistic branch. This is of course dependent on the detection threshold, and de-blending settings in the \textsc{SExtractor} configuration file, which for our setup we chose as $\text{\texttt{DETECT\_THRESH}}=1.5$, $\text{\texttt{DETECT\_MINAREA}}=6$, $\text{\texttt{DEBLEND\_NTHRESH}}=32$, $\text{\texttt{DEBLEND\_MINCONT}}=0.005$, and $\text{\texttt{CLEAN\_PARAM}}=1.0$.

\clearpage
\textcolor{white}{XXXX}
\end{appendix}
\label{LastPage}

\end{document}